\documentclass[aps,prd,floatfix,superscriptaddress,balancelastpage,nofootinbib,amsmath,showpacs,twocolumn]{revtex4-1} 
\usepackage{graphicx,epstopdf,epsfig,bm,dcolumn,natbib,amsmath,amssymb,multirow,cancel,slashed,color}

\newcommand{\Eq}[1]{Eq.~\ref{#1}}  
 
\newcommand{\beq}{\begin{equation}} \newcommand{\eeq}{\end{equation}}
\newcommand{\bea}{\begin{eqnarray}} \newcommand{\eea}{\end{eqnarray}}
     \newcommand{\cL}{{\cal L}}   \newcommand{\cO}{{\cal O}}   
\newcommand{\pL}{\left(} \newcommand{\pR}{\right)}     \newcommand{\mL}{\left|} \newcommand{\mR}{\right|}

\def\lsim{\mathrel{\raise.3ex\hbox{$<$\kern-.75em\lower1ex\hbox{$\sim$}}}}
\def\gsim{\mathrel{\raise.3ex\hbox{$>$\kern-.75em\lower1ex\hbox{$\sim$}}}}
\newcommand\supsetsim{\mathrel{%
  \ooalign{\raise0.2ex\hbox{$\supset$}\cr\hidewidth\raise-0.8ex\hbox{\scalebox{0.9}{$\sim$}}\hidewidth\cr}}}

\begin{document}

\hspace{13cm} \parbox{5cm}{FERMILAB-PUB-15-359-A YITP-SB-15-29}~\\
\vspace{0.2cm}

\title{Everything You Always Wanted to Know About Dark Matter Elastic Scattering Through Higgs Loops But Were Afraid to Ask}

\author{Asher Berlin}
\affiliation{Department of Physics, University of Chicago, Chicago, IL 60637}

\author{Dan Hooper}
\affiliation{Center for Particle Astrophysics, Fermi National Accelerator Laboratory, Batavia, IL 60510}
\affiliation{Department of Astronomy and Astrophysics, University of Chicago, Chicago, IL 60637}

\author{Samuel D.~McDermott}
\affiliation{C.~N.~Yang Institute for Theoretical Physics, Stony Brook University, Stony Brook, NY 11794}

\begin{abstract}

We consider a complete list of simplified models in which Majorana dark matter particles annihilate at tree level to $hh$ or $hZ$ final states, and calculate the loop-induced elastic scattering cross section with nuclei in each case. Expressions for these annihilation and elastic scattering cross sections are provided, and can be easily applied to a variety of UV complete models. We identify several phenomenologically viable scenarios, including dark matter that annihilates through the $s$-channel exchange of a spin-zero mediator or through the $t$-channel exchange of a fermion. Although the elastic scattering cross sections predicted in this class of models are generally quite small, XENON1T and LZ should be sensitive to significant regions of this parameter space. Models in which the dark matter annihilates to $hh$ or $hZ$ can also generate a gamma-ray signal that is compatible with the excess observed from the Galactic Center.

\end{abstract}


\maketitle

\section{Introduction}

Weakly interacting massive particles (WIMPs) represent the most widely studied class of dark matter candidates. WIMPs are motivated in large part by the fact that stable particles with weak-scale masses and interactions are generally predicted to freeze-out in the early universe with a density similar to the measured dark matter abundance. Although neutralinos perhaps represent the most well-known example, viable WIMP candidates appear within a wide range of popular and well-motivated extensions of the Standard Model.

In order for a stable particle species that was in thermal equilibrium in the early universe to avoid being produced in excess of the measured dark matter abundance, its number density must be efficiently depleted through self-annihilations (or through coannihilations). For particles in the GeV-TeV mass range, this typically requires an annihilation cross section of approximately $\sigma v \sim 2 \times 10^{-26}$ cm$^3$/s, as evaluated at the temperature of thermal freeze-out~\cite{Steigman:2012nb,Drees:2015exa}. In many cases, the Feynman diagrams responsible for dark matter annihilation can be related via a crossing symmetry to diagrams for dark matter elastic scattering with Standard Model particles. This has motivated experiments designed to observe the elastic scattering of WIMPs with atomic nuclei. Over the past fifteen years or so, the sensitivity of the direct detection experimental program has increased exponentially with time, strengthening constraints on average by a factor of two every 15 months. As a consequence of this progress, many otherwise viable dark matter candidates have been ruled out, as well as many classes of diagrams for dark matter annihilation. 

There are many scenarios, however, in which the dark matter can avoid being overproduced in the early universe while yielding very low rates in direct detection experiments. Such possibilities include:
\begin{itemize}
\item{WIMPs that are depleted in the early universe through an efficient resonance~\cite{Griest:1990kh,Hooper:2013qjx}.}
\item{WIMPs that are depleted primarily through coannihilations with another particle species~\cite{Griest:1990kh,Edsjo:1997bg}.}
\item{Low-mass WIMPs ($m_{\chi}$\,$\lsim$\,$5$-10 GeV), which generate signals below the detection thresholds of most direct detection experiments.}
\item{WIMPs with elastic scattering amplitudes that are kinematically suppressed by powers of velocity or momentum transfer. Dark matter mediated by a pseudoscalar is one well known example~\cite{Boehm:2014hva,Berlin:2014tja,Izaguirre:2014vva,Ipek:2014gua}.}
\item{Dark matter that annihilates to metastable particles which subsequently decay to Standard Model states~\cite{Pospelov:2007mp,Pospelov:2008jd,ArkaniHamed:2008qn,Cholis:2008qq,Morrissey:2009ur,Meade:2009rb,Cohen:2010kn,Hooper:2012cw,Berlin:2014pya,Ko:2014gha,Abdullah:2014lla,Boehm:2014bia}. In such scenarios, the dark matter's elastic scattering cross section can be almost arbitrarily suppressed.}
\item{WIMPs whose annihilations take place only through interactions with an odd number of Standard Model states~\cite{Boehm:2014bia,Fonseca:2015rwa,McDermott:2014rqa}.}
\item{Scenarios in which the abundance of dark matter is depleted as a result of a non-standard cosmological history, such as moduli domination and decay~\cite{Fornengo:2002db,Gelmini:2006pq,Hooper:2013nia,Kane:2015jia,Patwardhan:2015kga} or a period of late-time inflation~\cite{Lyth:1995ka,Cohen:2008nb,Boeckel:2011yj,Boeckel:2009ej,Davoudiasl:2015vba}.}
\item{WIMPs that annihilate to Standard Model states which are not present in the nucleons that constitute the targets of direct detection experiments. In particular, the amplitudes for annihilations to leptons, heavy quarks, gauge bosons, and/or Higgs bosons lead to elastic scattering only at the loop-level.}
\end{itemize}

It is the last of these possibilities that we consider in this paper, in particular focusing on Majorana dark matter that annihilates to $hh$ or $hZ$ final states. Although the elastic scattering of dark matter through loop-induced processes has been discussed extensively in the literature~\cite{Drees:1992rr,Hisano:2010ct,Hisano:2011cs,Hisano:2015rsa,Hisano:2015bma,Ibarra:2015fqa,Hill:2014yka,Hill:2014yxa,Agrawal:2011ze,Ipek:2014gua}, and especially in the case of dark matter that annihilates to electroweak gauge bosons, such as wino-like neutralinos~\cite{Hisano:2010ct,Hisano:2011cs,Hisano:2015rsa}, final states including Higgs bosons have been comparatively unexplored. Such phenomenology can arise in a wide variety of theoretical frameworks, and dark matter candidates with these characteristics have become more attractive as direct detection constraints have become more stringent. Furthermore, such models can be motivated by the spectrum of the Galactic Center gamma-ray excess~\cite{Goodenough:2009gk,Hooper:2010mq,Boyarsky:2010dr,Hooper:2011ti,Abazajian:2012pn,Gordon:2013vta,Hooper:2013rwa,Huang:2013pda,Abazajian:2014fta,Daylan:2014rsa,Calore:2014xka,fermigc}, which appears to be compatible with dark matter annihilating to $hh$ or $hZ$ final states~\cite{Calore:2014xka}. 

In the following section, we describe the approach taken in this paper, which is based on a simplified model framework rather than any particular top-down theory. In Sec.~\ref{gcsec}, we briefly discuss the Galactic Center gamma-ray excess and use this observation to motivate dark matter candidates that annihilates to $hh$ or $hZ$ final states. In Sec.~\ref{formalism} we describe the formalism for dark matter elastic scattering that we use throughout this paper.  In Secs.~\ref{hh} and~\ref{hz}, we consider dark matter particles that annihilate to $hh$ (through the $s$-channel exchange of a spin-zero mediator or the $t$-channel exchange of a fermion) or $hZ$ (through the $s$-channel exchange of a spin-zero mediator, the $s$-channel exchange of a spin-one mediator, or the $t$-channel exchange of a fermion), respectively. We summarize our results and their implications in Sec.~\ref{summary}.





\section{A Simplified Model Approach}
\label{simplified}

In this paper, we do not adopt any particular ultraviolet (UV) complete theory, but instead follow a bottom-up approach that makes use of simplified models. More specifically, we have assembled an exhaustive list of tree-level diagrams for the annihilation of Majorana fermion dark matter particles to $hh$ or $hZ$ final states, and calculated the annihilation cross section and the (loop-induced) elastic scattering cross section with nuclei in each case. This a{pproach is similar to that taken in Ref.~\cite{Berlin:2014tja}, in which we considered dark matter that annihilates to Standard Model fermions. In the case of annihilations to $hh$ or $hZ$, however, it is not as obvious that this could be the dominant annihilation channel in a UV complete theory, or that the same couplings responsible for the annihilation cross section will provide the dominant contribution to the elastic scattering rate (see, for example, the discussion in Sec.~4.2 of Ref.~\cite{Agrawal:2014oha}). In this section, we will discuss this issue and argue that such phenomenology could arise within the context of physically realistic models.

We begin by considering a Majorana dark matter candidate, $\chi$, which we take to be an electroweak singlet. The following effective field theory describes its interactions with the Higgs $SU(2)$ doublet field, $H$:
\begin{align} \label{Higgs portal}
\cL &\supset \frac1 {2\Lambda} \bar \chi (\lambda_s + i \lambda_p \gamma_5) \chi \pL H^\dagger H - \frac{v^2}2 \pR 
\nonumber \\
&+  \frac1{2\Lambda^3} \bar \chi (\lambda_s' + i \lambda_p' \gamma_5) \chi \pL H^\dagger H - \frac{v^2}2 \pR^2 + \cdots,
\end{align}
where $\Lambda$ is a scale at which some electroweak-charged particle has been integrated out. 
%
%
After electroweak symmetry breaking, the dark matter has the following interactions with the physical Higgs boson, $h$:
\beq
\cL \supset \frac{v}{2\Lambda} \bar \chi (\lambda_s + i \lambda_p \gamma_5) \chi h +  \frac{v^2}{2\Lambda^3} \bar \chi (\lambda_s' + i \lambda_p' \gamma_5) \chi h^2 + \cdots
\label{Higgs portal2}
\eeq
If we were to assume that $\lambda_s \sim \lambda_p \sim \lambda_s' \sim \lambda_p' \sim \cO(1)$ and that $v^2 \lesssim \Lambda^2$, the first term would yield large elastic scattering and annihilation cross sections through the CP-even and CP-odd vertices, respectively~\cite{Berlin:2014tja}. Such a large elastic scattering cross section would be in conflict with existing direct detection constraints. If, however, the couplings in \Eq{Higgs portal} were to respect the following hierarchy:
\beq \label{par tuning}
\lambda_s,\lambda_p \ll \lambda_s',\lambda_p' \times \frac{v^2}{\Lambda^2},
\eeq
then the trilinear coupling between the dark matter and the Higgs will be subdominant, leading to rather different phenomenology.  The leading interactions with the physical Higgs particle will be a four-particle vertex in this case, leading to a low-velocity annihilation cross section (to $hh$) that scales as $\langle \sigma v \rangle \sim (\lambda_s' \lambda_p' /m_\chi)^2$, and an elastic scattering cross section that is induced by a diagram involving a Higgs loop.

In Eq.~\ref{Higgs portal}, one would also expect additional dimension-7 operators involving the $W^{\pm}$ and $Z$, for example coupling the dark matter bilinear to $\mL D^a_\mu H \mR^2$ and $W^a_{\mu \nu}W^{a\mu \nu}$. This would lead to terms in Eq.~\ref{Higgs portal2} that couple $\bar \chi \chi$ to $ZZ$, $hZ$ and $W^+ W^-$. As these interactions are expected to be generated at the same scale, $\Lambda$, the effective coupling to $hh$ could plausibly dominate without significant tunings or any large hierarchy between dimensionful scales.

\begin{figure*}[htb]
\includegraphics[width=0.6\textwidth]{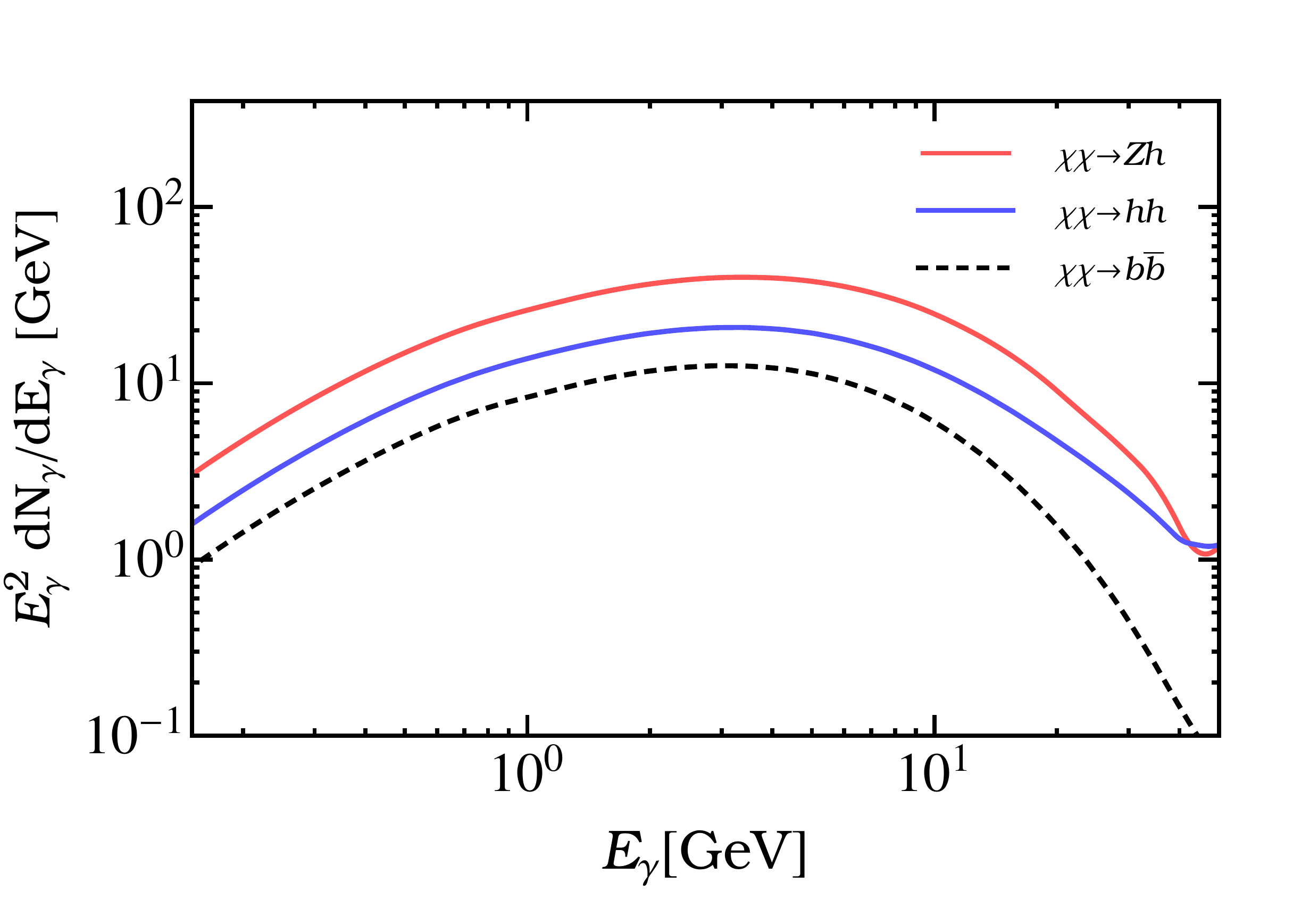} 
\caption{The differential gamma-ray spectra for dark matter, $\chi$, annihilating to $hh$, $Zh$ and $b\bar{b}$ with masses of $m_\chi = 130$ GeV, 110 GeV, and 65 GeV, respectively. It has been shown in the past that the Galactic Center gamma-ray excess can be well-fit by dark matter annihilating to $b\bar{b}$ or $hh$ for these values of the mass~\cite{Calore:2014xka}. Given the similarity of the shapes (and not necessarily normalization) of these spectra, we expect that annihilations to $Zh$ would provide a similarly good fit.}
\label{spectra}
\end{figure*}

As an explicit example, we consider a simple supersymmetric scenario in which the dark matter is a $\sim$100 GeV bino-like neutralino, $\chi^0_1$, and there exist two significantly heavier, quasi-degenerate higgsino-like neutralinos, $\chi^0_2$ and $\chi^0_3$.  Taking the large $\tan \beta$ limit, these collectively couple to the Higgs through the following interactions: $\cL \supset (\sqrt{2} g_1/4)\, h \bar{\chi}^0_1  \chi^0_2 + (\sqrt{2} g_1/4)\, h \bar{\chi}^0_1 i \gamma^5 \chi^0_3$. As there is no Higgs-bino-bino coupling, tree-level elastic scattering is suppressed. The gauge eigenstates mix, however, leading to a non-zero Higgs$-\chi^0_1-\chi^0_1$ coupling. 
After mass mixing, the relevant couplings are as follows:
\begin{eqnarray}
\cL &\supsetsim& (\sqrt{2} g_1/4) \, h \bar{\chi}^0_1  \chi^0_2 + (\sqrt{2} g_1/4)\, h \bar{\chi}^0_1 i \gamma^5 \chi^0_3 \nonumber \\
&+&    \frac{g_1^2 v (m_{\chi^0_1}-m_{\chi^0_{2,3}}  \sin (2 \beta ))}{2\sqrt{2}  m_{\chi^0_{2,3}} ^2}  \, h \bar \chi_1^0  \chi_1^0.
\end{eqnarray}

For a modest hierarchy between the bino and higgsino masses, $m_{\chi_2^0}\gg m_{\chi^0_1}$, the $h \bar \chi_1^0  \chi_1^0$ coupling can be efficiently suppressed (as can the $Z \bar \chi_1^0  \chi_1^0$ coupling). In this case, annihilations will proceed dominantly to $hh$, and the loop-level diagram for elastic scattering will dominate over that from from the tree-level process (details of this calculation will be presented in Sec.~\ref{hh-fermion}): 
\begin{eqnarray}
\frac{\sigma_{\rm loop}}{\sigma_{\rm tree}} &\sim& 1.3 \times 10^{-4} \, \bigg(\frac{m_{\chi^0_2}}{m_{\chi^0_1}}\bigg)^4 \, \ln \bigg(\frac{m_{\chi^0_2}}{m_{\chi^0_1}}\bigg) \\
&\sim& \mathcal{O}(3) \, \bigg(\frac{m_{\chi^0_2}}{1 \,{\rm TeV}}\bigg)^4 \, \bigg(\frac{100 \, {\rm GeV}}{m_{\chi^0_1}}\bigg)^4. \nonumber
\end{eqnarray}

Furthermore, it is well known that the supersymmetric parameter space contains significant regions, referred to as ``blind spots'', in which the tree-level elastic scattering amplitude is suppressed by accidental cancellations~\cite{Cheung:2012qy,Huang:2014xua,Cheung:2014lqa}; loop-level processes could easily dominate in such regions.

This example and others like it make it evident that UV complete and gauge invariant models can contain regions of parameter space in which dark matter annihilations proceed largely to pairs of Higgs bosons and elastic scattering is dominated by diagrams involving a Higgs loop. Furthermore, as direct detection experiments gain in sensitivity over the coming years, it is precisely these types of models (those with suppressed tree-level interactions) that will be the most difficult to test, and that will be the most interesting if no signal is detected. 

In light of these considerations, we proceed with an approach that is purely phenomenological in its philososphy. In the following sections, we explore simplified models that introduce a minimal number of interactions, requiring only that they respect Lorentz invariance. Of course, UV complete models will generally introduce additional interactions, but since we parametrize our models in a sufficiently general way, our results can be easily mapped onto a given UV-complete scenario. In this sense, the cross sections calculated and presented here correspond to a phenomenological lower bound on the elastic scattering rate of WIMPs which annihilate directly to Higgs bosons (in the absence of any strong cancellations).

\section{The Galactic Center Gamma-Ray Excess}
\label{gcsec}

An excess of gamma rays from the region surrounding the Galactic Center has been reported~\cite{Goodenough:2009gk,Hooper:2010mq,Boyarsky:2010dr,Hooper:2011ti,Abazajian:2012pn,Gordon:2013vta,Hooper:2013rwa,Huang:2013pda,Abazajian:2014fta,Daylan:2014rsa,Calore:2014xka,fermigc} . The morphology and spectrum of this excess are each in good agreement with that predicted from annihilating dark matter. More specifically, the best fits to the observed spectrum are found for dark matter that annihilates to quarks. For example, Ref.~\cite{Calore:2014xka} find $p$-values of 0.35, 0.37 and 0.22 for the fit of dark matter annihilating to $b\bar{b}$, $c\bar{c}$ and $q\bar{q}$, respectively. Dark matter annihilating to Higgs bosons also yields a fairly good fit to the data, with a $p$-values of 0.13~\cite{Calore:2014xka}. Annihilations to $W^+W^-$ or $ZZ$ provide a significantly poorer fit~\cite{Calore:2014xka} (see, however, Ref.~\cite{Caron:2015wda}). 

As far as we are aware, the Galactic Center gamma-ray excess has not been previously considered within the context of dark matter that annihilates to $hZ$. In Fig.~\ref{spectra}, we present the differential gamma-ray spectra from dark matter, $\chi$, annihilating to $hh$ and $Zh$, with masses of $m_\chi = 130$ GeV and 110 GeV, respectively. We note that for larger masses, the increased Lorentz boost between the dark matter and the $hh$ or $Zh$ frames would result in significant broadening of the photon spectrum, worsening the fit to the Galactic Center excess~\cite{Berlin:2014pya,Ko:2014gha,Abdullah:2014lla,Boehm:2014bia}. In Fig.~\ref{spectra}, we present these spectra alongside that generated by direct annihilations to bottom quarks for $m_\chi=65$ GeV, which is known to give an adequate fit to the excess~\cite{Calore:2014xka}. This figure illustrates that the spectral shape from the $hh$ and $hZ$ final states is very similar to that from direct annihilations to $b\bar{b}$. Furthermore, the greatest departures are found at the highest energies, where the spectrum of the excess is least well measured. 

\section{Direct Detection Formalism}
\label{formalism}

For each simplified model considered in this paper, we will derive the effective Lagrangian describing the interactions of Majorana dark matter with quarks and gluons (following the formalism of, for example, Refs.~\cite{Hisano:2010ct,Hisano:2011cs,Hisano:2015rsa,Hisano:2015bma,Ibarra:2015fqa,Hill:2014yka,Hill:2014yxa}):
\begin{eqnarray}
\label{lag}
\mathcal{L} &=&  \sum_{u,d,s}  \bigg[ f_q^{(0)} m_q \bar{\chi} \chi ~\bar{q} q 
+f_q^{(1)} \bar{\chi} \gamma^{\mu} \gamma^5 \chi ~\bar{q} \gamma_{\mu}\gamma^5 q \nonumber \\
&+&\frac{f_q^{(2)}}{m^2_{\chi}} \bar{\chi} i \partial^{\mu}i \partial^{\nu} \chi ~ \mathcal{O}^q_{\mu \nu} \bigg]
+f^{(0)}_g \bar{\chi} \chi~ G^a_{\mu \nu} G^{a \mu \nu}  \nonumber \\
&+&\frac{f^{(2)}_g}{m^2_{\chi}} \bar{\chi} i \partial^{\mu}i \partial^{\nu} \chi~ \mathcal{O}^g_{\mu \nu}, 
\end{eqnarray}
where $f_q^{(0)}$ is the scalar quark coupling, $f_q^{(1)}$ is the axial-vector quark coupling, $f^{(0)}_g$ is the scalar gluon coupling, and $f^{(2)}_q$ and $f^{(2)}_g$ are the quark and gluon couplings from twist-2 operators, respectively.  The twist-2 operators are the symmetric traceless parts of the energy-momentum tensors and are given by:
\begin{eqnarray}
\mathcal{O}^q_{\mu \nu} &=& \frac{1}{2}\bar{q} \bigg(iD_{\mu}\gamma_{\nu} + iD_{\nu}\gamma_{\mu} - \frac{1}{2} g_{\mu \nu}\, i \slashed{D} \bigg) q, \\
\mathcal{O}^g_{\mu \nu} &=& G_{\mu}^{a \rho} G^a_{\rho \nu} + \frac{1}{4} g_{\mu \nu} G^a_{\alpha \beta} G^{a \alpha \beta}.\nonumber
\end{eqnarray}
In Eq.~\ref{lag}, we have included all of the lowest-dimension terms that enable Majorana dark matter to scatter with nuclei without suppression by powers of momentum or velocity. Although the twist-2 gluon operator appears in this expression, its contribution to the effective nucleon coupling is subdominant compared to the scalar gluon coupling in all the cases that we consider throughout this study. We will neglect this contribution from here on.

\begin{figure*}[htb]
\includegraphics[width=0.5\textwidth]{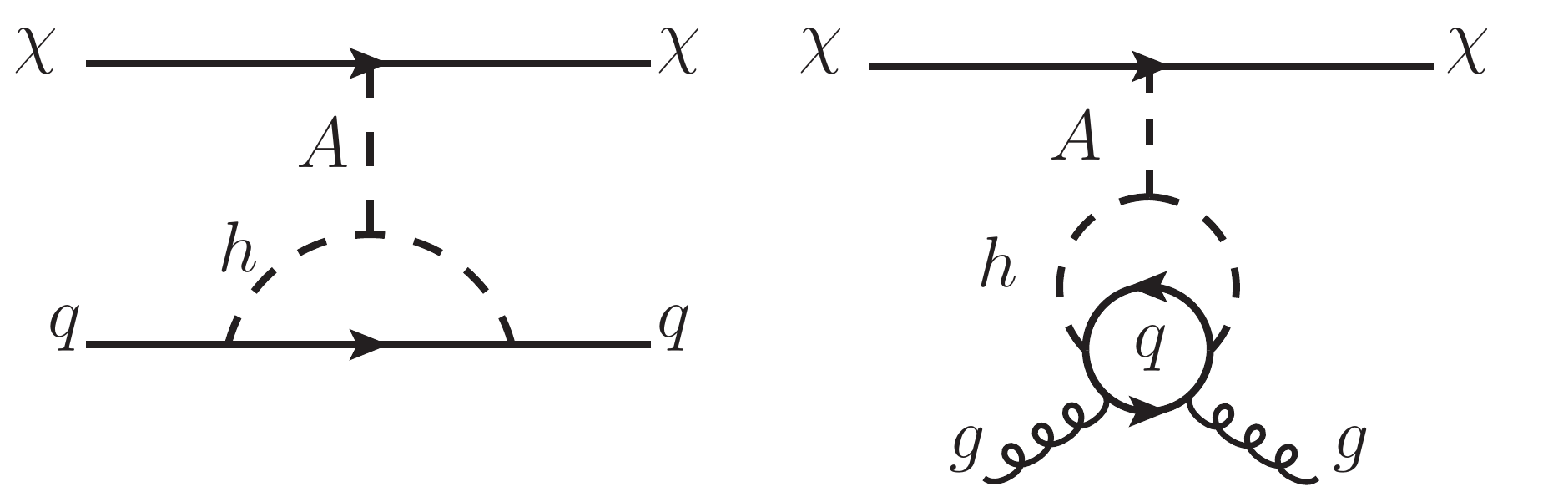} 
\caption{The diagrams for dark matter elastic scattering with nuclei corresponding to the case in which the dark matter annihilates to $hh$ through the $s$-channel exchange of a spin-zero mediator, $A$ (see Sec.~\ref{hh-scalar}).}
\label{diagram1}
\end{figure*}

From the Wilson coefficients $f_q^{(0)}$, $f_q^{(1)}$, $f_q^{(2)}$, and $f^{(0)}_g$, we can derive the spin-independent and spin-dependent scattering cross sections relevant for direct detection. The spin-independent cross section per nucleon is given by:
\begin{equation}
\sigma_{SI} = \frac{4}{\pi} \bigg(\frac{m_{\chi}m_n}{m_{\chi}+m_n}\bigg)^2 \, \left[\frac{Z f_p + (A-Z) f_n}{A} \right]^2, 
\end{equation}
where $m_n$ is the nucleon mass and $Z$ and $A$ are the atomic number and atomic mass of the target nucleus, respectively. The effective couplings to protons and neutrons are given by:
\begin{align}
\frac{f_{p,n}}{m_{p,n}} &= \sum_{q=u,d,s} \bigg[ f^{(p,n)}_{T_q} f^{(0)}_q  + \frac{3}{4}Q^{(p,n)}_{q}  f^{(2)}_q  \bigg] \nonumber \\
&-\frac{8\pi}{9 \alpha_s} f^{(0)}_g f^{(p,n)}_{TG}, \,\,\,\,\,\,\,\,\,\,
\end{align}
where $\alpha_s$ is the strong coupling. For the mass fraction of light quarks and gluons, we adopt~\cite{Young:2009zb,Oksuzian:2012rzb}:\footnote{Other studies have favored significantly larger values for the nucleon's strange quark content~\cite{Durr:2011mp,Junnarkar:2013ac}. The elastic scattering cross sections presented in this paper, however, are only impacted by this difference at the $\sim$10\% level.}
\begin{eqnarray}
f^{(p)}_{T_u} &=& 0.019, \,\,\,\, f^{(p)}_{T_d} = 0.027,  \,\,\,\,f^{(p)}_{T_s} = 0.009, \\
f^{(n)}_{T_u} &=& 0.013,\,\,\,\, f^{(n)}_{T_d} = 0.04,\,\, \,\,f^{(n)}_{T_s}= 0.009. \nonumber
\end{eqnarray}

From these quantities, we can calculate the gluon content, $f^{(p,n)}_{TG} = 1-\sum\limits_{q=u,d,s} f_{T_q}^{(p,n)}$. The quantities, $Q^{(p,n)}_q$, are the second moments of the proton and neutron parton distribution functions, which are scale-dependent. Since we will be performing our matching calculations at $m_h$ or $m_Z$, we will take the values as determined at these scales:
\begin{eqnarray} 
m_h:  Q^{(p)}_{u}&=&0.251, ~Q^{(p)}_{d}=0.155,~Q^{(p)}_{s}=0.055, \\
m_Z:  Q^{(p)}_{u}&=&0.254,~Q^{(p)}_{d}=0.156,~Q^{(p)}_{s}=0.054. \nonumber
\end{eqnarray}
Values for the neutron are obtained by exchanging the up and down quarks in the proton case~\cite{Martin:2009iq}. These quantities were derived using the parameterization and analysis of Ref.~\cite{Martin:2009iq}.

The spin-dependent cross section of the dark matter with a proton or neutron is given by:
\begin{equation}
\sigma_{SD} = \frac{12}{\pi} \bigg(\frac{m_{\chi}m_{p,n}}{m_{\chi}+m_{p,n}}\bigg)^2 \, \bigg[\sum_{q=u,d,s} f_q^{(1)} \Delta_q^{(p,n)} \bigg]^2,
\end{equation}
where the spin fractions of the proton and neutron are given by: $\Delta^{(p)}_u = 0.77$, $\Delta^{(p)}_d = -0.49$, $\Delta^{(p)}_s = -0.15$, $\Delta^{(n)}_u = -0.49$, $\Delta^{(n)}_d = 0.77$, and $\Delta^{(n)}_s = -0.15$~\cite{Cheng:2012qr}. 

A complete treatment of these interactions would involve the following procedure. First, at the electroweak scale, one integrates out the $Z$, $W^\pm$, $t$, and $h$ by matching onto a complete basis of non-renormalizable operators (including ones that are kinematically suppressed and that couple the dark matter to leptons). These operators are then evolved down to the nuclear scale of roughly $\mathcal{O}$(GeV), including threshold corrections from integrating out the $b$ and $c$ quarks and the $\tau$ lepton. Finally, once the operators are scaled down to $\sim 1$ GeV (at which the nucleon matrix elements are defined), one uses the nucleon matrix elements of the parton operators to generate the effective nucleon couplings. For the simplified models considered in this study, however, the operators of Eq.~$\ref{lag}$ are generated at leading order at the level of weak scale matching without strong cancellations, and thus the mixing of operators via the renormalization group flow is expected to only introduce a mild correction at the level of a few percent (see for example Ref.~\cite{Hisano:2015bma}). With this in mind, we derive the elastic scattering cross sections in this paper by performing a simple matching calculation at the weak scale (taken as $m_h$ or $m_Z$).


\section{Dark Matter Annihilating to $hh$}
\label{hh}

Majorana fermion dark matter can annihilate to a pair of Higgs bosons through two types of diagrams: the $s$-channel exchange of a spin-zero mediator, or the $t$-channel exchange of a fermionic mediator. In this section, we will calculate the cross sections for annihilation and elastic scattering in each of these scenarios. 

\subsection{$s$-channel spin-zero mediator}
\label{hh-scalar}

We begin with the case of dark matter that annihilates to a pair of Higgs bosons through the $s$-channel exchange of a spin-zero state. The Lagrangian for the simplified model describing this process is given as follows:
\begin{equation}
\mathcal{L} \supset A~\bar{\chi} \left( \lambda_s + \lambda_p i \gamma^5 \right) \chi + \mu_h A h^2,
\end{equation}
where $A$ denotes the mediator and $\chi$ is the (Majorana fermion) dark matter candidate. The annihilation cross section for this process is given by:
\begin{equation}
\sigma v = \frac{\mu _h^2}{2 \pi  }\left(1-\frac{4 m_h^2}{s}\right)^{1/2}~\frac{ \lambda_s^2 \left(1-\frac{4 m_{\chi }^2}{s} \right) + \lambda_p^2}{\left(s-m_A^2\right)^2 + m_A^2 \Gamma_A^2},
\end{equation}
where, $m_A$ is the mass of the mediator and $\sqrt{s}$ is the center-of-mass energy. The width of the mediator, $\Gamma_A$, potentially receives contributions from decays to $hh$ and $\chi \chi$, given by:
\begin{eqnarray}
\Gamma(A\rightarrow hh) &=& \frac{\mu_h^2}{8 \pi m^2_A} (m^2_A-4m^2_h)^{1/2}, \\
\Gamma(A\rightarrow \chi \chi) &=& \frac{(m^2_A-4m^2_{\chi})^{1/2}}{16 \pi m^2_A}  [(m^2_A-4 m^2_{\chi})\lambda^2_s+m^2_A\lambda^2_p]. \nonumber
\end{eqnarray}

%
%

\begin{figure*}[t]
\includegraphics[width=0.49\textwidth]{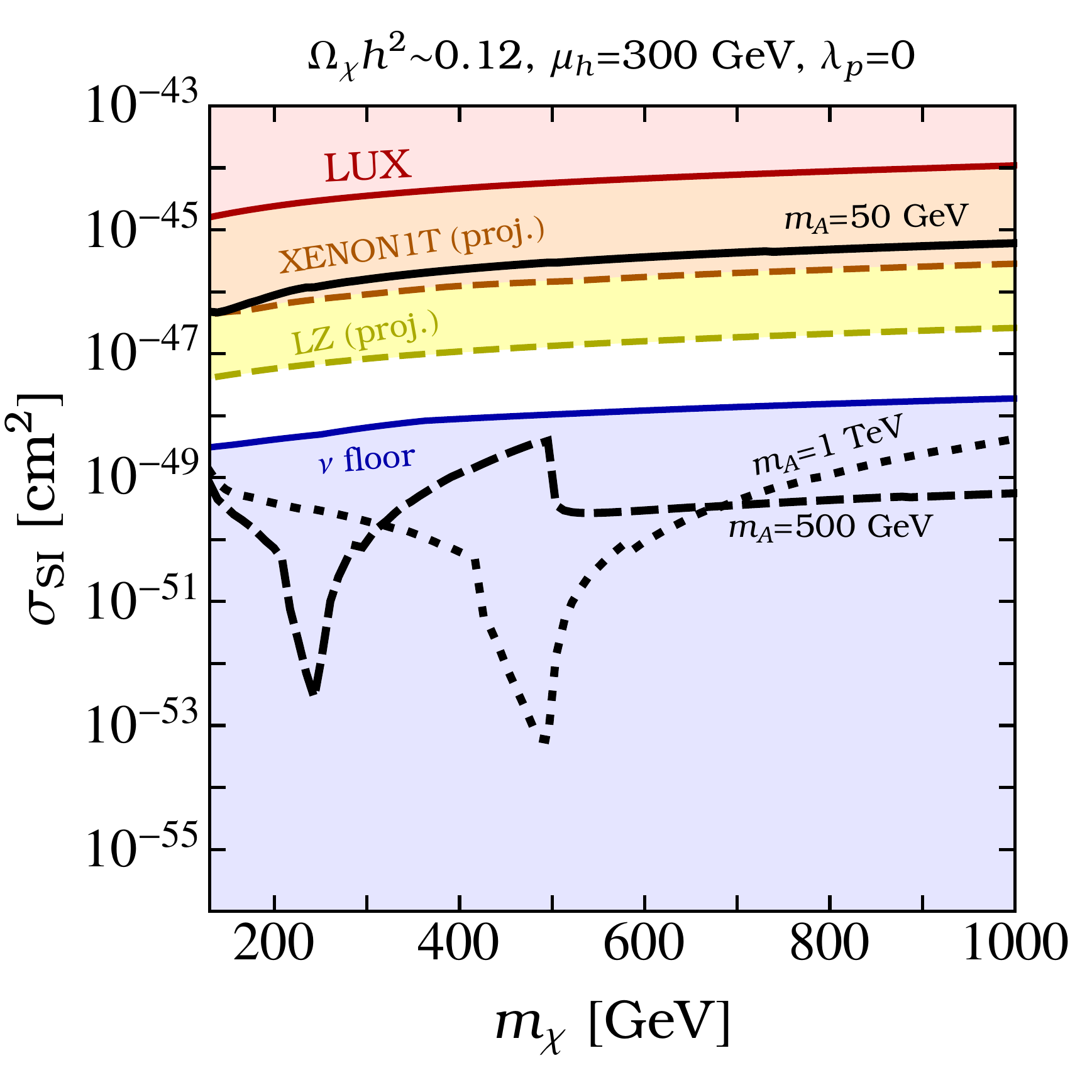} 
\includegraphics[width=0.49\textwidth]{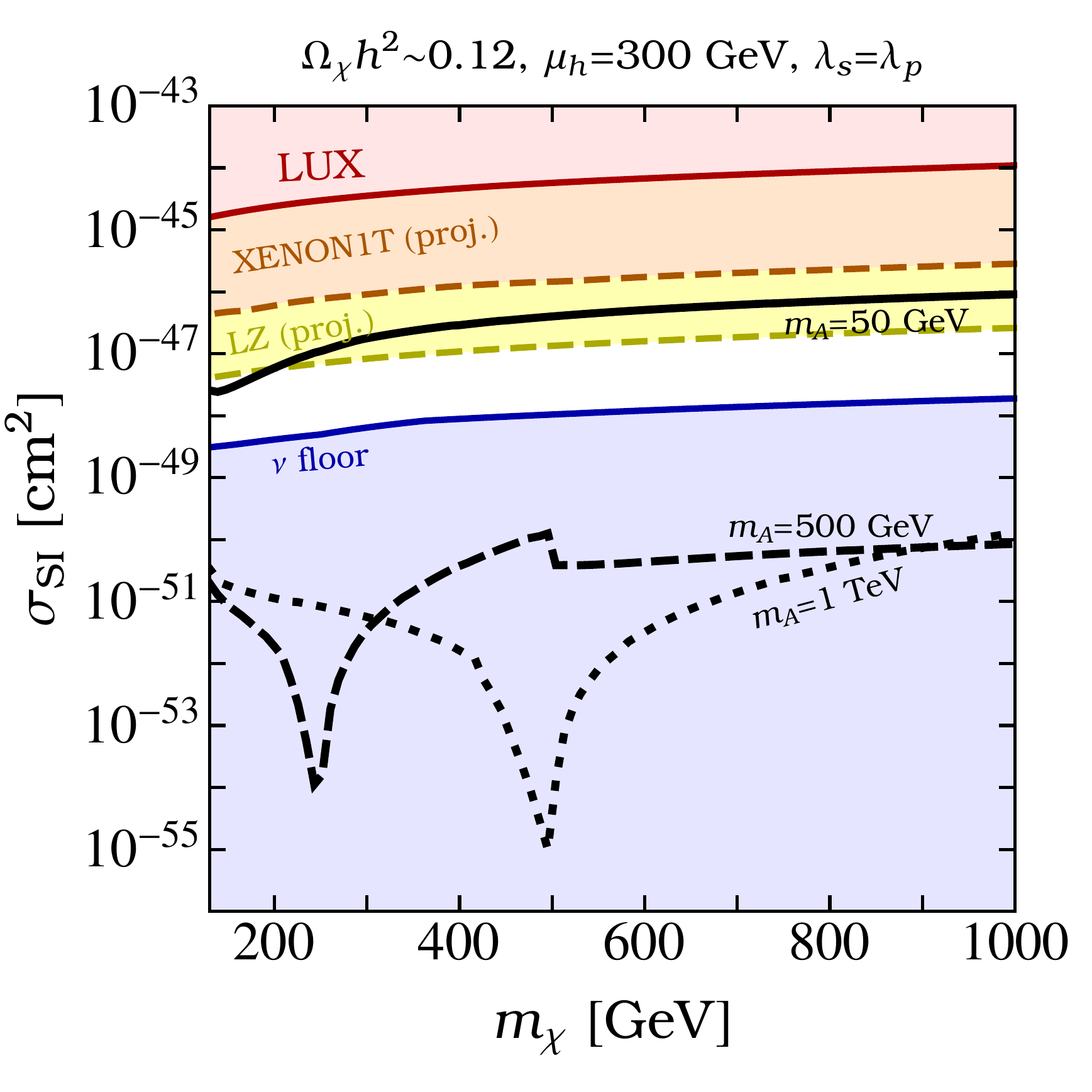}
\caption{For Majorana dark matter annihilating to $hh$ through a spin-zero mediator, we plot the spin-independent elastic scattering cross section per nucleon as a function of the dark matter mass, for various values of the mediator mass, $m_A$. In each case, the couplings, $\lambda_{s,p}$, have been set to obtain the desired thermal relic abundance, $\Omega_{\chi} h^2 =0.12$. In the left and right frames, we assume a purely scalar interaction ($\lambda_p=0$) and a mixed scalar-pseudoscalar interaction ($\lambda_s=\lambda_p$), respectively. We do not show results for $\lambda_s=0$, as the elastic scattering cross section is suppressed by powers of velocity in this case. The red regions in the upper portion of the frames are currently excluded by the LUX direct detection experiment~\cite{Akerib:2013tjd}, whereas in the blue regions we predict a cross section that is below the neutrino floor, making it difficult for dark matter to be detected by any planned direct detection experiment. Also shown are the regions within the projected reach of XENON1T (orange) and LZ (yellow)~\cite{Cushman:2013zza}.}
\label{case1}
\end{figure*}

\begin{figure*}[h]
\includegraphics[width=0.49\textwidth]{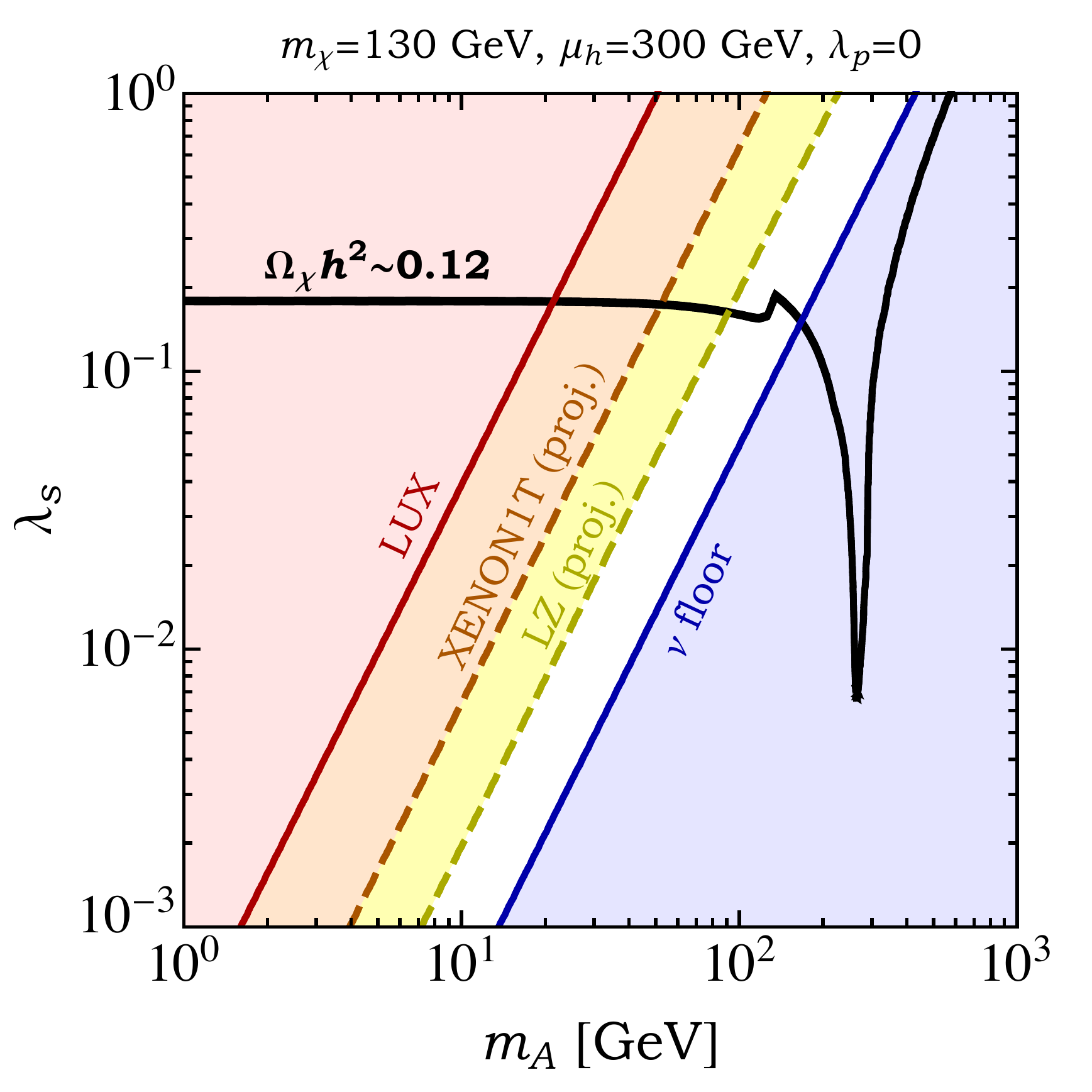}
\includegraphics[width=0.49\textwidth]{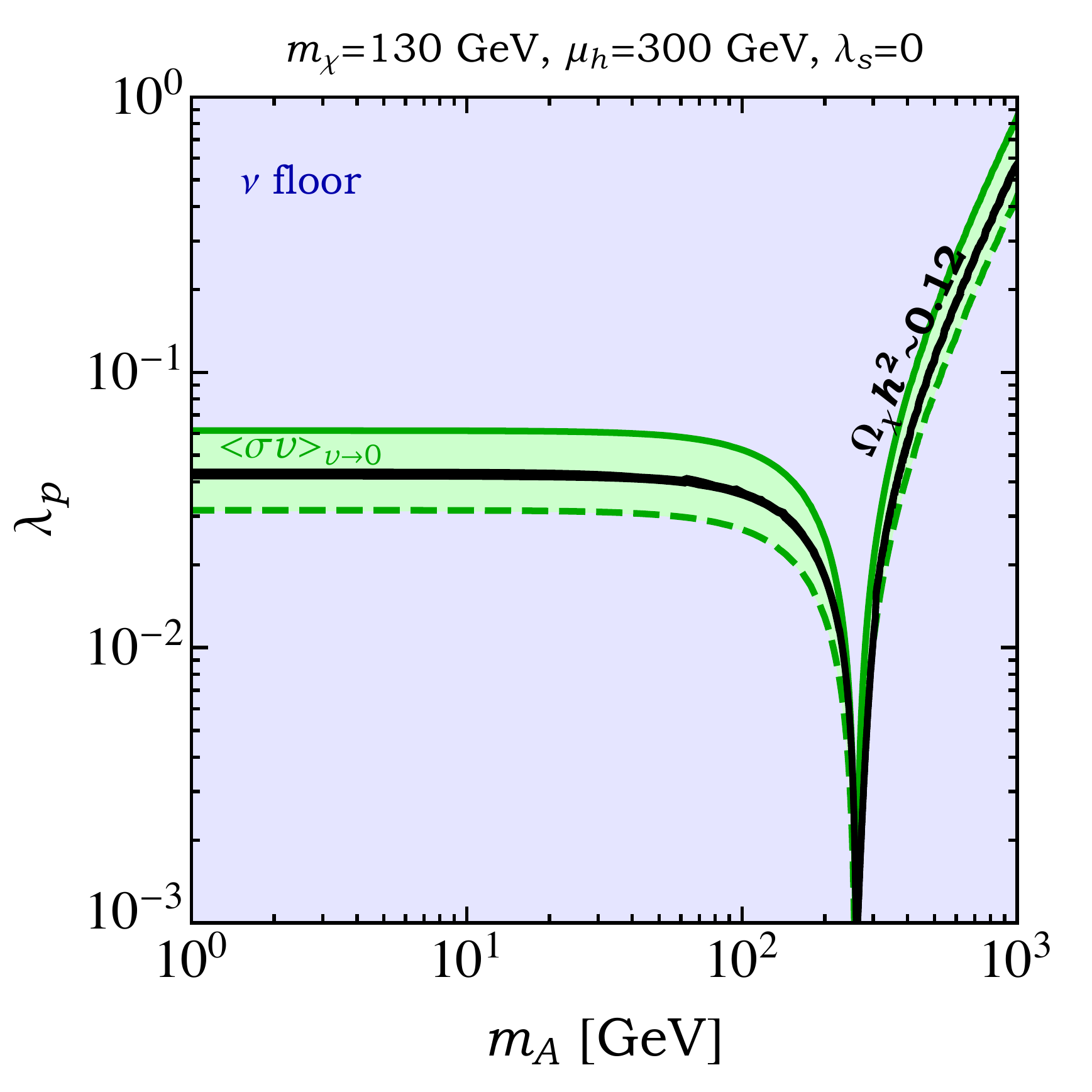} 
\includegraphics[width=0.49\textwidth]{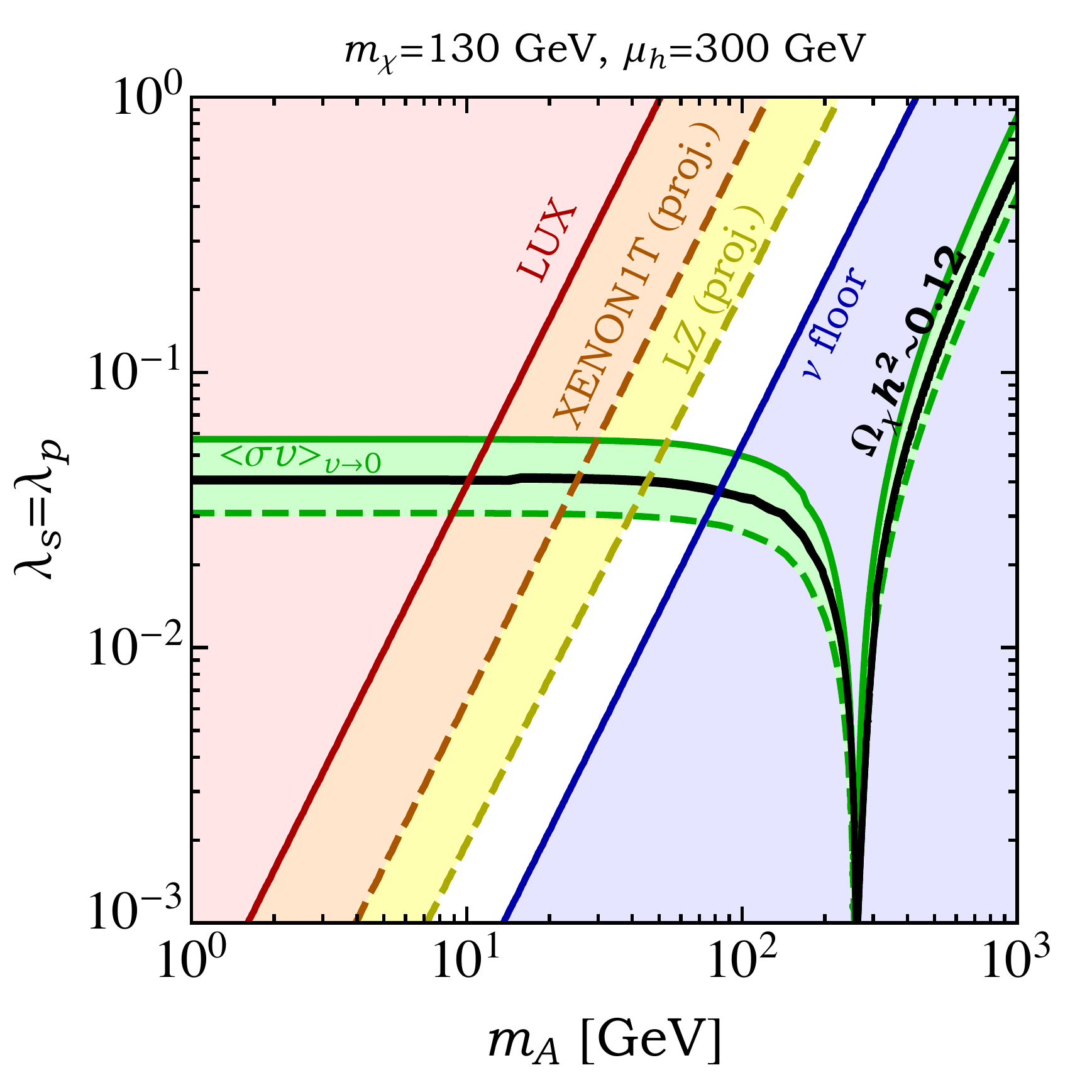} 
\caption{For Majorana dark matter with a mass of 130 GeV, annihilating to $hh$ through a spin-zero mediator, we plot the parameter space in which the predicted thermal relic abundance matches the observed dark matter density (solid black line) and the region in which the Galactic Center gamma-ray excess can be generated (green shaded band). The upper boundary of this region corresponds to the upper limit on the annihilation cross section from Fermi's observations of dwarf galaxies~\cite{Ackermann:2015zua} (green solid). The red regions in the upper-left portion of the first two frames are currently excluded by the LUX direct detection experiment~\cite{Akerib:2013tjd}, whereas in the blue regions we predict a cross section that is below the neutrino floor, making it difficult for dark matter to be detected by any planned direct detection experiment. Also shown are the regions within the projected reach of XENON1T (orange) and LZ (yellow)~\cite{Cushman:2013zza}.}
\label{case1pheno}
\end{figure*}

Far from any resonances or particle production thresholds, the low-velocity annihilation cross section can be well-approximated by the first two terms of its Taylor series expansion, which in this case yields:
\begin{widetext}
\begin{eqnarray}
\label{eq:SChAnnih}
\sigma v &\approx& \frac{\mu _h^2}{2 \pi} \left(1-\frac{m_h^2}{m_{\chi }^2}\right)^{1/2} \frac{\lambda_p^2}{\left(4 m_{\chi }^2-m_A^2\right)^2} \nonumber \\
&-& v^2 \frac{\mu _h^2}{16 \pi} \left(1-\frac{m_h^2}{m_{\chi }^2}\right)^{-1/2} \frac{ \lambda_p^2 \left(16 m_{\chi }^4-20 m_h^2 m_{\chi }^2+m_A^2 m_h^2\right)-2 \lambda_s^2 \left(4 m_{\chi }^2-m_A^2\right) \left(m_{\chi }^2-m_h^2\right) }{m_{\chi }^2 \left(4 m_{\chi }^2-m_A^2\right)^3}.
\end{eqnarray}

Dark matter scattering with nuclei is dominated by the loop-diagrams shown in Fig.~\ref{diagram1}. These diagrams induce the following effective couplings between the dark matter and Standard Model quarks and gluons (see Sec.~\ref{formalism}):
%
\begin{eqnarray}
f_q^{(0)} &=&  \frac{3 \lambda_s \mu_h y_s^2}{16 \pi ^2 m_A^2 m_h^2}, \\
f_g^{(0)} &=& \sum\limits_q \frac{\alpha_s \lambda_s y_s^2 \mu_h}{32 \pi^3 m_A^2} \int_0^1 dx_1 dx_2~ \Bigg( \frac{1}{\Delta \xi^2} (x_1-1) x_1 x_2 + \frac{1}{\Delta \xi^3} (x_1-1)^2 x_1^2 (x_2-1) x_2 \nonumber \\
&+& \frac{1}{3\Delta^2 \xi^2} m_q^2 \Big[3 (x_1-1) x_1+2\Big] (x_2-1) x_2 + \frac{1}{\Delta^2 \xi^3} \frac{1}{3} m_q^2 (1-2 x_1)^2 (x_1-1) x_1 (x_2-1)^2 x_2 \Bigg), \nonumber
\end{eqnarray}
\end{widetext}
where $\Delta \equiv x_2 m_h^2 + (x_2-1)m_q^2$ and $\xi \equiv x_2 + x_1(x_1-1)(1-x_2)$. Throughout this paper, $y_s=-m_q/v$, where $v=246$ GeV is the vacuum expectation value of the Standard Model Higgs boson.


\begin{figure*}[htb]
\includegraphics[width=1\textwidth]{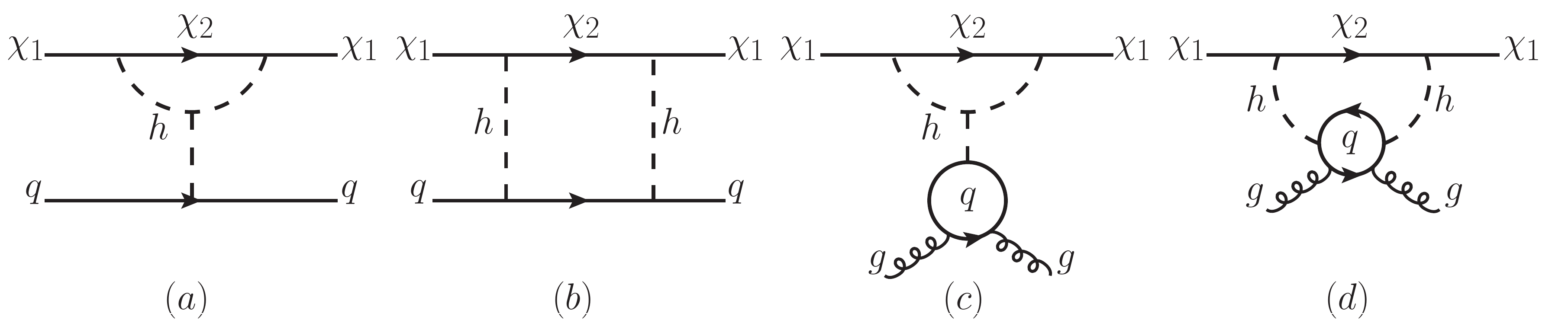} 
\caption{The diagrams for dark matter elastic scattering with nuclei corresponding to the case in which the dark matter annihilates to $hh$ through the $t$-channel exchange of a fermonic mediator mediator, $\chi_2$ (see Sec.~\ref{hh-fermion}). Crossed diagrams are not shown.}
\label{diagram2}
\end{figure*}

In Fig.~\ref{case1}, we plot the spin-independent elastic scattering cross section per nucleon as a function of the dark matter mass, for three values of the mediator mass, $m_A$. In each case, the couplings, $\lambda_{s,p}$, have been set to obtain the desired thermal relic abundance, $\Omega_{\chi} h^2 =0.12$. We compare these results to the current constraints from the LUX experiment~\cite{Akerib:2013tjd} and the projected sensitivity of XENON1T (orange) and LUX-ZEPLIN (LZ, yellow)~\cite{Cushman:2013zza}. The blue regions fall below the so-called ``neutrino floor'' (see, for example, Ref.~\cite{Ruppin:2014bra}), where it unlikely that any direct detection experiment will be sensitive in the foreseeable future.

In Fig.~\ref{case1pheno}, we plot some of the phenomenological features of this model for the case of $m_{\chi}=130$ GeV and $\mu_h=300$ GeV. For scenarios with a non-negligible value of $\lambda_p$, the annihilation is dominantly $s$-wave (as seen in the first line of Eq.~\ref{eq:SChAnnih}), which allows this model to generate the observed normalization of the Galactic Center gamma-ray excess (this criteria for this is taken to be that the low-velocity annihilation cross section is greater than $10^{-26}$ cm$^3$/s~\cite{Calore:2014xka} and below the upper limit from Fermi's observations of dwarf galaxies~\cite{Ackermann:2015zua}, appropriately rescaled for the case of annihilations to $hh$). In contrast, the annihilation cross section is velocity suppressed in the case of a purely scalar coupling ($\lambda_p=0$), leading to no appreciable indirect detection signals. In none of these cases are the current constraints from direct detection experiments very restrictive, and unless $m_A$ is quite small, such scenarios predict elastic scattering cross sections that are below the neutrino floor. Furthermore, if the coupling between the dark matter and the mediator is purely pseudoscalar ($\lambda_s=0$), no detectable elastic scattering cross section is generated.

\bigskip

\subsection{$t$-channel fermionic mediator}
\label{hh-fermion}

In this subsection, we consider Majorana dark matter, $\chi_1$, that annihilates to $hh$ through the $t$-channel exchange of an additional fermion, $\chi_2$:
\begin{equation}
\mathcal{L} \supset h~\bar{\chi}_1 \left( \lambda_s + \lambda_p i \gamma^5 \right) \chi_2. 
\end{equation}
The expanded low-velocity annihilation cross section in this case is given by:

\begin{widetext}
\begin{eqnarray}
\label{sigmavHHTch}
\sigma v  &\simeq& \frac{m_{\chi_2}^2 \lambda _p^2 \lambda _s^2}{2 \pi  \left(-m_h^2+m_{\chi_1}^2+m_{\chi_2}^2\right)^2}   \bigg(1-\frac{m_h^2}{m_{\chi_1}^2} \bigg)^{1/2} \nonumber
\nonumber \\
& +& \frac{v^2}{96 \pi  m_{\chi_1}^2 \left(-m_h^2+m_{\chi_1}^2+m_{\chi_2}^2\right)^4}  \bigg(1-\frac{m_h^2}{m_{\chi_1}^2} \bigg)^{-1/2}
\Bigg\{3 \bigg[\left(\lambda _p^2-\lambda _s^2\right)^2 m_{\chi_1}^2 -\left(\lambda _p^4-4 \lambda _s^2 \lambda _p^2+\lambda _s^4\right)m_h^2 \bigg] m_{\chi_2}^6
\nonumber \\
&+6& \left(\lambda _p^4-\lambda _s^4\right) \left(m_h^2-m_{\chi_1}^2\right) m_{\chi_1} m_{\chi_2}^5+3 \left(m_h^2-m_{\chi_1}^2\right) \bigg[2 m_h^2 \left(\lambda _p^4-4 \lambda _s^2 \lambda _p^2+\lambda _s^4\right)-3 m_{\chi_1}^2 \left(\lambda _p^4-6 \lambda _s^2 \lambda _p^2+\lambda _s^4\right)\bigg] m_{\chi_2}^4
\nonumber \\
&-8& \left(\lambda _p^4-\lambda _s^4\right) m_{\chi_1} \left(m_h^2-m_{\chi_1}^2\right)^2  m_{\chi_2}^3-\left(m_h^2-m_{\chi_1}^2\right)^2 \bigg[3 \left(\lambda _p^4-4 \lambda _s^2 \lambda _p^2+\lambda _s^4\right) m_h^2+m_{\chi_1}^2 \left(-5 \lambda _p^4+18 \lambda _s^2 \lambda _p^2-5 \lambda _s^4\right)\bigg] m_{\chi_2}^2
\nonumber \\
&+2& \left(\lambda _p^4-\lambda _s^4\right) m_{\chi_1} \left(m_h^2-m_{\chi_1}^2\right)^3 m_{\chi_2}+3 \left(\lambda _p^2+\lambda _s^2\right)^2m_{\chi_1}^2 \left(m_{\chi_1}^2-m_h^2\right)^3 \Bigg\} ~.
\end{eqnarray}

Scattering with quarks and gluons occurs in this simplified model at leading order through the diagrams presented in Fig.~\ref{diagram2}. Similar to Sec.~\ref{hh-scalar}, the full set of diagrams involves Higgs-vertex corrections, but now also includes the Higgs box diagram of Fig.~\ref{diagram2}(b). Although this latter process is suppressed by the Yukawa couplings of the light valence quarks and is subdominant, for completeness we include it in our numerical analysis. The quark and gluon Wilson coefficients are given by:
%
\begin{eqnarray}
f_q^{(0)} &=& \frac{y_s/m_q}{32 \pi ^2 v} \int_0^1 dx ~ \frac{3 (x-1) (\lambda^{(-)} m_{\chi_2}-\lambda^{(+)} m_{\chi_1} (x-1))}{ \Delta_1} \\
&+&\frac{y_s^2}{16 \pi^2}\int_0^1 dx_2~\int_0^{1-x_2} dx_1~\frac{(1-x_1-x_2)}{\Delta_2^2}\Bigg\{ \frac{1}{2} \Big[\lambda^{(+)} m_{\chi_1} \big(x_1 (3 x_2-4)-2 x_2+4\big)-2 \lambda^{(-)} m_{\chi_2} (x_2-2)\Big] \nonumber \\
&+&\frac{m_{\chi_1}^2}{\Delta_2} x_1^2 x_2\Big[ \lambda^{(-)} m_{\chi_2}-\lambda^{(+)} m_{\chi_1} (x_1-1)\Big]\Bigg\}, \nonumber 
\end{eqnarray}
\begin{eqnarray}
f_q^{(2)} &=& \frac{y_s^2}{4 \pi ^2 }\int_0^1 dx_2~\int_0^{1-x_2} dx_1 ~\frac{(1-x_1-x_2)}{\Delta_2^2} \Bigg\{ \frac{m_{\chi_1}^2}{\Delta_2}x_1^2 x_2 \Big[ (\lambda^{(-)} m_{\chi_2}-\lambda^{(+)} m_{\chi_1} (x_1-1))\Big]+\frac{1}{2}\lambda^{(+)} m_{\chi_1} x_1 x_2 \Bigg\}, \nonumber
\end{eqnarray}
\begin{eqnarray}
f_g^{(0)}&=& \sum_q \frac{-\alpha_s y_s}{384 \pi^3 v m_q} \int_0^1 dx ~ \frac{3 (x-1) (\lambda^{(-)} m_{\chi_2}-\lambda^{(+)} m_{\chi_1} (x-1))}{ \Delta_1} \nonumber \\
&+& \sum\limits_q \frac{\alpha_s y_s^2}{128 \pi^3} \int_0^1 dx_1 dx_2 dx_3 \Bigg( \frac{1}{\Delta_3^2 \xi^2} (x_1-1) x_1 (x_2-1) x_3^2 (\lambda^{(+)} m_{\chi_1}+\lambda^{(-)} m_{\chi_2}) \nonumber \\
&+& \frac{1}{ \Delta_3^2 \xi^3}(x_1-1) x_1 (x_2-1) x_3^2\bigg[ \lambda^{(+)} m_{\chi_1} \Big(x_1^2 (x_3-1)-x_1 x_3+x_1+x_2 x_3\Big)+\lambda^{(-)} m_{\chi_2} (x_1-1) x_1 (x_3-1)\bigg]
\nonumber \\
&+& \frac{3}{2 \Delta_3^2 \xi^4} \lambda^{(+)} m_{\chi_1} (x_1-1)^2 x_1^2 (x_2-1) x_2 (x_3-1) x_3^3 \nonumber \\
&+& \frac{2}{3 \Delta_3^3 \xi^2} m_q^2 \big(3 (x_1-1) x_1+2\big) (x_2-1) (x_3-1) x_3^2 (\lambda^{(+)} m_{\chi_1}+\lambda^{(-)}m_{\chi_2})
\nonumber \\
&+& \frac{2}{3 \Delta_3^3 \xi^3} m_q^2 (x_2-1) (x_3-1) x_3^2 \bigg[\lambda^{(+)} m_{\chi_1} \Big(x_3 \big[ (x_1-1) x_1 \left((1-2 x_1)^2+3x_2\right)+2 x_2\big]-(1-2 x_1)^2 (x_1-1) x_1\Big)
\nonumber \\
&+&\lambda^{(-)} m_{\chi_2} (x_1-1) x_1 (1-2 x_1)^2 (x_3-1)\bigg]
\nonumber \\
&-&\frac{1}{ \Delta_3^3 \xi^4} m_{\chi_1} (x_1-1) x_1 (x_2-1) x_2 (x_3-1) x_3^3 \bigg[\lambda^{(+)} m_{\chi_1}^2 (x_1-1) x_1 x_2 x_3+\lambda^{(-)} m_{\chi_1} m_{\chi_2} (x_1-1) x_1 x_2 x_3
\nonumber \\  
& -&\lambda^{(+)} m_q^2 (1-2 x_1)^2 (x_3-1)\bigg]
-\frac{1}{ \Delta_3^3 \xi^5} \lambda^{(+)} m_{\chi_1}^3 (x_1-1)^2 x_1^2 (x_2-1) x_2^3 (x_3-1) x_3^5
\nonumber \\
&-& \frac{1}{\Delta_3^4 \xi^4} m_{\chi_1}^2 m_q^2 (1-2 x_1)^2 (x_1-1) x_1 (x_2-1) x_2^2 (x_3-1)^2 x_3^4 (\lambda^{(+)} m_{\chi_1}+\lambda^{(-)} m_{\chi_2})
\nonumber \\
&-&\frac{1}{ \Delta_3^4 \xi^5} \lambda^{(+)} m_{\chi_1}^3 m_q^2 (1-2 x_1)^2 (x_1-1) x_1 (x_2-1) x_2^3 (x_3-1)^2 x_3^5 \Bigg),
\nonumber 
\end{eqnarray}
where we define the following:
\begin{eqnarray}
\lambda^{(\pm)} &\equiv& \lambda_s^2 \pm \lambda_p^2, \nonumber \\
\Delta_1 &\equiv& x(x-1)m_{\chi_1}^2 + x m_{\chi_2}^2 + (1-x) m_h^2,  \nonumber \\
\Delta_2 &\equiv& m_{\chi_1}^2 (x_1-1) x_1+m_{\chi_2}^2 x_1+m_h^2 (1-x_1-x_2), \nonumber \\
\Delta_3 &\equiv& \frac{x_2^2 x_3^2}{\xi} m_{\chi_1}^2 + x_2 x_3 (m_{\chi_1}^2 - m_{\chi_2}^2 ) + (x_2-1) x_3 m_h^2 + (x_3-1)m_q^2, \nonumber \\
\xi &\equiv& x_1 (x_1-1)(1-x_3)-x_3. \nonumber 
\end{eqnarray}
\end{widetext}

In the expression for $f_q^{(0)}$ above, the vertex diagram of Fig.~\ref{diagram2}(a) contributes to the first line, whereas the box diagrams of Fig.~\ref{diagram2}(b) contributes to the second and third lines. The first line of the expression for $f_g^{(0)}$ corresponds to diagram~\ref{diagram2}(c), while the rest of $f_g^{(0)}$ results from diagram~\ref{diagram2}(d). Although the box diagram in Fig.~\ref{diagram2}(b) contributes negligibly to the overall scattering cross section, its 2-loop gluon equivalent, shown in Fig.~\ref{diagram2}(d), has a significant effect due to the large top quark Yukawa coupling. 

\begin{figure*}[t]
\includegraphics[width=0.49\textwidth]{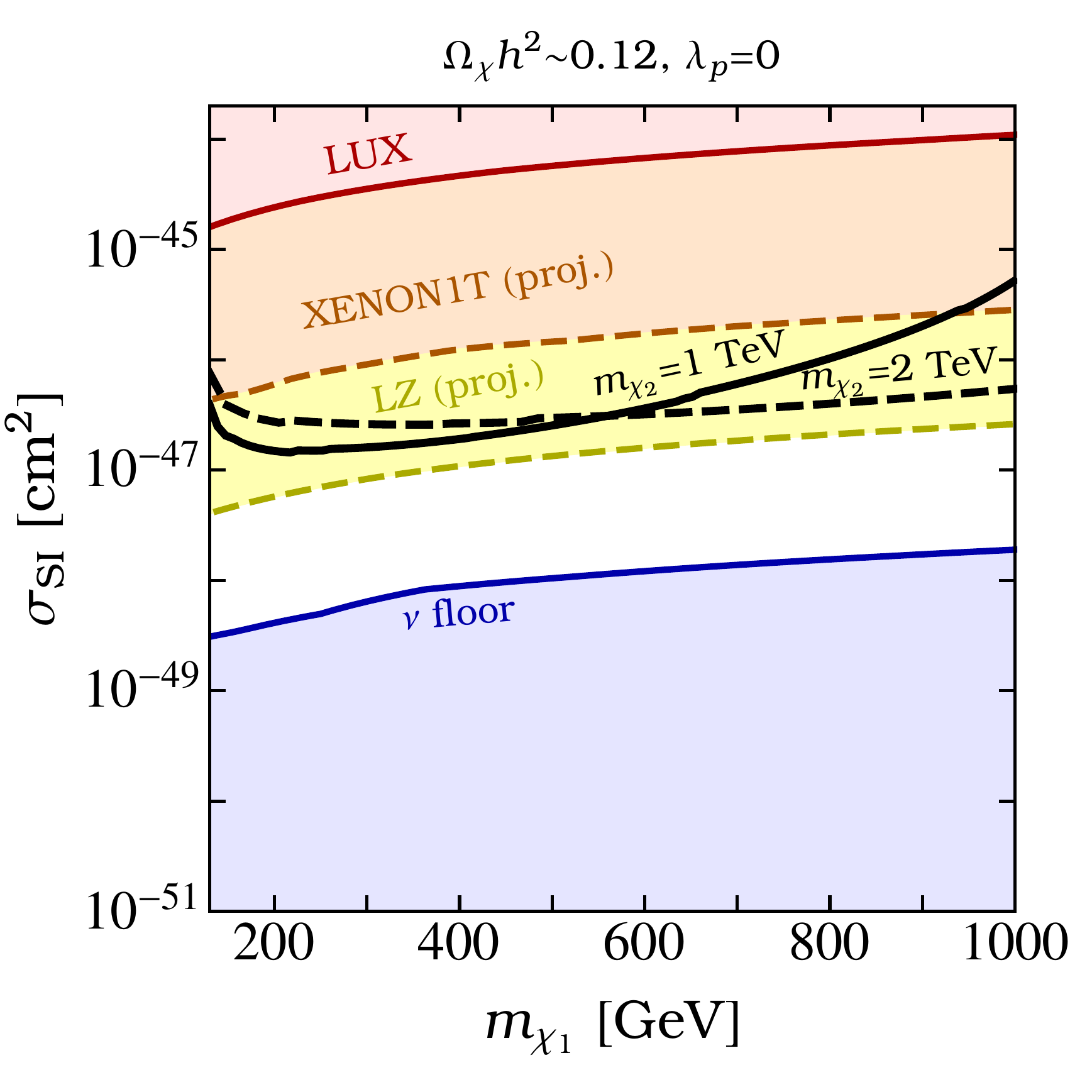} 
\includegraphics[width=0.49\textwidth]{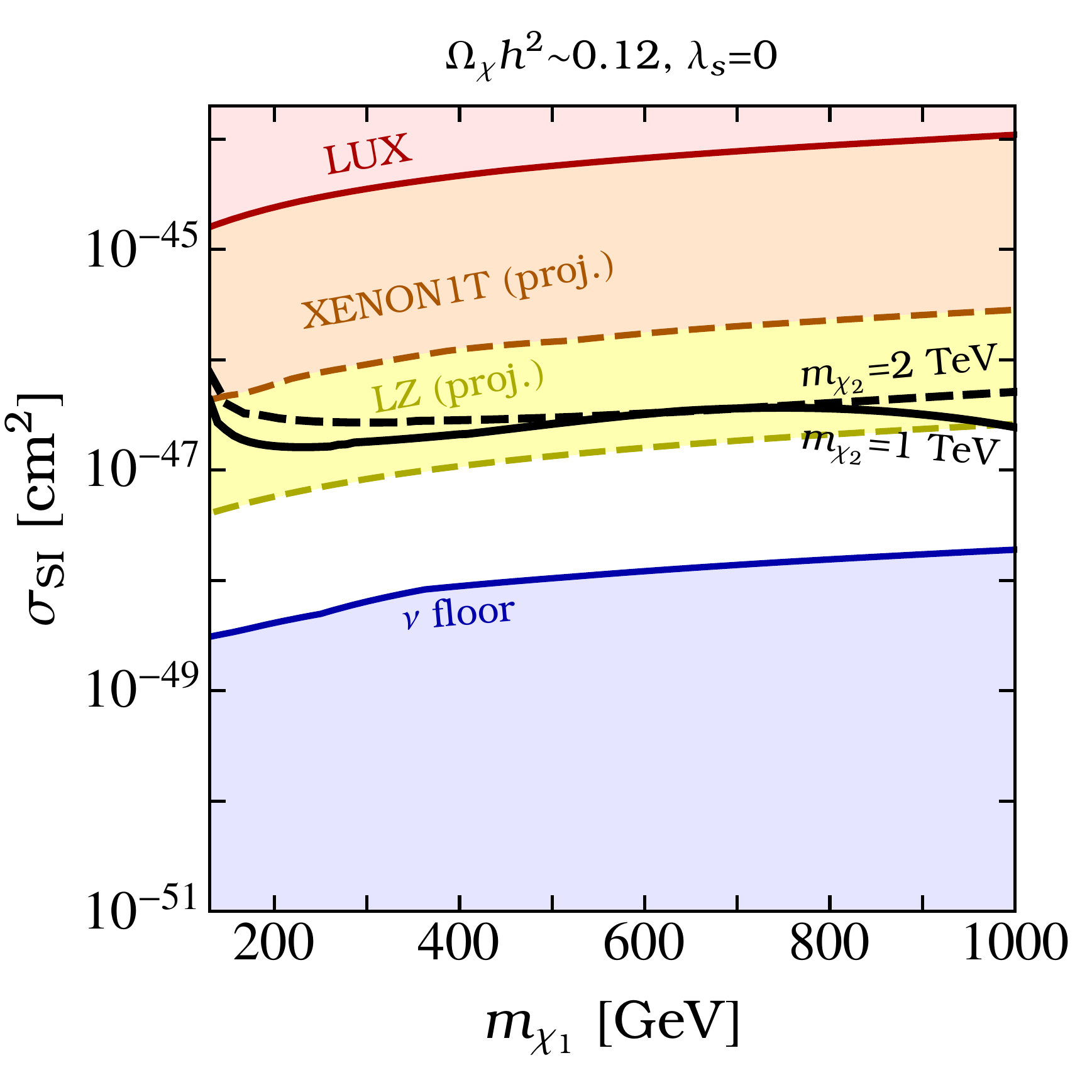}
\includegraphics[width=0.49\textwidth]{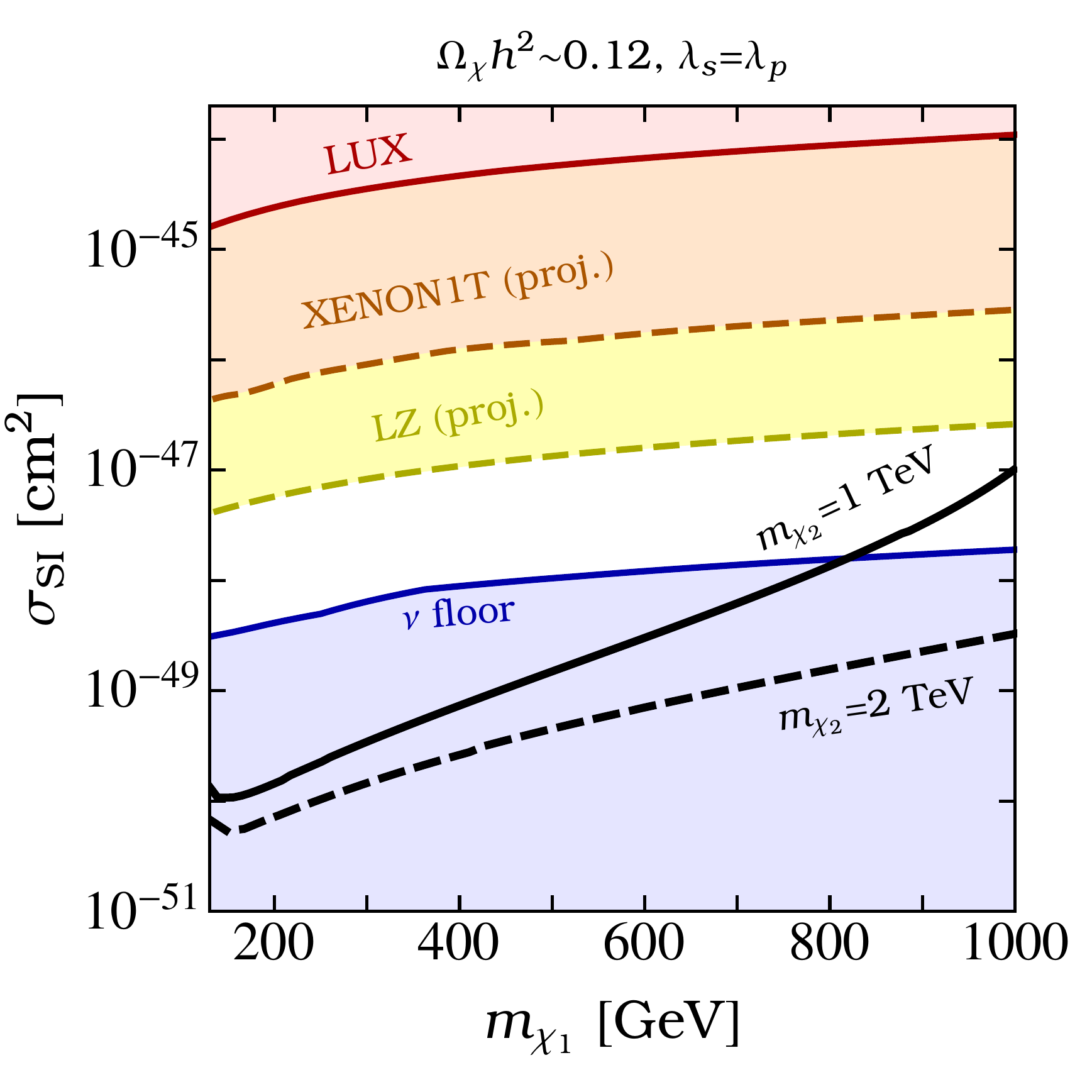}
\caption{For Majorana dark matter, $\chi_1$ interacting through a fermionic mediator, $\chi_2$, we plot the spin-independent elastic scattering cross section per nucleon as a function of the dark matter mass, for two values of the mediator mass. In each case, the couplings, $\lambda_{s,p}$, have been set to obtain the desired thermal relic abundance, $\Omega_{\chi} h^2 =0.12$. In the upper left, upper right and lower frames, we assume a purely scalar interaction ($\lambda_p=0$), a purely pseudoscalar interaction ($\lambda_s=0$)  and a mixed scalar-pseudoscalar interaction ($\lambda_s=\lambda_p$), respectively. The red regions in the upper portion of the frames are currently excluded by the LUX direct detection experiment~\cite{Akerib:2013tjd}, whereas in the blue regions we predict a cross section that is below the neutrino floor, making it difficult for dark matter to be detected by any planned direct detection experiment. Also shown are the regions within the projected reach of XENON1T (orange) and LZ (yellow)~\cite{Cushman:2013zza}. \\ }
\label{case2}
\end{figure*}
\begin{figure*}[htb]
\includegraphics[width=0.49\textwidth]{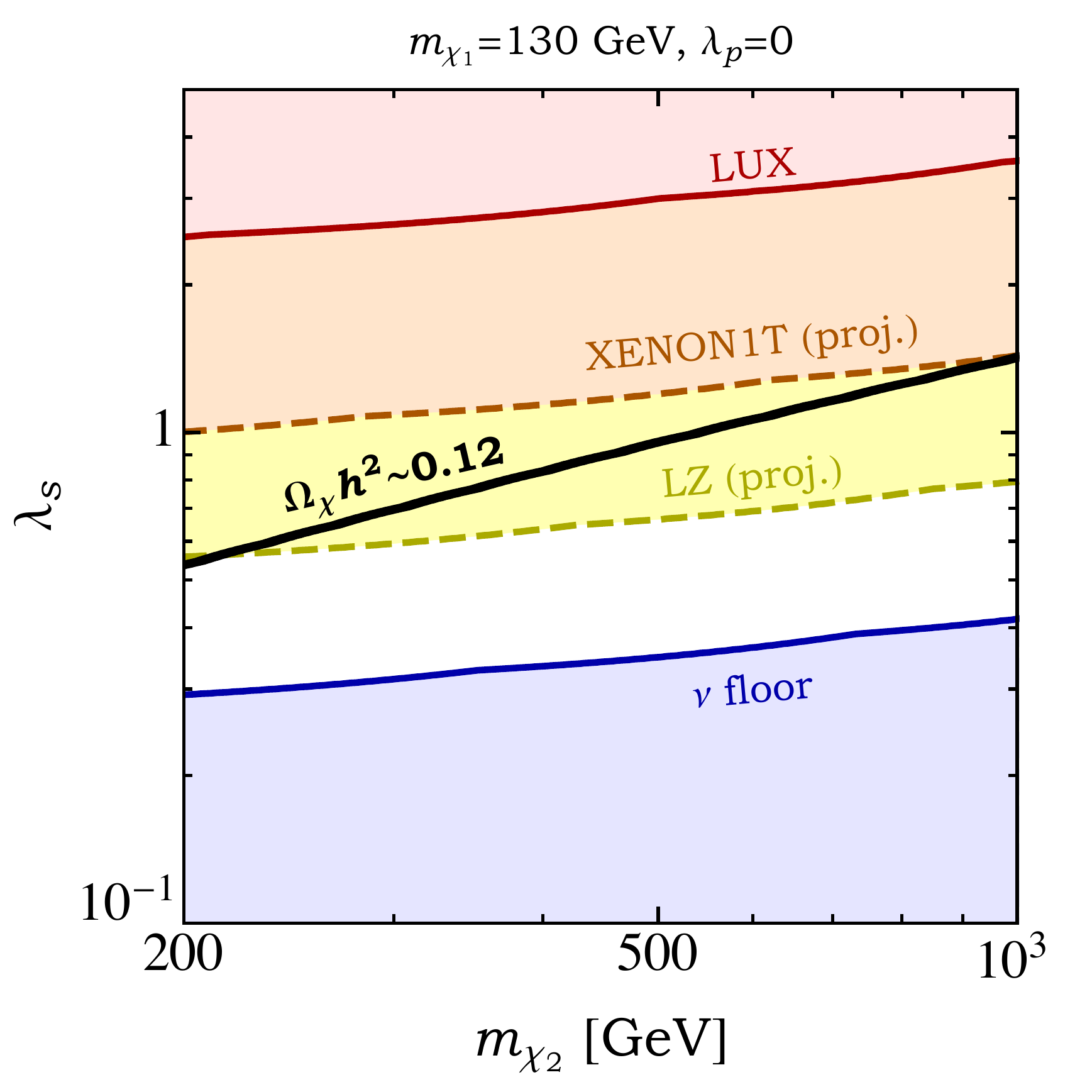}
\includegraphics[width=0.49\textwidth]{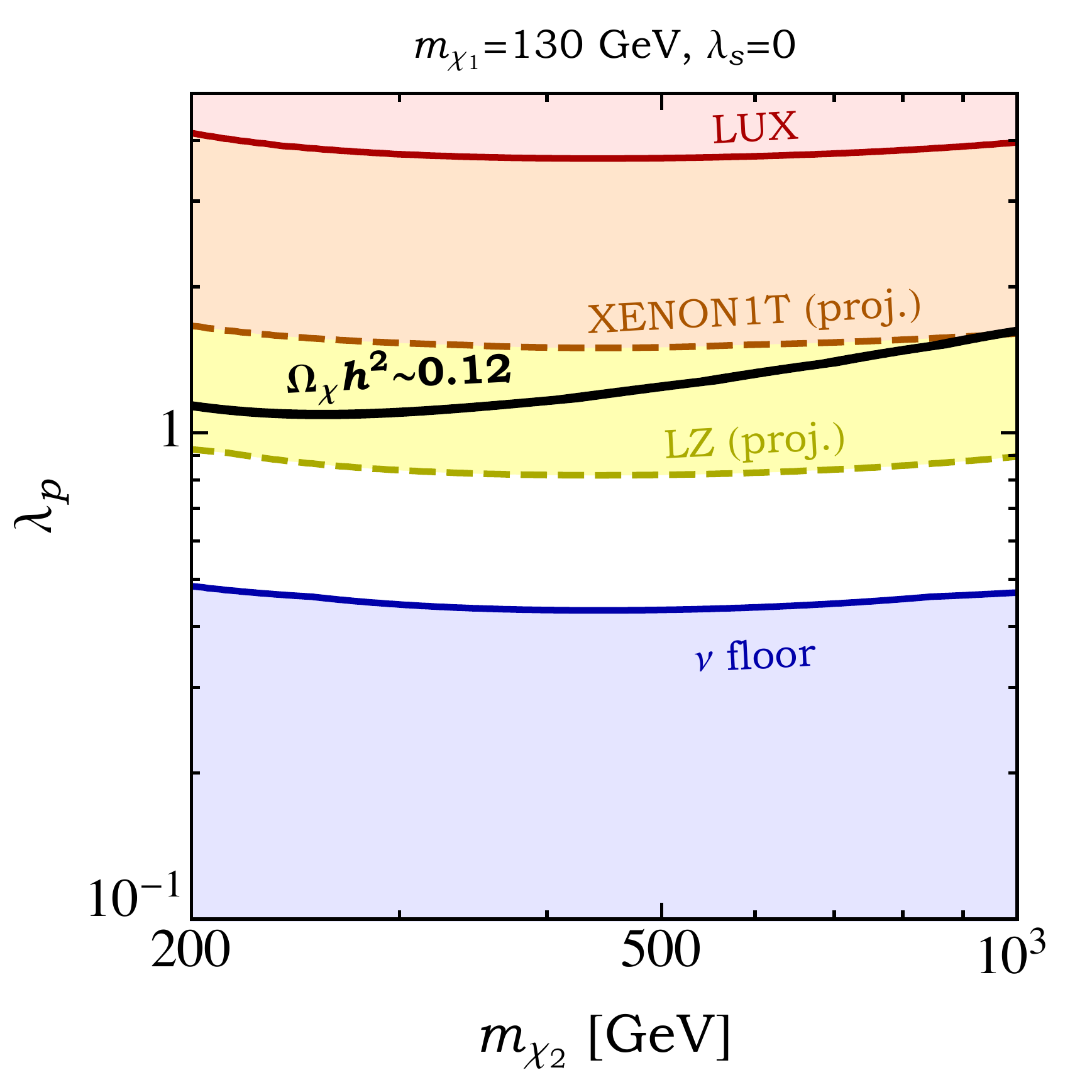} 
\includegraphics[width=0.49\textwidth]{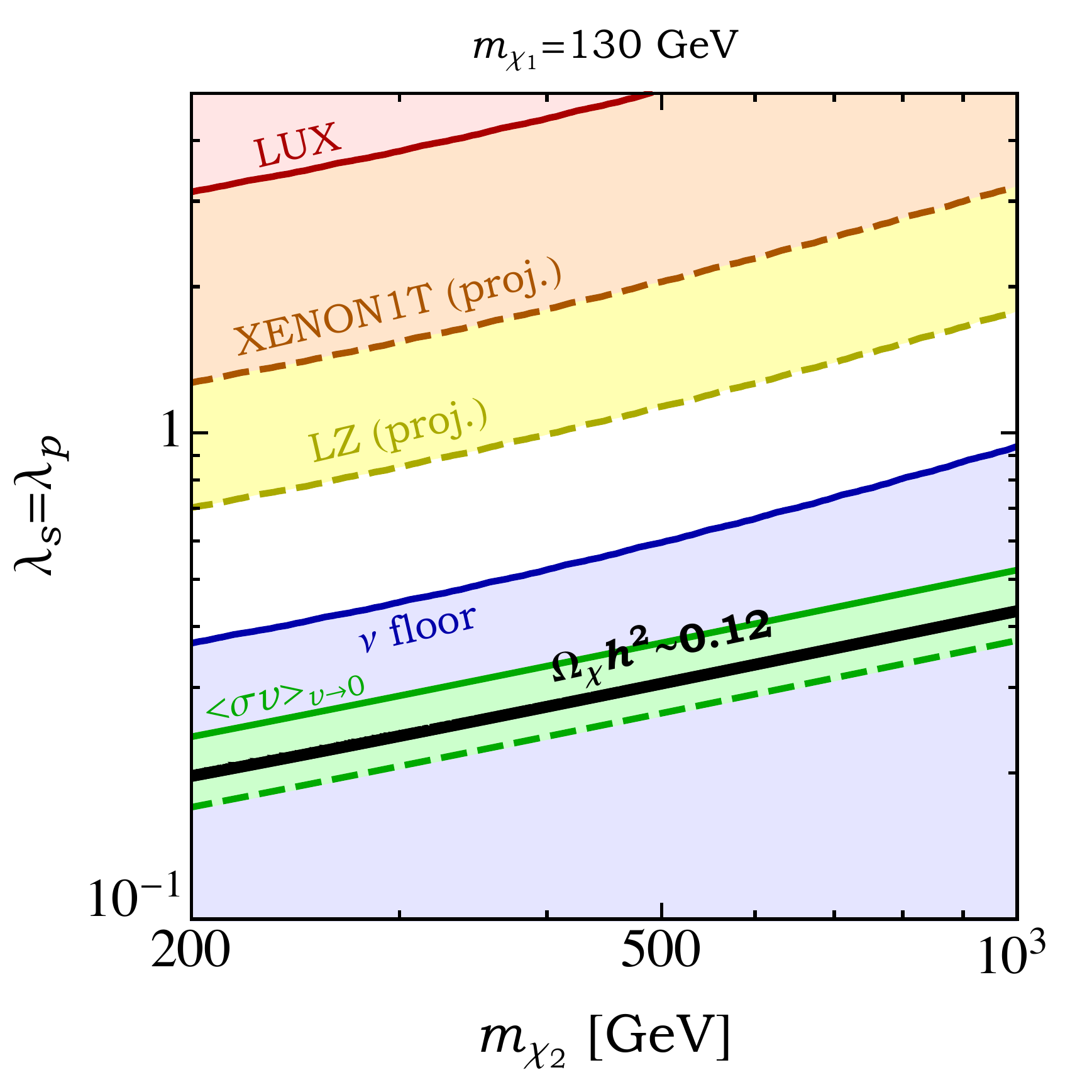} 
\includegraphics[width=0.49\textwidth]{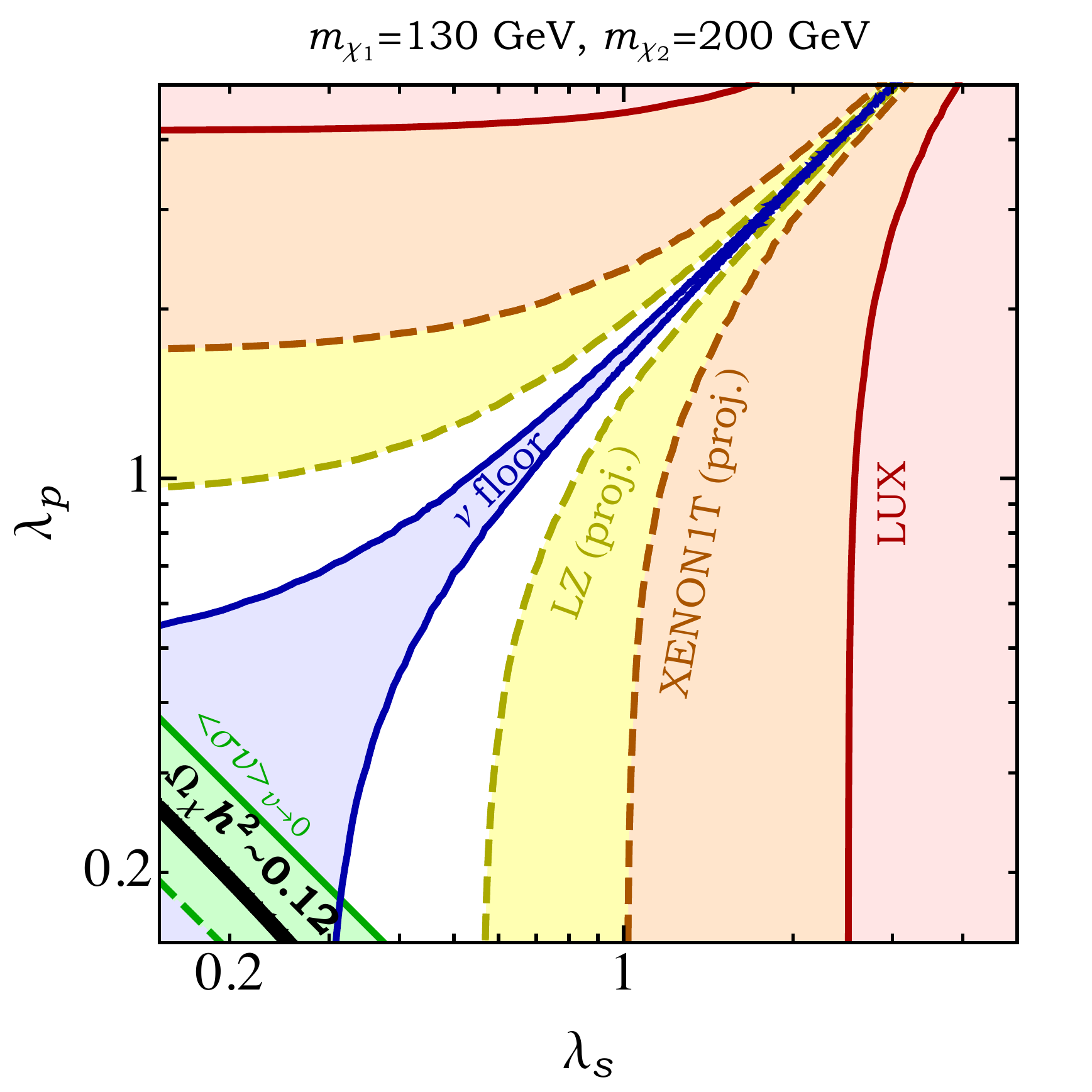} 
\caption{For Majorana dark matter with a mass of 130 GeV, annihilating to $hh$ through a fermionic mediator, we plot the parameter space in which the predicted thermal relic abundance matches the observed dark matter density (solid black line) and the region in which the Galactic Center gamma-ray excess can be generated (green shaded band). The upper boundary of this region corresponds to the upper limit on the annihilation cross section from Fermi's observations of dwarf galaxies~\cite{Ackermann:2015zua} (green solid). The red regions are currently excluded by the LUX direct detection experiment~\cite{Akerib:2013tjd}, whereas in the blue regions we predict a cross section that is below the neutrino floor, making it difficult for dark matter to be detected by any planned direct detection experiment. Also shown are the regions within the projected reach of XENON1T (orange) and LZ (yellow)~\cite{Cushman:2013zza}.}
\label{case2pheno}
\end{figure*}

In Fig.~\ref{case2}, we plot the spin-independent elastic scattering cross section per nucleon as a function of the dark matter mass, for two values of the mediator mass, $m_{\chi_2}$. In each case, the couplings, $\lambda_{s,p}$, have been set to obtain the desired thermal relic abundance, $\Omega_{\chi} h^2 =0.12$. Again, we compare these results to the current and projected constraints from direct detection experiments. 

In Fig.~\ref{case2pheno}, we plot the annihilation and scattering behavior for this model. As seen in Eq.~\ref{sigmavHHTch}, $s$-wave annihilation can only occur if both the scalar ($\lambda_s$) and pseudoscalar ($\lambda_p$) couplings are non-zero, and the annihilation rate is maximized in the case of $\lambda_s=\lambda_p$.  In this case, as seen in the bottom-left and bottom-right frames of Fig.~\ref{case2pheno}, the required couplings are around $\mathcal{O}(0.2)$, resulting in an elastic scattering cross section significantly below the neutrino floor. However, in the situation where either $\lambda_s$ or $\lambda_p$ vanishes, the annihilation amplitude is purely $p$-wave and necessitates $\mathcal{O}(1)$ couplings in order to generate an acceptable thermal relic abundance. As a result, these larger couplings lead to scattering rates that might be within the reach of next generation experiments such as LZ.

Among other features, the bottom-right frame of Fig.~\ref{case2pheno} illustrates that special relations among the couplings $\lambda_s$ and $\lambda_p$ can lead to ``blind-spots'' in which there is an accidental cancelation between the contributions from the diagrams in Fig.~\ref{diagram2}. Even for $\mathcal{O}(1)$ values of the dark matter Yukawa couplings, this cancellation can persist if the combination of $\lambda_s$ and $\lambda_p$ is tuned at the few percent level. 



\begin{widetext}
\section{Dark Matter Annihilating to $hZ$}
\label{hz}

Majorana fermion dark matter can annihilate to a Higgs boson and a $Z$ boson through three types of tree-level diagrams: the $s$-channel exchange of a spin-zero mediator, the $s$-channel exchange of a spin-one mediator, or the $t$-channel exchange of a fermionic mediator. In this section, we will present the cross sections for annihilation and elastic scattering in each of these scenarios.

\subsection{$s$-channel spin-zero mediator}
\label{Zh-scalar}

Next, we turn our attention to Majorana fermion dark matter, $\chi$, that annihilates to $hZ$ through the $s$-channel exchange of spin-zero mediator, $A$. The relevant interactions of this simplified model are described by the following Lagrangian:
\begin{equation}
\mathcal{L} \supset A~\bar{\chi} \left( \lambda_s + \lambda_p i \gamma^5 \right) \chi + g_A Z^\mu \left( A \partial_\mu h - h \partial_\mu A \right)~.
\end{equation}

The annihilation cross section in this case is given by:
%
\begin{align}
\sigma v = \frac{g_A^2}{16 \pi m_Z^2 }   \bigg(1-2 ~ \frac{m_h^2+m_Z^2}{s}+\frac{\left(m_h^2-m_Z^2\right)^2}{s^2}\bigg)^{1/2} \,\, \frac{\lambda _s^2\left(1-\frac{4 m_{\chi }^2}{s}\right) + \lambda _p^2}{ \left(s-m_A^2\right)^2 + m_A^2 \Gamma_A^2} \, \Big(9 m_h^4+\left(6 s-14 m_Z^2\right) m_h^2+5 m_Z^4+s^2-10 s m_Z^2\Big),
\end{align}
where $\sqrt{s}$ is the center-of-mass energy. The width of the mediator, $\Gamma_A$, potentially receives contributions from decays to $hZ$ and $\chi \chi$, given by:
\begin{eqnarray}
\Gamma{(A \to hZ)} &=& \frac{g_A^2}{16 \pi  m_A^3}\bigg(m_A^4-2 m_A^2 \left(m_h^2+m_Z^2\right)+\left(m_h^2-m_Z^2\right)^2 \bigg)^{1/2} \Big[ m_Z^2-2
   \left(m_A^2+m_h^2\right)+\frac{\left(m_A^2-m_h^2\right)^2}{m_Z^2}\Big], \nonumber \\
\Gamma{(A \to \chi \chi)} &=& \frac{1}{16 \pi m_A^2} \bigg(m_A^2-4m_\chi^2 \bigg)^{1/2} \Big[ \left( m_A^2-4m_\chi^2 \right) \lambda_s^2+m_A^2 \lambda_p^2 \Big].
\end{eqnarray}
At low velocities and sufficiently far from resonance, this cross section can be expanded in powers of velocity: 
%
\begin{eqnarray}
\sigma v &\approx& \frac{g_A^2 \lambda_p^2}{64 \pi m_Z^2 }  \bigg(m_h^4-2(m_Z^2+4m^2_{\chi}) m_h^2+(m_Z^2 -4m_{\chi}^2)^2 \bigg)^{1/2} \,\, \frac{5m_Z^4 -2(7m_h^2+20 m_{\chi}^2)m^2_Z +(3m_h^2+4m_{\chi}^2)^2}{(m_A^2-4m_{\chi}^2)^2} \\
&+&v^2 \frac{g_A^2}{256 \pi  m_Z^2 m_{\chi }^4 \left(m_A^2-4 m_{\chi }^2\right)^2} \Bigg\{ 8 \lambda _p^2 m_{\chi }^4 \left(3 m_h^2-5 m_Z^2+4 m_{\chi }^2\right) \Big(m_h^4-2 \left(m_Z^2+4 m_{\chi }^2\right) m_h^2+\left(m_Z^2-4 m_{\chi }^2\right)^2 \Big)^{1/2}
\nonumber \\
&-&\frac{2 \lambda _p^2 m_{\chi }^2}{\left(m_A^2-4 m_{\chi }^2\right) \Big( m_h^4-2 \left(m_Z^2+4 m_{\chi }^2\right) m_h^2+\left(m_Z^2-4 m_{\chi }^2\right)^2 \Big)^{1/2} }  \bigg[\Big(5 m_Z^4-2 \left(7 m_h^2+20 m_{\chi }^2\right) m_Z^2+\left(3 m_h^2+4 m_{\chi }^2\right)^2\Big)  \nonumber \\
&\times&\Big(m_A^2 \big(m_h^4-2 \left(m_Z^2+3 m_{\chi }^2\right) m_h^2+m_Z^4+8 m_{\chi }^4-6 m_Z^2 m_{\chi }^2\big)-8 m_{\chi }^2 \left(m_h^4-\left(2 m_Z^2+7 m_{\chi }^2\right) m_h^2+m_Z^4+12 m_{\chi }^4-7 m_Z^2 m_{\chi }^2\right)\Big)\bigg]
\nonumber \\
&+&\Big(m_h^4-2 \left(m_Z^2+4 m_{\chi }^2\right) m_h^2+\left(m_Z^2-4 m_{\chi }^2\right)^2\Big)^{1/2}  \left(\lambda _p^2+\lambda _s^2\right) m_{\chi }^2\Big( 5 m_Z^4-2 \left(7 m_h^2+20 m_{\chi }^2\right) m_Z^2+\left(3 m_h^2+4 m_{\chi }^2\right)^2\Big) \Bigg\}.\nonumber
\end{eqnarray}

In Fig.~\ref{diagramZhSchScalar}, we show the dominant diagrams for elastic scattering in this case. This interaction is described by:
\begin{eqnarray}
\mathcal{L} \sim \frac{g_A y_s g_q^a}{8 \pi ^2 m_A^2} \int_0^1 dx_1 ~ \int_0^{1-x_1} dx_2 ~ \left(1-2 \log{\frac{\mu ^2}{\Delta }}\right) \bar{\chi} \left( \lambda_s + \lambda_p i \gamma^5 \right) \chi ~ \bar{q} i \gamma^5 q,
\label{small}
\end{eqnarray}
where $\Delta \equiv x_1 m_h^2 + x_2 m_Z^2$ and $g^a_q=-g_2 T^3_q/ (2 \cos \theta_W)$ is the axial piece of the Standard Model $q$-$q$-$Z$ coupling. The $\gamma^5$ appearing in the quark part of this interaction guarantees that the resulting elastic scattering cross section will be both spin-dependent and suppressed by two or four powers of momentum for the $\lambda_s$ and $\lambda_p$ terms, respectively. As a consequence, the elastic scattering cross sections predicted in this model will be extremely small, and well below the neutrino floor. The reason that no $\bar q q$ term is present in Eq.~\ref{small} is that any interaction with two scalars and a $Z$ is anti-hermitian, and thus cancels in the Lagrangian. As a result, only the scalar-pseudoscalar-$Z$ vertex remains, providing a $\gamma_5$ in the effective quark bilinear.

In Fig.~\ref{zHfigs}, we plot some of the phenomenological features of this model for the case of $m_{\chi}=110$ GeV and $g_A=0.3$. For scenarios with a non-negligible value of $\lambda_p$, the annihilation is dominantly $s$-wave, allowing this model to generate the observed normalization of the Galactic Center gamma-ray excess. In contrast, the annihilation cross section is velocity suppressed in the case of a purely scalar coupling ($\lambda_p=0$), leading to no appreciable indirect detection signals.

\begin{figure*}[t]
\includegraphics[width=0.25\textwidth]{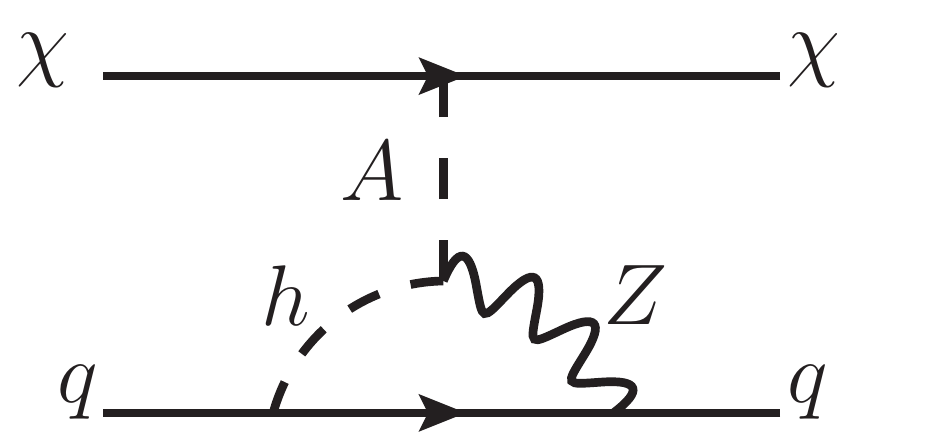}
\caption{The diagram for dark matter elastic scattering with nuclei corresponding to the case in which the dark matter annihilates to $hZ$ through the $s$-channel exchange of a spin-zero mediator, $A$ (see Sec.~\ref{Zh-scalar}).}
\label{diagramZhSchScalar}
\end{figure*}


\begin{figure*}[h]
\includegraphics[width=0.49\textwidth]{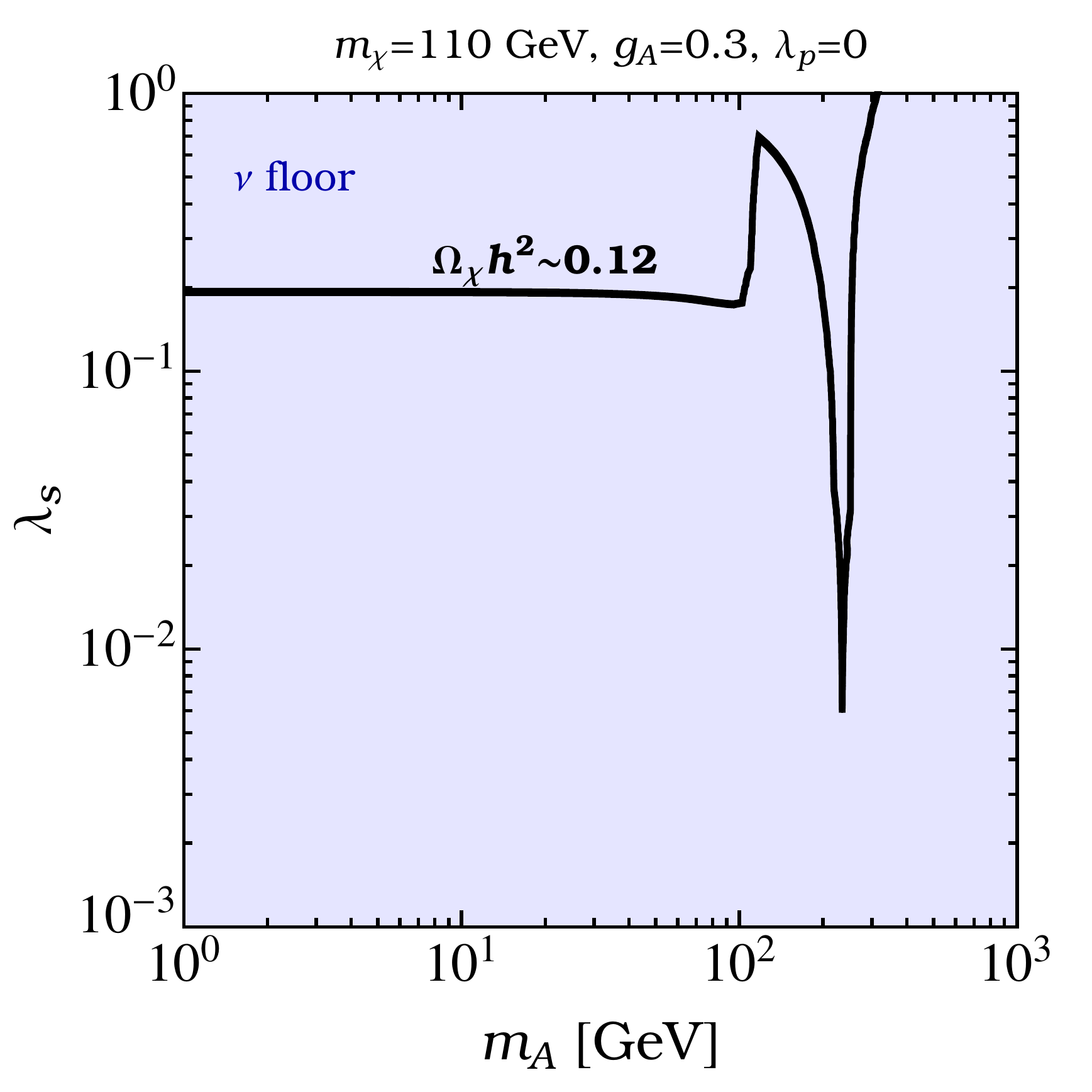}
\includegraphics[width=0.49\textwidth]{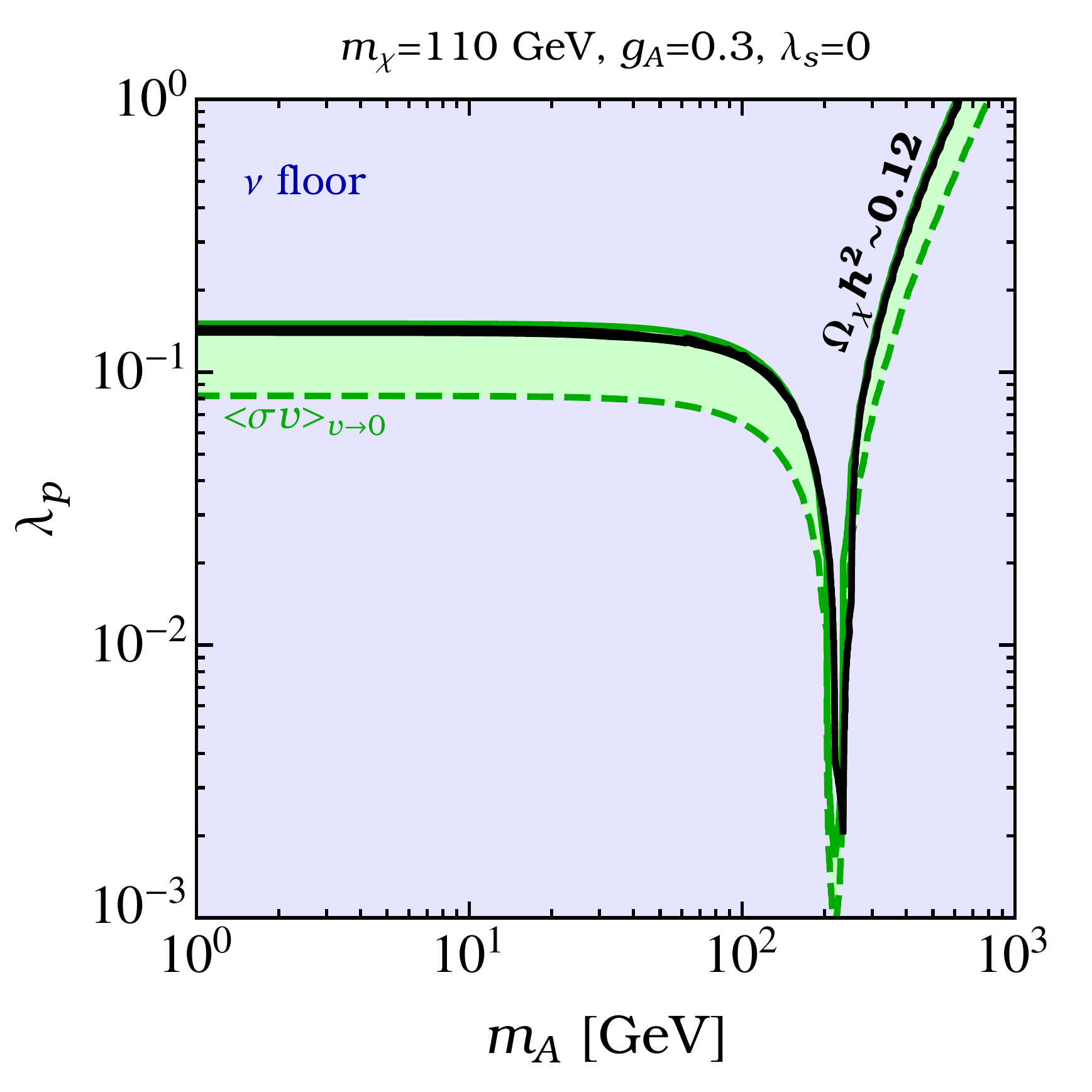} 
\includegraphics[width=0.49\textwidth]{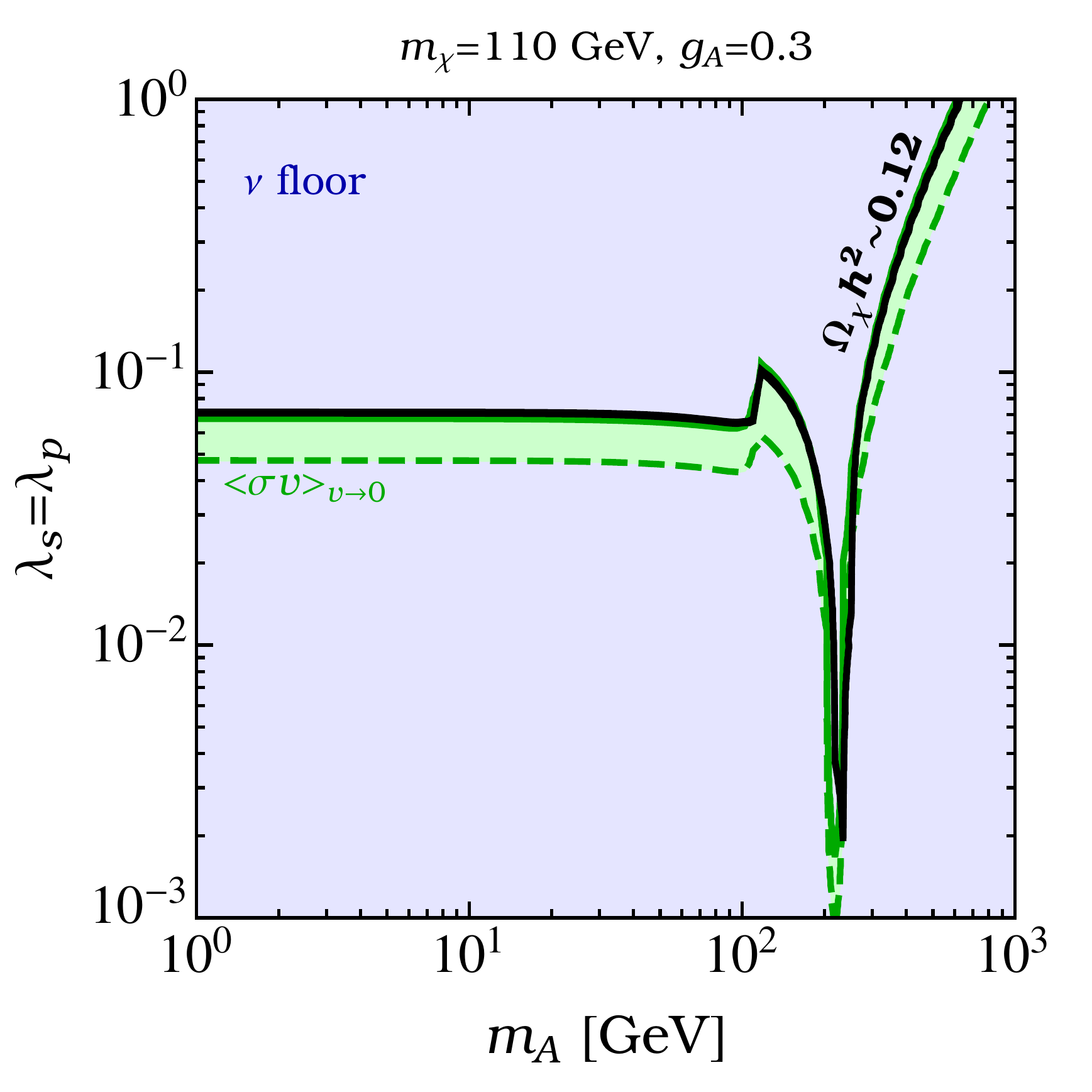}
\caption{For Majorana dark matter with a mass of 110 GeV, annihilating to $hZ$ through a spin-zero mediator, we plot the parameter space in which the predicted thermal relic abundance matches the observed dark matter density (solid black line) and the region in which the Galactic Center gamma-ray excess can be generated (green shaded band). The upper boundary of this region corresponds to the upper limit on the annihilation cross section from Fermi's observations of dwarf galaxies~\cite{Ackermann:2015zua} (green solid). In this case, we predict a cross section that is below the neutrino floor throughout the entire region shown.}
\label{zHfigs}
\end{figure*}





\subsection{$s$-channel spin-one mediator}
\label{Zh-vector}

In this subsection, we consider Majorana dark matter, $\chi$, that annihilates to $hZ$ through the $s$-channel exchange of a spin-one particle, $Z'$:
\begin{equation}
\mathcal{L} \supset  g_\chi Z^\prime_\mu~\bar{\chi} \gamma^\mu \gamma^5 \chi + \mu_h Z_\mu^\prime Z^\mu h. 
\label{lag1}
\end{equation}

\begin{figure*}[t]
\includegraphics[width=0.5\textwidth]{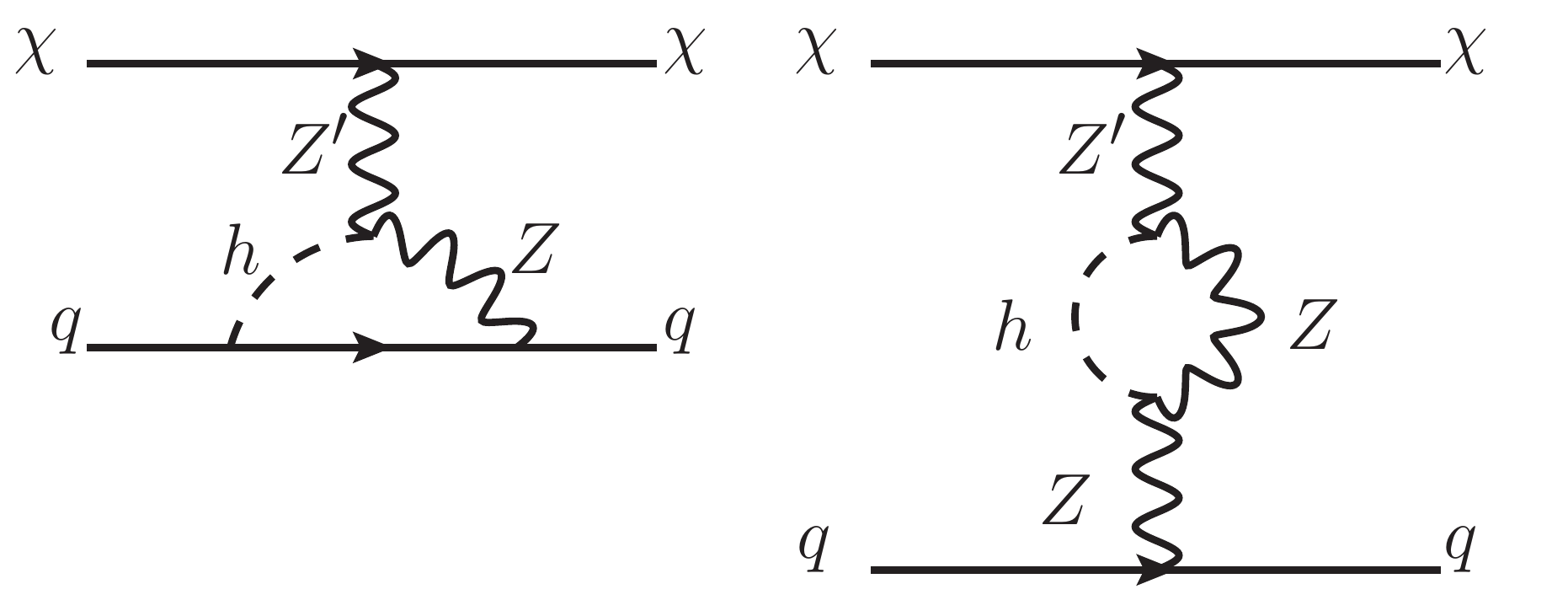} 
\caption{The diagrams for dark matter elastic scattering with nuclei corresponding to the case in which the dark matter annihilates to $hZ$ through the $s$-channel exchange of a spin-one mediator, $Z'$ (see Sec.~\ref{Zh-vector}).}
\label{diagramZhSchVec}
\end{figure*}

The annihilation cross section in this case is given by:
\begin{eqnarray}
\sigma v &=&  \frac{g_{\chi }^2 \mu _h^2  \Big(m_h^4-2 \left(s+m_Z^2\right) m_h^2+\left(s-m_Z^2\right)^2\Big)^{1/2} 
}{24 \pi  s^3 m_Z^2 m_{Z^\prime}^4 \Big( (s-m_{Z^\prime}^2 )^2 +m_{Z^\prime}^2 \Gamma_{Z^\prime}^2 \Big) } 
\Bigg\{ \Big( s m_{Z^\prime}^4+2 m_{\chi }^2 (m_{Z^\prime}^4-6 s m_{Z^\prime}^2+3 s^2 ) \Big) m_h^4
\nonumber \\
&-&2 \left(s+m_Z^2\right) \Big( s m_{Z^\prime}^4+2 m_{\chi }^2 (m_{Z^\prime}^4-6 s m_{Z^\prime}^2+3 s^2) \Big)  m_h^2+s \left(m_Z^4+10 s m_Z^2+s^2\right) m_{Z^\prime}^4 \nonumber \\
&+&2 m_{\chi }^2 \bigg[\left(m_{Z^\prime}^4-6 s m_{Z^\prime}^2+3 s^2\right) m_Z^4
-2 s \left(13 m_{Z^\prime}^4-6 s m_{Z^\prime}^2+3 s^2\right) m_Z^2+s^2 \left(m_{Z^\prime}^4-6 s m_{Z^\prime}^2+3 s^2\right)\bigg] \Bigg\},
\end{eqnarray}
where $\sqrt{s}$ is the center-of-mass energy. The width of the mediator, $\Gamma_{Z'}$, potentially receives contributions from decays to $hZ$ and $\chi \chi$:
\begin{eqnarray}
\Gamma(Z'\rightarrow hZ) &=& \frac{      \mu_h^2}{48 \pi  m_{Z'}^3} 
 \Big(  m_{Z'}^4+\left(m_h^2-m_Z^2\right)^2-2 m_{Z'}^2 \left(m_h^2+m_Z^2\right) \Big)^{1/2}   \, \,
\Bigg[2-\frac{       (m_{Z'}^2-m_h^2+m_Z^2 )^2        }{4 m_Z^2 m_{Z'}^2} \Bigg], \nonumber \\
\Gamma(Z'\rightarrow \chi \chi) &=& \frac{g_{\chi}^2 \left(m_{Z'}^2-4 m_{\chi}^2\right)^{3/2}}{24 \pi m_{Z'}^2}.
\end{eqnarray}
Expanding the annihilation cross section in powers of velocity yields:
%
\begin{eqnarray}
\sigma v &\approx&  \frac{g_{\chi }^2 \mu _h^2}{256 \pi  m_{\chi }^4 m_Z^2 m_{Z'}^4} \Big(m_h^4-2 \left(m_Z^2+4 m_{\chi }^2\right) m_h^2+\left(m_Z^2-4 m_{\chi }^2\right)^2\Big)^{3/2} \\
&+&v^2 \frac{g_{\chi }^2 \mu _h^2 \Big(m_h^4-2 \left(m_Z^2+4 m_{\chi }^2\right) m_h^2+\left(m_Z^2-4 m_{\chi }^2\right)^2\Big)^{1/2} }{3072 \pi  m_Z^2 m_{\chi }^4 m_{Z'}^4 \left(m_{Z'}^2-4 m_{\chi }^2\right)^2} \nonumber \\
&\times& \Bigg\{\bigg[-7 m_h^4+20 m_{\chi }^2 m_h^2-7 m_Z^4+32 m_{\chi }^4+2 m_Z^2 \left(7 m_h^2+58 m_{\chi }^2\right)\bigg] m_{Z'}^4 \nonumber \\
&+&72 \bigg[m_h^4-2 \left(m_Z^2+2 m_{\chi }^2\right) m_h^2 + m_Z^4-4 m_Z^2 m_{\chi }^2\bigg] m_{Z'}^2 m_{\chi }^2
-144 \bigg[m_h^4-2 \left(m_Z^2+2 m_{\chi }^2\right) m_h^2+m_Z^4-4 m_Z^2 m_{\chi }^2\bigg] m_{\chi }^4 \Bigg\}. \nonumber 
\end{eqnarray}

Scattering with nuclei occurs through the diagrams presented in Fig.~\ref{diagramZhSchVec}. The dominant Wilson coefficient associated with these diagrams is given by:
\begin{eqnarray}
\label{f1}
f_q^{(1)} = \frac{g_\chi g_q^a \mu_h}{16 \pi ^2 m_{Z^\prime}^2 v} \int_0^1 dx \log{\frac{\mu ^2}{x~m_h^2+(1-x)~m_Z^2 }} + \frac{g_q^a g_\chi \mu_h m_q y_s}{8 \pi ^2  m_{Z^\prime}^2} \int_0^1 dx_1~ \int_0^{1-x_1}dx_2~\frac{(x_1+x_2+1)}{\Delta}, \\ \nonumber 
\end{eqnarray}
where $\Delta \equiv x_1 m_h^2+x_2 m_Z^2$ and $\mu \simeq m_Z$.

If we were to strictly adhere to our simplified model framework, we would use the Wilson coefficient given in Eq.~\ref{f1} to calculate the elastic scattering cross section in this model. In a realistic model with an interaction of the form $Z^\prime_\mu Z^\mu h$, however, mixing between the $Z$ and $Z'$ is an inevitable consequence of gauge invariance, leading to elastic scattering through tree-level $Z$ exchange. For example, one might consider extending the Standard Model gauge group by a new broken $U(1)$, under which the Standard Model Higgs is charged. After electroweak symmetry breaking, the kinetic term for the Higgs leads to both an effective $Z'$-$Z$-$h$ coupling as well as to mass mixing between the $Z$ and $Z'$. As a result, these two effects cannot be decoupled, and a non-zero value for $\mu_h$ generically induces mass mixing, as well as an effective coupling between the dark matter and the Standard Model $Z$. This results in the following tree-level contribution to the axial-vector Wilson coefficient:
\begin{eqnarray}
f_q^{(1)} &=& \frac{g_\chi g_q^a}{m_{Z^\prime}^2 } \epsilon, 
\end{eqnarray}
where the mixing parameter, $\epsilon$, is given by:
\begin{equation}
\epsilon = \frac{\mu_h v}{4 (m^2_{Z'} - m^2_Z)}.
\end{equation}
\begin{figure*}[htb]
\includegraphics[width=0.49\textwidth]{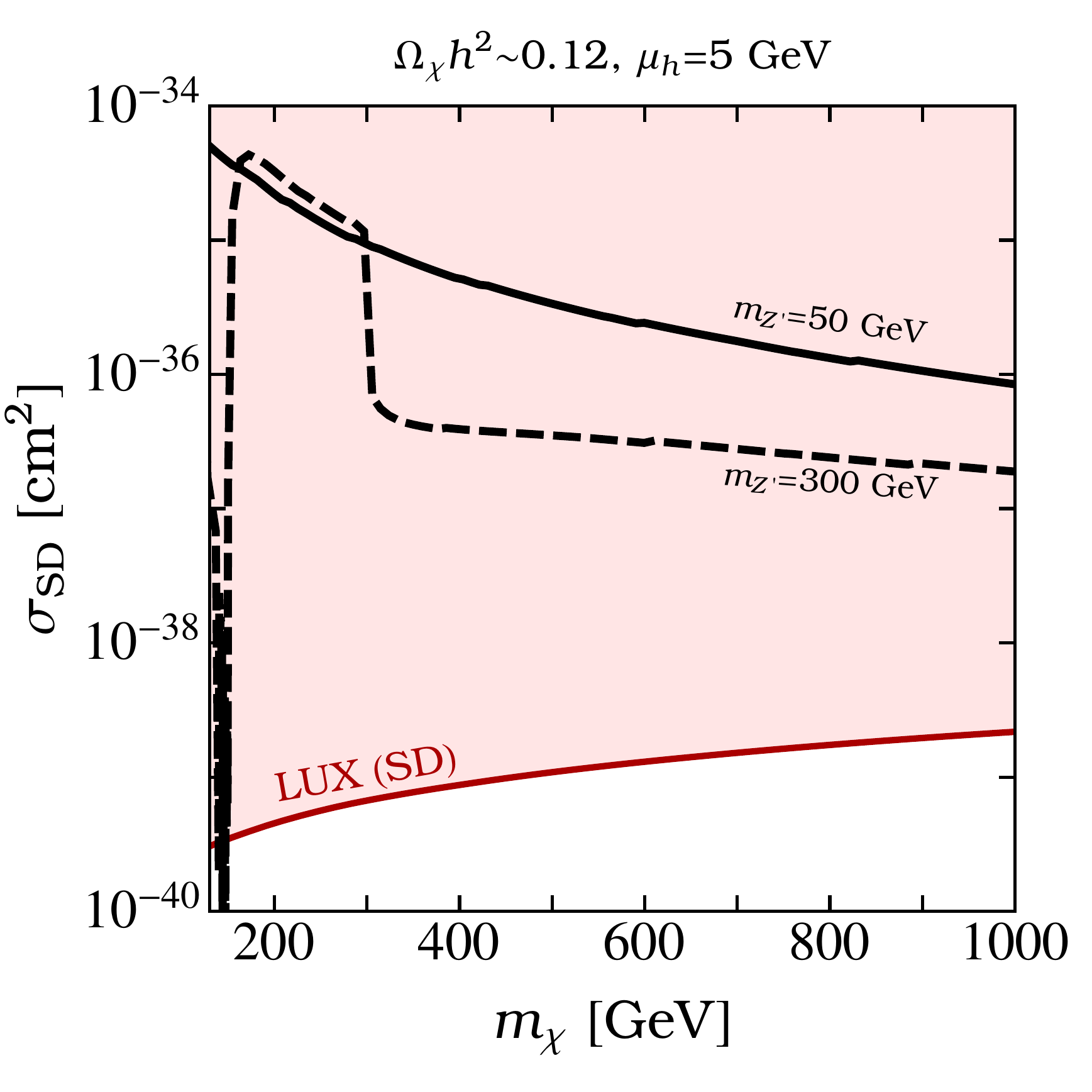} 
\includegraphics[width=0.49\textwidth]{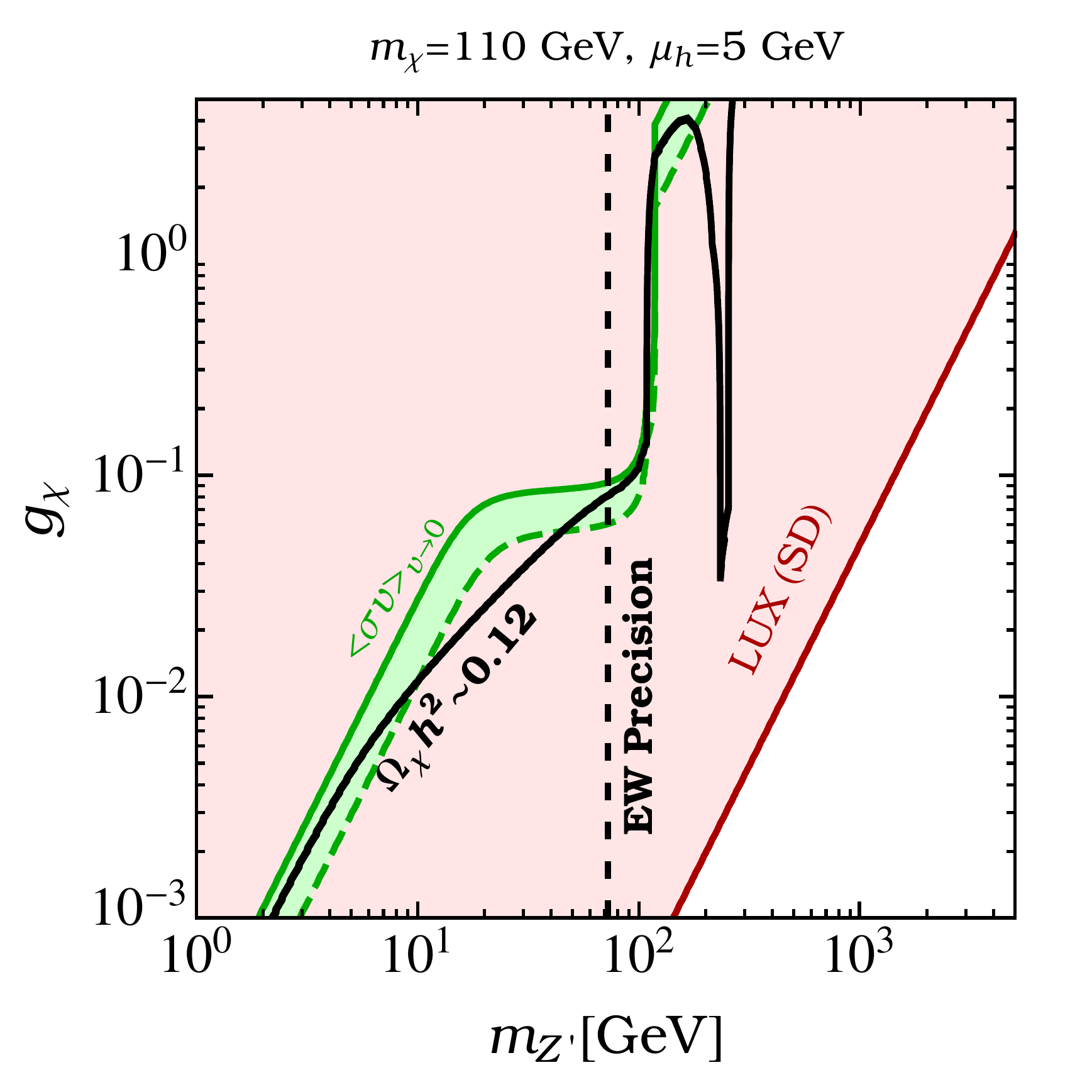} 
\caption{Phenomenology in the case of Majorana dark matter, $\chi$, annihilating to $hZ$ through a spin-one mediator, $Z'$. In the left frame, we plot the dark matter's spin-dependent elastic scattering cross section per nucleon as a function of the dark matter mass, for two values of the mediator mass. In each case, we have adopted a value of $\mu_h=5$ GeV and the coupling, $g_{\chi}$, has been set to obtain the desired thermal relic abundance, $\Omega_{\chi} h^2 =0.12$. In the right frame, we plot for the case of $m_{\chi}=110$ GeV the parameter space in which the predicted thermal relic abundance matches the observed dark matter density (solid black line) and the region in which the Galactic Center gamma-ray excess can be generated (green shaded band). The upper boundary of this region corresponds to the upper limit on the annihilation cross section from Fermi's observations of dwarf galaxies~\cite{Ackermann:2015zua} (green solid). The red region of this frame is currently excluded by the LUX direct detection experiment, when converted to apply to the case of spin-dependent scattering~\cite{Akerib:2013tjd}. This constraint excludes the overwhelming majority of the parameter space for this simplified model. We also plot as a vertical dashed line the constraint from precision measurements of the $Z$ mass, as described in the text.}
\label{zHfigs2}
\end{figure*}

Mass mixing between the $Z$ and $Z'$ can also alter the value of the $Z$ mass, which is strongly constrained by electroweak precision measurements. In particular, in the simple model described in the previous paragraph, the $\rho$ parameter is shifted from its Standard Model value of unity as follows:
\begin{eqnarray}
 \rho = 1 +  \epsilon^2 \bigg(\frac{m_{Z^\prime}^2-m_Z^2}{m^2_Z}\bigg).
 \label{rho}
 \end{eqnarray}
Experimentally, $\rho$ is bounded to be less than 1.0009 at the 95\% confidence level~\cite{pdg}, which translates into a fairly strong lower bound on $m_{Z'}$, even for modest values of $\mu_h$.

As discussed in Sec.~\ref{formalism}, axial-vector quark couplings, $f_q^{(1)}$, lead to spin-dependent scattering with nuclei. Although spin-dependent scattering is comparatively weakly constrained, the cross sections predicted in this model are very large and well within the reach of current experiments. In the left frame of Fig.~\ref{zHfigs2}, we plot the spin-dependent elastic scattering cross section in this model, finding that it is strongly constrained by LUX~\cite{Akerib:2013tjd}. In the right frame, we consider the case of $m_{\chi}=110$ GeV, and plot the parameter space in which the predicted thermal relic abundance matches the observed dark matter density  and the region in which the Galactic Center gamma-ray excess can be generated. The constraint from LUX excludes this simplified model, unless one considers dark matter that is very near resonance, $2m_{\chi} \simeq m_{Z'}$. We also include in the right frame the constraint from the measurement of the $Z$ mass, as described in Eq.~\ref{rho}.

\subsection{$t$-channel fermionic mediator}
\label{zht}

Lastly, we consider Majorana dark matter, $\chi_1$, that annihilates to $hZ$ (and necessarily to $hh$ and $ZZ$) through the $t$-channel exchange of a fermion, $\chi_2$:

\begin{eqnarray}
\mathcal{L} \supset Z_\mu \bar{\chi}_1 \gamma^\mu \left( i g_\chi^v + g_\chi^a \gamma^5 \right) \chi_2 + h \bar{\chi}_1 \left( \lambda_s + \lambda_p i \gamma^5 \right) \chi_2. \nonumber \\ 
\end{eqnarray}


The expanded low-velocity annihilation cross section to $hZ$ is given by:
%
\begin{eqnarray}
\sigma_{hZ} v &\approx&  \frac{1}{64 \pi  m_{\chi _1}^4 m_Z^2} \frac{\Big[m_h^4+(m_Z^2-4 m_{\chi _1}^2)^2 -2 m_h^2 (4 m_{\chi _1}^2+m_Z^2) \Big]^{3/2}}{(m_h^2-2 m_{\chi _1}^2-2 m_{\chi_2}^2+m_Z^2)^2}  \Big[(m_{\chi _1}-m_{\chi _2}) \lambda _s g_{\chi }^a+(m_{\chi _1}+m_{\chi _2}) \lambda _p  g_{\chi }^v\Big]^2 \nonumber \\
&+&v^2  \frac{ \Big[m_h^4-2 \left(m_Z^2+4 m_{\chi _1}^2\right) m_h^2+\left(m_Z^2-4 m_{\chi _1}^2\right)^2 \Big]}{6 \pi  m_Z^2 m_{\chi _1}^4 \left(m_h^2+m_Z^2-2 m_{\chi _1}^2-2 m_{\chi _2}^2\right)^4} \bigg(A_{hZ} (g^a_{\chi})^2 +B_{hZ} (g^v_{\chi})^2+C_{hZ} g^v_{\chi}g^a_{\chi}\bigg). \nonumber
\end{eqnarray}
In the limit of $m_{\chi_2} \gg m_h, m_Z$:
\begin{eqnarray}
A_{hZ} &\simeq&  m_{\chi_2}^6 m^2_{\chi_1} \bigg[ m_{\chi_1}^2 \bigg(1-4x+3x^2+8x^3+3x^4+4x^5+x^6 \bigg)+ \frac{5}{8}m_h^2 +\frac{29}{8} m_Z^2  \bigg] \lambda^2_s \\
&+&m_{\chi_2}^4 m^4_{\chi_1} \bigg[ m_{\chi_1}^2 \bigg(6-4x+10x^2+4x^4 \bigg)+ 3m_h^2 +3 m_Z^2  \bigg] \lambda^2_p, \nonumber \\
%
%
B_{hZ} &\simeq&  m_{\chi_2}^4 m^4_{\chi_1} \bigg[ m_{\chi_1}^2 \bigg(6+4x+10x^2+4x^4\bigg) -3m_h^2 -3m_Z^2\bigg]  \lambda^2_s  \nonumber \\
&+& m_{\chi_2}^6 m^2_{\chi_1} \bigg[ m_{\chi_1}^2 \bigg(1+4x+3x^2 -8x^3 +3x^4 -4x^5+x^6\bigg) +\frac{5}{8}m_h^2 +\frac{29}{8}m_Z^2\bigg]  \lambda^2_p, \nonumber \\
%
C_{hZ} &\simeq& m_{\chi_2}^6 m_{\chi_1}^4  \bigg[ m_{\chi_1}^2 \bigg(-2+10x^2-2x^4-6x^6\bigg) -\frac{5}{4}m^2_h -\frac{29}{4} m^2_Z\bigg]  \lambda_s  \lambda_p, \nonumber 
\end{eqnarray}
where $x\equiv m_{\chi_1}/m_{\chi_2}$.

The velocity-expanded annihilation cross section to $ZZ$ is given by:
\begin{eqnarray}
\sigma_{ZZ} v &\approx& \frac{1}{4 \pi} \bigg(1-\frac{m_Z^2}{m_{\chi_1}^2}\bigg)^{1/2} \frac{1}{\left(m_{\chi_2}^2+m_{\chi_1}^2-m_Z^2\right)^2}\Bigg\{ \left( g_\chi^a g_{\chi}^v \right)^2  \Bigg[ \frac{8m_{\chi_2}^2 m_{\chi_1}^4}{m^4_Z}+ 6 m_{\chi_2}^2 - \frac{8m_{\chi_2}^2 m_{\chi_1}^2}{m_Z^2}-m_Z^2 \Bigg] \\
&+& \left(\left( g_\chi^a \right)^2+\left( g_\chi^v \right)^2\right)^2 ( m^2_{\chi_1} - m^2_Z ) \Bigg\}
\nonumber \\
&+&v^2  \frac{1}{24 \pi  m_{\chi_1} m_Z^4 (m^2_{\chi_1} -m^2_Z)^{1/2} \left(m_{\chi_2}^2+m_{\chi_1}^2-m_Z^2\right)^4} \bigg( A_{ZZ} \left( g_\chi^a \right)^4 + B_{ZZ} \left( g_\chi^v \right)^4 + C_{ZZ} \left( g_\chi^v g_\chi^a\right)^2 \bigg),\nonumber
\end{eqnarray}
where in the limit of $m_{\chi_2} \gg m_Z$:
\begin{eqnarray}
A_{ZZ} &\simeq&  m_{\chi_2}^6 m^6_{\chi_1} \bigg(3+4x+4x^2+8x^3+7x^4+4x^5+2x^6 \bigg), \\
B_{ZZ} &\simeq&  m_{\chi_2}^6 m^6_{\chi_1} \bigg(3-4x+4x^2-8x^3+7x^4-4x^5+2x^6 \bigg), \nonumber \\
C_{ZZ} &\simeq&  2 m_{\chi_2}^6 m^4_{\chi_1} \bigg(m^2_{\chi_1} (9-8x^2-3x^4+2x^6) -9 m^2_Z \bigg). \nonumber
\end{eqnarray}
Again, $x\equiv m_{\chi_1}/m_{\chi_2}$. The annihilation cross section to $hh$ in this model is the same as that given in Eq.~\ref{sigmavHHTch}, and we do not repeat this expression here.

\begin{figure*}[t]
\includegraphics[width=1.0\textwidth]{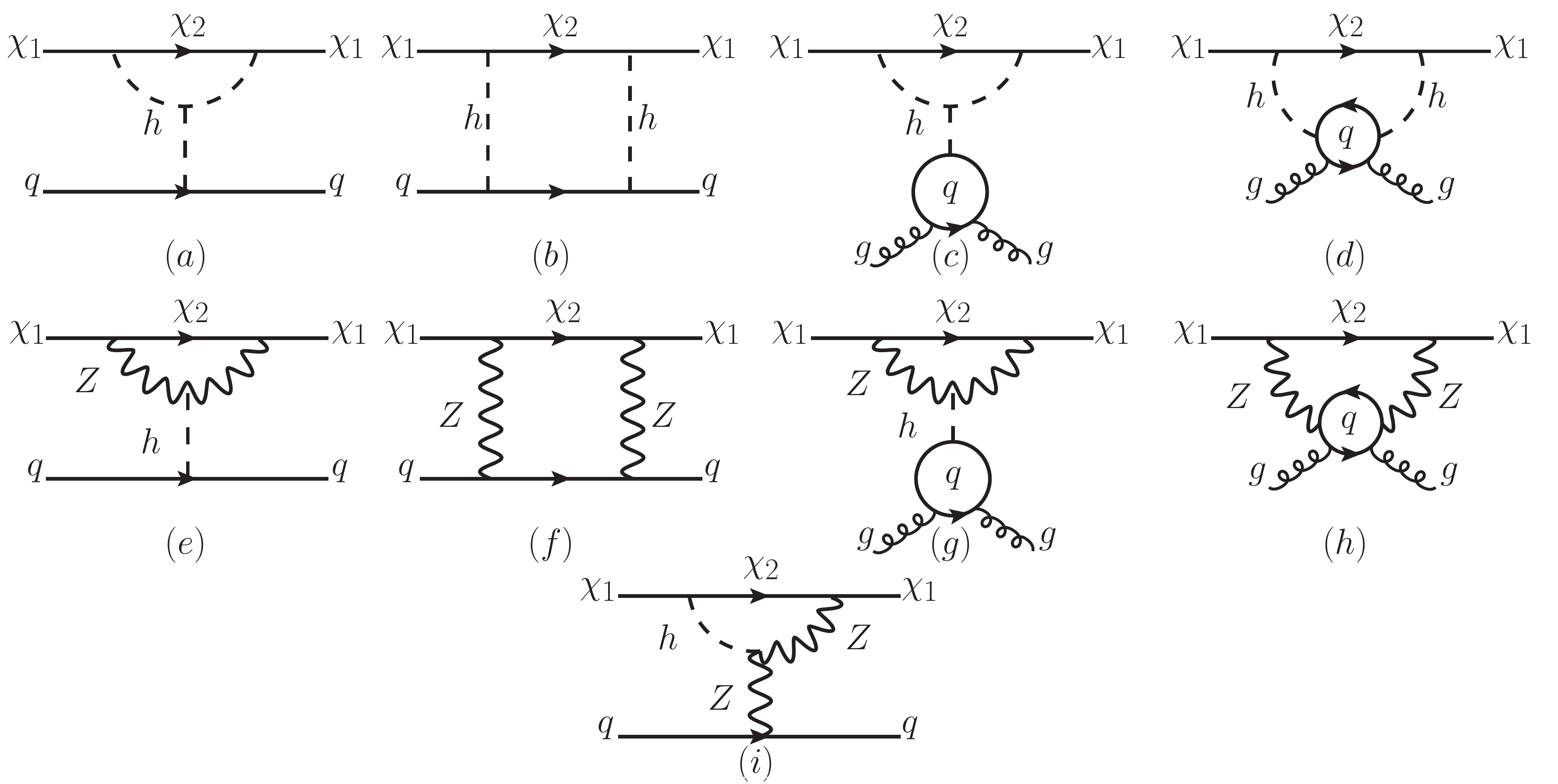} 
\caption{The diagrams for dark matter elastic scattering with nuclei corresponding to the case in which the dark matter annihilates to $hZ$ through the $t$-channel exchange of a fermionic mediator, $\chi_2$ (see Sec.~\ref{zht}).}
\label{ZhTchDiagrams}
\end{figure*}

Turning now to elastic scattering, there are many diagrams to consider in this case, including the $hh$ vertex correction (Fig.~\ref{ZhTchDiagrams}$a$, \ref{ZhTchDiagrams}$c$), the $hh$ box diagram (\ref{ZhTchDiagrams}$b$, \ref{ZhTchDiagrams}$d$), the $ZZ$ vertex correction (Fig.~\ref{ZhTchDiagrams}$e$, \ref{ZhTchDiagrams}$g$), the $ZZ$ box diagram (\ref{ZhTchDiagrams}$f$, \ref{ZhTchDiagrams}$h$), and the $hZ$ vertex correction (\ref{ZhTchDiagrams}$i$). The contribution from the $hZ$ box diagram vanishes~\cite{Hill:2014yka}. The resulting Wilson coefficients are given by:
\begin{eqnarray}
f_q^{(0)} &=& \frac{m_Z^2 (y_s/m_q)}{8 \pi ^2 m_h^2 v} \int_0^1 dx \frac{1-x}{\Delta_1}\Big[(g_\chi^a)^2 \big(m_{\chi_1} (x-1)-2 m_{\chi_2}\big)+(g_\chi^v)^2 \big(m_{\chi_1} (1-x)-2 m_{\chi_2}\big)\Big] \\
&+& \frac{y_s/m_q}{32 \pi ^2 v} \int_0^1 dx ~ \frac{3 (x-1) (\lambda^{(-)} m_{\chi_2}-\lambda^{(+)} m_{\chi_1} (x-1))}{ \Delta_1} \nonumber \\
&+&  \frac{1}{8 \pi ^2} \int_0^1 dx_2 \int_0^{1-x_2} dx_1 (x_1+x_2-1)\Bigg\{ \frac{1}{\Delta_2^2}\Bigg[-g_\chi^{(-)} m_{\chi_2} \Big(2 g_q^{(-)}+g_q^{(+)} (x_2-1)\Big) \nonumber \\
&+&\frac{1}{2} g_\chi^{(+)} m_{\chi_1}\Big(g_q^{(+)} (-6 x_1 x_2+x_1+x_2-1)-2 g_q^{(-)} (x_1-1)\Big)\Bigg] \nonumber \\
&+&\frac{1}{\Delta_2^3}\Bigg[g_q^{(+)} g_\chi^{(-)} m_{\chi_1}^2 m_{\chi_2}x_1^2 x_2+2 g_q^{(+)} g_\chi^{(+)} m_{\chi_1}^3 (x_1-1) x_1^2 x_2\Bigg]\Bigg\} \nonumber  \\
&+&\frac{y_s^2}{16 \pi^2}\int_0^1 dx_2~\int_0^{1-x_2} dx_1~\frac{(1-x_1-x_2)}{\Delta_3^2}\Bigg\{ \frac{1}{2} \Big[\lambda^{(+)} m_{\chi_1} \big(x_1 (3 x_2-4)-2 x_2+4\big)-2 \lambda^{(-)} m_{\chi_2} (x_2-2)\Big] \nonumber \\
&+&\frac{m_{\chi_1}^2}{\Delta_3} x_1^2 x_2\Big[ \lambda^{(-)} m_{\chi_2}-\lambda^{(+)} m_{\chi_1} (x_1-1)\Big]\Bigg\}, \nonumber \\
\nonumber \\
\nonumber \\
\nonumber \\
\nonumber \\
f_q^{(1)}&=&\frac{g_q^{(+)}}{8 \pi ^2} \int_0^1 dx_2 \int_0^{1-x_2} dx_1 (x_1+x_2-1)\Bigg\{ -\frac{3 }{2 \Delta_3 } g_\chi^{(+)}+\frac{1}{\Delta_3^2}\Big[ g_\chi^{(+)} m_{\chi_1}^2 x_1 (x_1+1)- g_\chi^{(-)} m_{\chi_1} m_{\chi_2} x_1\Big]\Bigg\}
\nonumber \\
&+& \frac{g_q^a}{8 \pi^2 v} \int_0^1 dx_2 ~\int_0^{1-x_2} dx_1~\frac{1}{ \Delta_5} \Bigg\{ \lambda_s g_\chi^a \Big[m_{\chi_1} (1-x_2)+m_{\chi_2}\Big] + \lambda_p g_\chi^v \Big[m_{\chi_1} (1-x_2)-m_{\chi_2}\Big] \Bigg\}, \nonumber
\nonumber \\
\nonumber \\
\nonumber \\
\nonumber \\
\nonumber \\
%
f_q^{(2)}&=& \frac{g_q^{(+)} m_{\chi_1} (x_1+x_2-1)}{4 \pi ^2} \Bigg\{ \frac{1}{\Delta^2_2} g_\chi^{(+)} (-4 x_1 x_2+x_1+x_2-1) + \frac{1}{\Delta^3_2} \Bigg[2 g_\chi^{(-)} m_{\chi_1} m_{\chi_2} x_1^2 x_2+4 g_\chi^{(+)} m_{\chi_1}^2 (x_1-1)x_1^2 x_2\Bigg]\Bigg\}, \nonumber \\
\nonumber \\
\nonumber \\
\nonumber \\
\nonumber \\
\nonumber \\
\nonumber \\
\nonumber \\
\nonumber \\
\nonumber \\
\nonumber \\
\nonumber \\
\nonumber \\
\nonumber \\
\nonumber 
\end{eqnarray}
\begin{eqnarray}
f_g^{(0)}&=& \sum_{q=c,b,t} \frac{\alpha_s}{128 \pi^3} \int_0^1 dx_1 dx_2 dx_3 ~\Bigg( -\frac{1}{\Delta^2_4 \xi^2}  (x_1-1) x_1 (x_2-1) x_3^2 ~g_q^{(+)}\left(g_\chi^{(+)} m_{\chi_1}-2 g_\chi^{(-)} m_{\chi_2}\right)
\nonumber \\
&-& \frac{1}{\Delta^2_4 \xi^3}  (x_1-1) x_1 (x_2-1) x_3^2~g_q^{(+)} \bigg[2 g_\chi^{(-)} m_{\chi_2} x_1 \Big(-x_1 x_3+x_1+x_3-1\Big) \nonumber \\
&+&g_\chi^{(+)}m_{\chi_1} \Big(x_1^2 (x_3-1)-x_1 x_3+x_1+x_2 x_3\Big)\bigg]
\nonumber \\
&-&\frac{2}{3 \Delta^3_4 \xi^2}m_q^2 \Big(3 (x_1-1) x_1+1\Big) (x_2-1) (x_3-1) x_3^2 \left(2 g_q^{(-)}-g_q^{(+)}\right) \left(g_\chi^{(+)} m_{\chi_1}-2 g_\chi^{(-)}m_{\chi_2}\right)
\nonumber \\
&&\frac{2}{3\Delta^3_4 \xi^3} m_q^2 (x_2-1) (x_3-1) x_3^2 \Bigg[x_3 \bigg\{(x_1-1) x_1 \Big(g_q^{(+)} (3 (x_1-1) x_1+1)-2 g_q^{(-)}(x_1-1) x_1\Big) \left(g_\chi^{(+)} m_{\chi_1}-2 g_\chi^{(-)} m_{\chi_2}\right)
\nonumber \\
&+&g_\chi^{(+)} m_{\chi_1} \left(3 (x_1-1) x_1+1\right) x_2 \left(g_q^{(+)}-2g_q^{(-)}\right)\bigg\} \nonumber \\
&-&(x_1-1) x_1 \Big(g_q^{(+)} (3 (x_1-1) x_1+1)-2 g_q^{(-)} (x_1-1) x_1\Big) \left(g_\chi^{(+)} m_{\chi_1}-2 g_\chi^{(-)}m_{\chi_2}\right)\Bigg]
\nonumber \\
&-&\frac{2}{\Delta^3_4 \xi^4} m_{\chi_1} (x_1-1) x_1 (x_2-1) x_2 (x_3-1) x_3^3 \Bigg[g_q^{(-)} g_\chi^{(+)} m_q^2 (x_1-1) x_1(x_3-1)+g_q^{(+)} \bigg\{g_\chi^{(-)} m_{\chi_1} m_{\chi_2} (x_1-1) x_1 x_2 x_3
\nonumber \\
&-&2 g_\chi^{(+)} m_q^2 \Big(3 (x_1-1)x_1+1\Big) (x_3-1)\bigg\}\Bigg]
\nonumber \\
&&\frac{2}{\Delta^4_4 \xi^4} m_{\chi_1}^2 m_q^2 (x_1-1) x_1 (x_2-1) x_2^2 (x_3-1)^2 x_3^4 \Bigg[g_q^{(-)} (x_1-1) x_1 \left(g_\chi^{(+)} m_{\chi_1}-2 g_\chi^{(-)} m_{\chi_2}\right)
\nonumber \\
&-&g_q^{(+)} \Big(3 (x_1-1) x_1+1\Big) \left(2 g_\chi^{(+)} m_{\chi_1}-g_\chi^{(-)} m_{\chi_2}\right)\Bigg]
\nonumber \\
&&\frac{2}{\Delta^4_4 \xi^5} g_\chi^{(+)} m_{\chi_1}^3 m_q^2 (x_1-1) x_1 (x_2-1) x_2^3 (x_3-1)^2 x_3^5 \Bigg[g_q^{(-)} (x_1-1) x_1-g_q^{(+)}\Big(6 (x_1-1) x_1+2\Big)\Bigg] \Bigg), \nonumber 
\end{eqnarray}
where
\begin{eqnarray}
\Delta_1 &\equiv& x (x-1) m^2_{\chi_1} +xm^2_{\chi_2} +(1-x) m^2_Z, \nonumber \\
\Delta_2 &\equiv& x_1 (x_1-1) m^2_{\chi_1} +x_1 m^2_{\chi_2} +(1-x_1-x_2) m^2_Z, \nonumber\\
\Delta_3 &\equiv& m_{\chi_1}^2 (x_1-1) x_1+m_{\chi_2}^2 x_1+m_h^2 (1-x_1-x_2), \nonumber \\
\Delta_4 &\equiv& \frac{x_2^2 x_3^2}{\xi} m_{\chi_1}^2 + x_2 x_3 (m_{\chi_1}^2 - m_{\chi_2}^2 ) + (x_2-1) x_3 m_Z^2 + (x_3-1)m_q^2, \nonumber\\
\Delta_5 &\equiv& x_2(x_2-1)m_{\chi_1}^2 + x_2 m_{\chi_2}^2 + (1-x_1-x_2)m_h^2+x_1 m_Z^2, \nonumber \\
\xi &\equiv& x_1 (x_1-1)(1-x_3)-x_3 \nonumber\\
g_{\chi}^{(\pm)} &\equiv& (g_{\chi}^v)^2 \pm (g_{\chi}^a)^2 \nonumber\\
g_q^{(\pm)} &\equiv& (g_q^v)^2 \pm (g_q^a)^2, \nonumber \\
\lambda^{(\pm)} &\equiv& \lambda_s^2 \pm \lambda_p^2, \nonumber
\end{eqnarray}
\end{widetext}
and the vector and axial pieces of the Standard Model $q$-$q$-$Z$ coupling are given by $g_q^v=-g^a_q=g_2 T^3_q/ (2 \cos \theta_W)$.

\begin{figure*}[htb]
\includegraphics[width=0.49\textwidth]{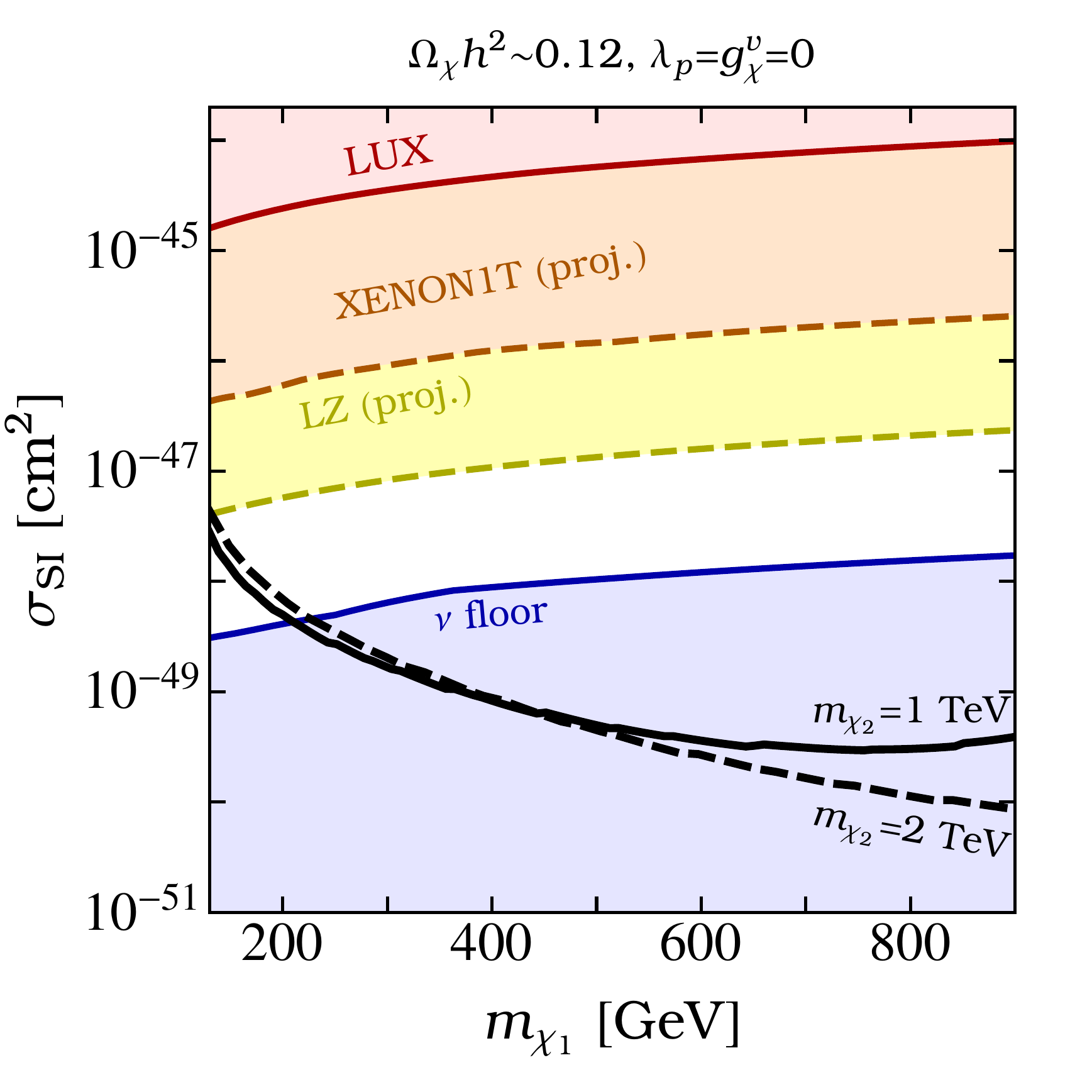} 
\includegraphics[width=0.49\textwidth]{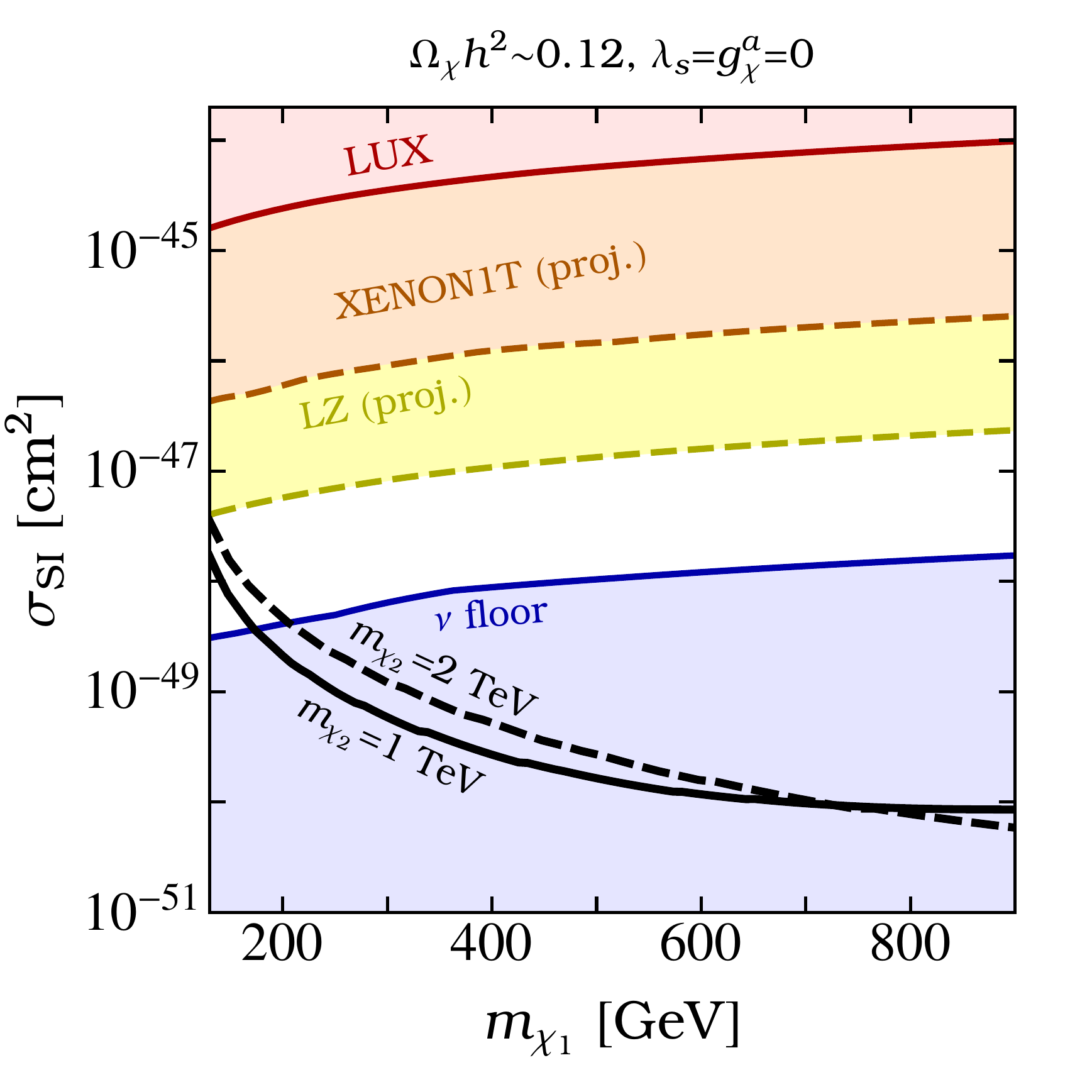} 
\includegraphics[width=0.49\textwidth]{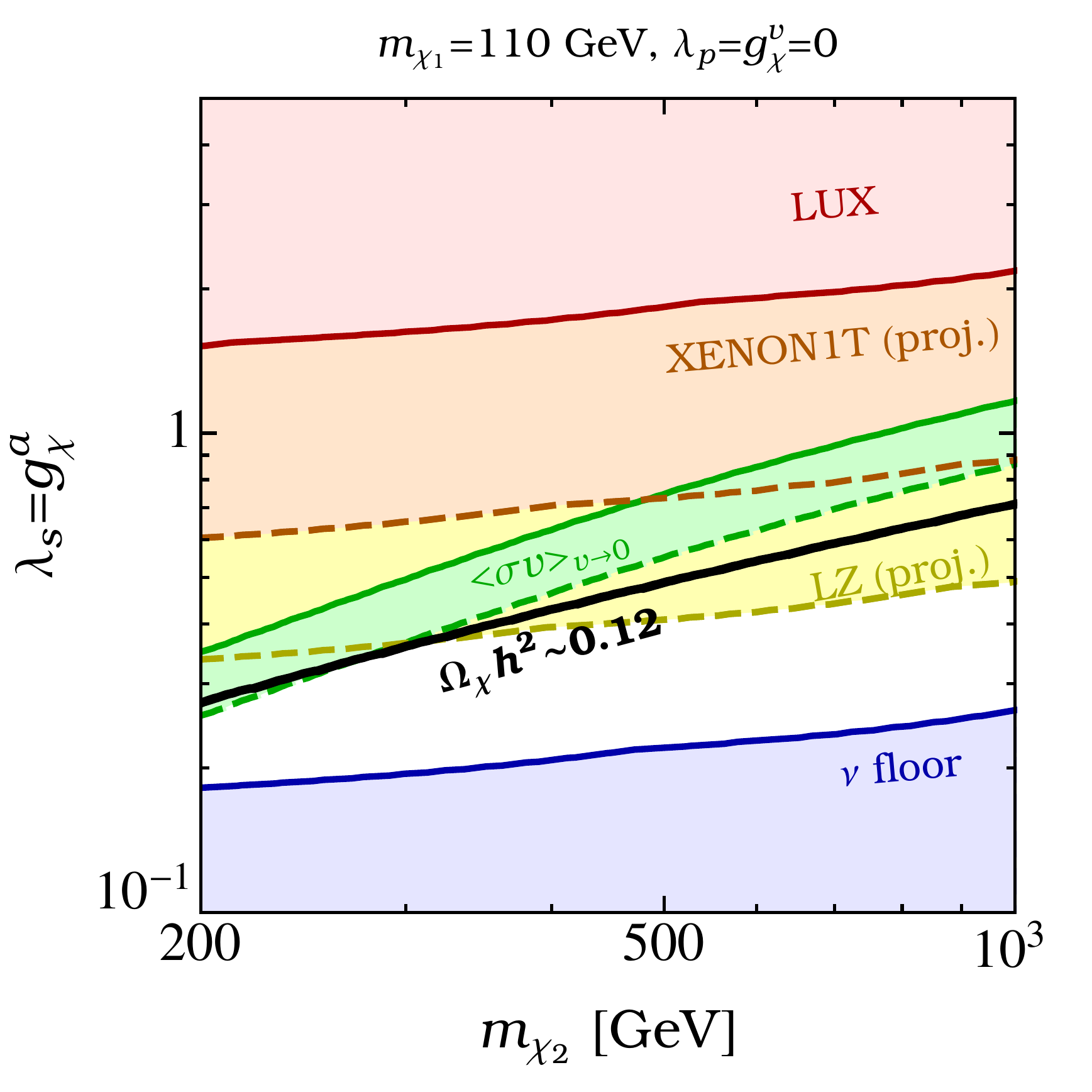} 
\includegraphics[width=0.49\textwidth]{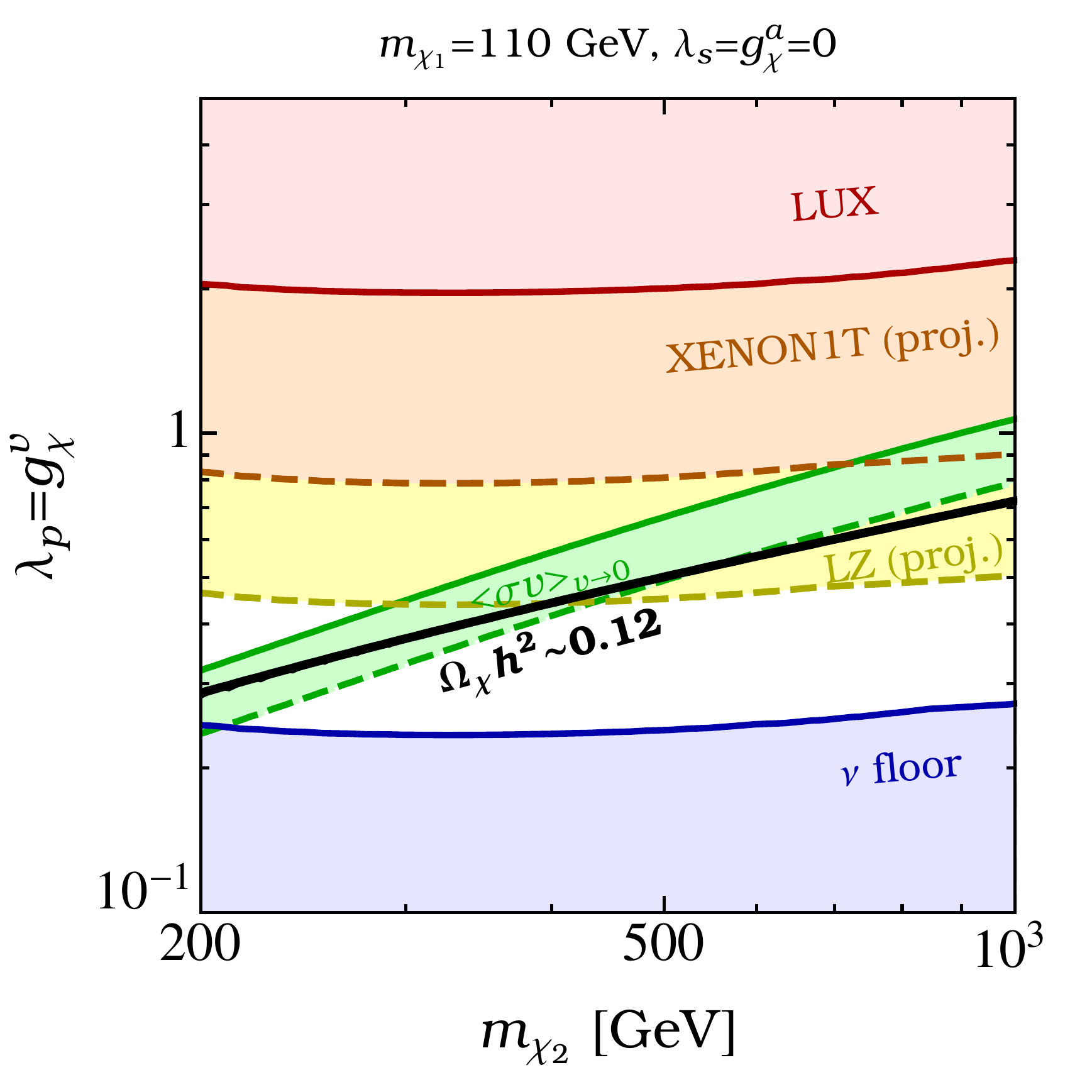} 
\caption{Phenomenology in the case of Majorana dark matter, $\chi_1$, annihilating to $hZ$ through a fermionic mediator, $\chi_2$. In the upper frames, we plot the dark matter's spin-independent elastic scattering cross section per nucleon as a function of the dark matter mass, for two values of the mediator mass, and for either $\lambda_p=g_{\chi}^v=0$ or $\lambda_s=g_{\chi}^a=0$. In each case, we have adopted a value for the couplings to obtain the desired thermal relic abundance, $\Omega_{\chi} h^2 =0.12$. In the lower frames, we plot for the case of $m_{\chi}=110$ GeV the parameter space in which the predicted thermal relic abundance matches the observed dark matter density (solid black line) and the region in which the Galactic Center gamma-ray excess can be generated (green shaded band). The upper boundary of this region corresponds to the upper limit on the annihilation cross section from Fermi's observations of dwarf galaxies~\cite{Ackermann:2015zua} (green solid). The red regions are currently excluded by the LUX direct detection experiment~\cite{Akerib:2013tjd}, while orange and yellow regions are within the projected reach of XENON1T and LZ, respectively~\cite{Cushman:2013zza}. Unless the dark matter is quite light, this model predicts cross sections that lie below the neutrino floor (blue regions), making it difficult for dark matter to be detected by any planned direct detection experiment.}
\label{ZhTchThermal}
\end{figure*}

In the expression for the scalar quark coupling, $f_q^{(0)}$, the first and second lines correspond to $ZZ$ and $hh$ vertex corrections, respectively. Lines 3-5 and 6-7 of this expression result from the $ZZ$ and $hh$ box diagrams, respectively. The first and second lines of the expression for $f_q^{(1)}$ arise from the $ZZ$ box and $hZ$ vertex correction, respectively. The quantities $f_q^{(2)}$ and $f_g^{(0)}$ each result from the $ZZ$ box diagram.

In the upper frames of Fig.~\ref{ZhTchThermal}, we plot the spin-independent elastic scattering cross section per nucleon as a function of the dark matter mass, for two values of the mediator mass, $m_{\chi_2}$ and for two selections of couplings: $\lambda_s=g_{\chi}^a\ne 0$, $\lambda_p=g_{\chi}^v=0$ and $\lambda_p=g_{\chi}^v \ne 0$, $\lambda_s=g_{\chi}^a=0$ in the left and right frames, respectively. By setting these combinations of couplings to zero, we both minimize CP violation and approximately maximize the fraction of annihilations that proceed to $hZ$, as opposed to $ZZ$ or $hh$ final states. In particular, these combinations assure that the annihilation cross section to $hh$ is $p$-wave suppressed, and that over much of the parameter space $\sigma_{hZ} v/\sigma_{ZZ} v \gsim (m_{\chi_2}/m_{\chi_1})^2 \, (\lambda_{s,p}/g^{a,v}_{\chi})^2$. In each case, the non-zero couplings have been set to obtain the desired thermal relic abundance, $\Omega_{\chi} h^2 =0.12$. While much of this parameter space predicts an elastic scattering cross section that lies below the neutrino floor, this model could potentially lead to observable rates in future direct detection experiments if the dark matter is lighter than $\sim$200 GeV (which is also required if the dark matter is to account for the Galactic Center gamma-ray excess).

In the lower frames of Fig.~\ref{ZhTchThermal}, we plot some of the phenomenological features of this model for the case of $m_{\chi}=110$ GeV (again for  either $\lambda_p=g_{\chi}^v=0$ or $\lambda_s=g_{\chi}^a=0$). For scenarios with a relatively light mediator, the annihilation cross section is dominantly $s$-wave, allowing this model to generate the observed normalization of the Galactic Center gamma-ray excess. Although current constraints do not yet restrict this model, future direct detection experiments are expected to be sensitive to this scenario. 

\begin{table*}[thb]
\begin{centering}
\renewcommand{\arraystretch}{1.1}
\begin{tabular}{| c | c | c | c |} 
\hline
Final State & Mediator & Annihilation Cross Section & Direct Detection Prospects  
\\
\hline\hline 
 \multirow{4}{*}{$hh$ (Sec. V)} & \multirow{2}{*}{$s$-channel, spin-zero} & \multirow{2}{*}{$s$-wave if $\lambda_s \ne 0$} & Difficult, but possible if $\lambda_s \ne 0$ \\ &&& and $m_{\rm med} \lesssim \mathcal O(100)$ GeV \\ \cline{2-4}
 & \multirow{2}{*}{$t$-channel, fermion} & \multirow{2}{*}{$s$-wave if $\lambda_s \ne 0$ and $\lambda_p \ne 0$} & Possible if $\lambda_s \gg \lambda_p$ \\&&& or $\lambda_s \ll \lambda_p$ \\ \hline
\hline
 \multirow{5}{*}{$hZ$ (Sec. VI)} & $s$-channel, spin-zero & $s$-wave if $\lambda_p \ne 0$ & Very Difficult \\ \cline{2-4}
 &  \multirow{2}{*}{$s$-channel, spin-one} & $s$-wave with comparable  & Largely excluded by LUX in
 \\ & & $p$-wave contribution &  a UV-complete theory \\ \cline{2-4}
 &  \multirow{2}{*}{$t$-channel, fermion} & $s$-wave with comparable & Typically observable if
 \\ & & $p$-wave contribution &  $m_{\rm DM} \lsim 200$ GeV
 \\ \hline
 \end{tabular}
\caption{A brief summary of the prospects for indirect and direct detection in the case of Majorana dark matter that annihilates to $hh$ or $hZ$.}
\label{final table} 
\end{centering}
\end{table*}

\section{Discussion and Summary}
\label{summary}

Direct detection experiments have become increasingly powerful in recent years. We are now able to test and exclude many models in which the dark matter couples at tree-level to Standard Model quarks. The null results of these experiments motivate us to consider dark matter candidates that instead couple primarily to gauge and/or Higgs bosons. Although such interactions can lead to efficient tree-level annihilations in the early universe, they only generate a cross section for elastic scattering with nuclei at the loop-level.

Past studies have considered the elastic scattering of dark matter through $W^{\pm}$ and $Z$ loops in some detail. In this paper, we have instead focused on scenarios in which the dark matter is a Majorana fermion that couples and annihilates to $hh$ and/or $hZ$. In addition to generating small elastic scattering cross sections (consistent with existing direct detection constraints), dark matter that annihilates to $hh$ or $hZ$ final states can provide an acceptable fit to the spectrum of the Galactic Center gamma-ray excess. Within a simplified model framework, we have assembled an exhaustive list of possible descriptions for a Majorana dark matter particle that annihilates at tree level to $hh$ or $hZ$. For each simplified model, we have calculated the annihilation cross section and the loop-induced elastic scattering cross section with nuclei. We have presented these cross sections in a format that can be straightforwardly applied to UV-complete models of interest.

We have identified several particularly interesting scenarios, including those in which the dark matter annihilates to $hh$ or $hZ$ through the $s$-channel exchange of a spin-zero mediator, or through the $t$-channel exchange of a fermion. Although we also considered spin-one mediated annihilations to $hZ$, we found that such models are excluded by LUX unless the dark matter is very near resonance, $m_Z' \simeq 2 m_{\chi}$. Future direct detection experiments such as  XENON1T and LZ are expected to be sensitive to several of the cases considered here. In particular, the fermion mediated models often yield cross sections that are within the anticipated reach of LZ. In contrast, models in which the dark matter annihilates to $hh$ or $hZ$ through a spin-zero mediator could plausibly exhibit elastic scattering cross sections that lie below the neutrino floor, making it possible that they will remain beyond the reach of direct detection for the foreseeable future.

We note that several of the models considered here exhibit non-negligible low-velocity annihilation cross sections, allowing for potentially observable indirect detection signals (and for the ability to account for the Galactic Center gamma-ray excess).  More specifically, each of the phenomenologically viable models considered in this paper yield significant low-velocity annihilation cross sections, so long as the mediator's coupling to the dark matter has a non-negligible pseudoscalar component. Models with purely scalar couplings, however, can lead to $p$-wave annihilation amplitudes, and highly suppressed signals for indirect detection.

In Table~\ref{final table}, we summarize some of the most salient features of the models discussed in this paper. Given the breadth of the available parameter space, it is difficult to make simple statements that capture all of the possible phenomenology. Nonetheless, certain broad features of a given model can help us to project whether a future discovery with direct or indirect detection experiments is likely.


\bigskip \bigskip \bigskip

{\it Acknowledgements}: We would like to thank Mikhail Solon and Richard Hill for helpful discussions. AB is supported by the Kavli Institute for Cosmological Physics at the University of
Chicago through grant NSF PHY-1125897. DH is supported by the US Department of Energy under contract DE-FG02-13ER41958. Fermilab is operated by Fermi Research Alliance, LLC, under Contract No.~DE-AC02-07CH11359 with the US Department of Energy. This work was performed in part at the Aspen Center for Physics, which is supported by National Science Foundation grant PHY-1066293. SDM is supported by the National Science Foundation, PHY1316617.

\bibliography{higgsloops}

\begin{thebibliography}{10}%
\makeatletter
\providecommand \@ifxundefined [1]{%
 \ifx #1\undefined \expandafter \@firstoftwo
 \else \expandafter \@secondoftwo
\fi
}%
\providecommand \@ifnum [1]{%
 \ifnum #1\expandafter \@firstoftwo
 \else \expandafter \@secondoftwo
\fi
}%
\providecommand \enquote [1]{``#1''}%
\providecommand \bibnamefont  [1]{#1}%
\providecommand \bibfnamefont [1]{#1}%
\providecommand \citenamefont [1]{#1}%
\providecommand\href[0]{\@sanitize\@href}%
\providecommand\@href[1]{\endgroup\@@startlink{#1}\endgroup\@@href}%
\providecommand\@@href[1]{#1\@@endlink}%
\providecommand \@sanitize [0]{\begingroup\catcode`\&12\catcode`\#12\relax}%
\@ifxundefined \pdfoutput {\@firstoftwo}{%
 \@ifnum{\z@=\pdfoutput}{\@firstoftwo}{\@secondoftwo}%
}{%
 \providecommand\@@startlink[1]{\leavevmode\special{html:<a href="#1">}}%
 \providecommand\@@endlink[0]{\special{html:</a>}}%
}{%
 \providecommand\@@startlink[1]{%
  \leavevmode
  \pdfstartlink
   attr{/Border[0 0 1 ]/H/I/C[0 1 1]}%
   user{/Subtype/Link/A<</Type/Action/S/URI/URI(#1)>>}%
  \relax
 }%
 \providecommand\@@endlink[0]{\pdfendlink}%
}%
\providecommand \url  [0]{\begingroup\@sanitize \@url }%
\providecommand \@url [1]{\endgroup\@href {#1}{\urlprefix}}%
\providecommand \urlprefix [0]{URL }%
\providecommand \Eprint[0]{\href }%
\@ifxundefined \urlstyle {%
  \providecommand \doi [1]{doi:\discretionary{}{}{}#1}%
}{%
  \providecommand \doi [0]{doi:\discretionary{}{}{}\begingroup
  \urlstyle{rm}\Url }%
}%
\providecommand \doibase [0]{http://dx.doi.org/}%
\providecommand \Doi[1]{\href{\doibase#1}}%
\providecommand \bibAnnote [3]{%
  \BibitemShut{#1}%
  \begin{quotation}\noindent
    \textsc{Key:}\ #2\\\textsc{Annotation:}\ #3%
  \end{quotation}%
}%
\providecommand \bibAnnoteFile [2]{%
  \IfFileExists{#2}{\bibAnnote {#1} {#2} {\input{#2}}}{}%
}%
\providecommand \typeout [0]{\immediate \write \m@ne }%
\providecommand \selectlanguage [0]{\@gobble}%
\providecommand \bibinfo [0]{\@secondoftwo}%
\providecommand \bibfield [0]{\@secondoftwo}%
\providecommand \translation [1]{[#1]}%
\providecommand \BibitemOpen[0]{}%
\providecommand \bibitemStop [0]{}%
\providecommand \bibitemNoStop [0]{.\EOS\space}%
\providecommand \EOS [0]{\spacefactor3000\relax}%
\providecommand \BibitemShut [1]{\csname bibitem#1\endcsname}%
\bibitem{Steigman:2012nb}%
  \BibitemOpen
  \bibfield{author}{%
  \bibinfo {author} {\bibfnamefont{G.}~\bibnamefont{Steigman}}, \bibinfo
  {author} {\bibfnamefont{B.}~\bibnamefont{Dasgupta}},\ and\ \bibinfo {author}
  {\bibfnamefont{J.~F.}\ \bibnamefont{Beacom}},\ }%
  \bibfield{journal}{%
  \Doi{10.1103/PhysRevD.86.023506}{\bibinfo {journal} {Phys.Rev.}}\ }%
  \textbf{\bibinfo {volume} {D86}},\ \bibinfo {pages} {023506} (\bibinfo {year}
  {2012}),\ \Eprint{http://arxiv.org/abs/1204.3622}{arXiv:1204.3622 [hep-ph]}%
  \bibAnnoteFile{NoStop}{Steigman:2012nb}%
\bibitem{Drees:2015exa}%
  \BibitemOpen
  \bibfield{author}{%
  \bibinfo {author} {\bibfnamefont{M.}~\bibnamefont{Drees}}, \bibinfo {author}
  {\bibfnamefont{F.}~\bibnamefont{Hajkarim}},\ and\ \bibinfo {author}
  {\bibfnamefont{E.~R.}\ \bibnamefont{Schmitz}},\ }%
  \bibfield{journal}{%
  \Doi{10.1088/1475-7516/2015/06/025}{\bibinfo {journal} {JCAP}}\ }%
  \textbf{\bibinfo {volume} {1506}},\ \bibinfo {pages} {025} (\bibinfo {year}
  {2015}),\ \Eprint{http://arxiv.org/abs/1503.03513}{arXiv:1503.03513
  [hep-ph]}%
  \bibAnnoteFile{NoStop}{Drees:2015exa}%
\bibitem{Griest:1990kh}%
  \BibitemOpen
  \bibfield{author}{%
  \bibinfo {author} {\bibfnamefont{K.}~\bibnamefont{Griest}}\ and\ \bibinfo
  {author} {\bibfnamefont{D.}~\bibnamefont{Seckel}},\ }%
  \bibfield{journal}{%
  \Doi{10.1103/PhysRevD.43.3191}{\bibinfo {journal} {Phys.Rev.}}\ }%
  \textbf{\bibinfo {volume} {D43}},\ \bibinfo {pages} {3191} (\bibinfo {year}
  {1991})%
  \bibAnnoteFile{NoStop}{Griest:1990kh}%
\bibitem{Hooper:2013qjx}%
  \BibitemOpen
  \bibfield{author}{%
  \bibinfo {author} {\bibfnamefont{D.}~\bibnamefont{Hooper}}, \bibinfo {author}
  {\bibfnamefont{C.}~\bibnamefont{Kelso}}, \bibinfo {author}
  {\bibfnamefont{P.}~\bibnamefont{Sandick}},\ and\ \bibinfo {author}
  {\bibfnamefont{W.}~\bibnamefont{Xue}},\ }%
  \bibfield{journal}{%
  \Doi{10.1103/PhysRevD.88.015010}{\bibinfo {journal} {Phys.Rev.}}\ }%
  \textbf{\bibinfo {volume} {D88}},\ \bibinfo {pages} {015010} (\bibinfo {year}
  {2013}),\ \Eprint{http://arxiv.org/abs/1304.2417}{arXiv:1304.2417 [hep-ph]}%
  \bibAnnoteFile{NoStop}{Hooper:2013qjx}%
\bibitem{Edsjo:1997bg}%
  \BibitemOpen
  \bibfield{author}{%
  \bibinfo {author} {\bibfnamefont{J.}~\bibnamefont{Edsjo}}\ and\ \bibinfo
  {author} {\bibfnamefont{P.}~\bibnamefont{Gondolo}},\ }%
  \bibfield{journal}{%
  \Doi{10.1103/PhysRevD.56.1879}{\bibinfo {journal} {Phys. Rev.}}\ }%
  \textbf{\bibinfo {volume} {D56}},\ \bibinfo {pages} {1879} (\bibinfo {year}
  {1997}),\ \Eprint{http://arxiv.org/abs/hep-ph/9704361}{arXiv:hep-ph/9704361
  [hep-ph]}%
  \bibAnnoteFile{NoStop}{Edsjo:1997bg}%
\bibitem{Boehm:2014hva}%
  \BibitemOpen
  \bibfield{author}{%
  \bibinfo {author} {\bibfnamefont{C.}~\bibnamefont{Boehm}}, \bibinfo {author}
  {\bibfnamefont{M.~J.}\ \bibnamefont{Dolan}}, \bibinfo {author}
  {\bibfnamefont{C.}~\bibnamefont{McCabe}}, \bibinfo {author}
  {\bibfnamefont{M.}~\bibnamefont{Spannowsky}},\ and\ \bibinfo {author}
  {\bibfnamefont{C.~J.}\ \bibnamefont{Wallace}}}%
   (\bibinfo {year} {2014}),\
  \Eprint{http://arxiv.org/abs/1401.6458}{arXiv:1401.6458 [hep-ph]}%
  \bibAnnoteFile{NoStop}{Boehm:2014hva}%
\bibitem{Berlin:2014tja}%
  \BibitemOpen
  \bibfield{author}{%
  \bibinfo {author} {\bibfnamefont{A.}~\bibnamefont{Berlin}}, \bibinfo {author}
  {\bibfnamefont{D.}~\bibnamefont{Hooper}},\ and\ \bibinfo {author}
  {\bibfnamefont{S.~D.}\ \bibnamefont{McDermott}}}%
   (\bibinfo {year} {2014}),\
  \Eprint{http://arxiv.org/abs/1404.0022}{arXiv:1404.0022 [hep-ph]}%
  \bibAnnoteFile{NoStop}{Berlin:2014tja}%
\bibitem{Izaguirre:2014vva}%
  \BibitemOpen
  \bibfield{author}{%
  \bibinfo {author} {\bibfnamefont{E.}~\bibnamefont{Izaguirre}}, \bibinfo
  {author} {\bibfnamefont{G.}~\bibnamefont{Krnjaic}},\ and\ \bibinfo {author}
  {\bibfnamefont{B.}~\bibnamefont{Shuve}},\ }%
  \bibfield{journal}{%
  \Doi{10.1103/PhysRevD.90.055002}{\bibinfo {journal} {Phys. Rev.}}\ }%
  \textbf{\bibinfo {volume} {D90}},\ \bibinfo {pages} {055002} (\bibinfo {year}
  {2014}),\ \Eprint{http://arxiv.org/abs/1404.2018}{arXiv:1404.2018 [hep-ph]}%
  \bibAnnoteFile{NoStop}{Izaguirre:2014vva}%
\bibitem{Ipek:2014gua}%
  \BibitemOpen
  \bibfield{author}{%
  \bibinfo {author} {\bibfnamefont{S.}~\bibnamefont{Ipek}}, \bibinfo {author}
  {\bibfnamefont{D.}~\bibnamefont{McKeen}},\ and\ \bibinfo {author}
  {\bibfnamefont{A.~E.}\ \bibnamefont{Nelson}},\ }%
  \bibfield{journal}{%
  \Doi{10.1103/PhysRevD.90.055021}{\bibinfo {journal} {Phys. Rev.}}\ }%
  \textbf{\bibinfo {volume} {D90}},\ \bibinfo {pages} {055021} (\bibinfo {year}
  {2014}),\ \Eprint{http://arxiv.org/abs/1404.3716}{arXiv:1404.3716 [hep-ph]}%
  \bibAnnoteFile{NoStop}{Ipek:2014gua}%
\bibitem{Pospelov:2007mp}%
  \BibitemOpen
  \bibfield{author}{%
  \bibinfo {author} {\bibfnamefont{M.}~\bibnamefont{Pospelov}}, \bibinfo
  {author} {\bibfnamefont{A.}~\bibnamefont{Ritz}},\ and\ \bibinfo {author}
  {\bibfnamefont{M.~B.}\ \bibnamefont{Voloshin}},\ }%
  \bibfield{journal}{%
  \Doi{10.1016/j.physletb.2008.02.052}{\bibinfo {journal} {Phys.Lett.}}\ }%
  \textbf{\bibinfo {volume} {B662}},\ \bibinfo {pages} {53} (\bibinfo {year}
  {2008}),\ \Eprint{http://arxiv.org/abs/0711.4866}{arXiv:0711.4866 [hep-ph]}%
  \bibAnnoteFile{NoStop}{Pospelov:2007mp}%
\bibitem{Pospelov:2008jd}%
  \BibitemOpen
  \bibfield{author}{%
  \bibinfo {author} {\bibfnamefont{M.}~\bibnamefont{Pospelov}}\ and\ \bibinfo
  {author} {\bibfnamefont{A.}~\bibnamefont{Ritz}},\ }%
  \bibfield{journal}{%
  \Doi{10.1016/j.physletb.2008.12.012}{\bibinfo {journal} {Phys. Lett.}}\ }%
  \textbf{\bibinfo {volume} {B671}},\ \bibinfo {pages} {391} (\bibinfo {year}
  {2009}),\ \Eprint{http://arxiv.org/abs/0810.1502}{arXiv:0810.1502 [hep-ph]}%
  \bibAnnoteFile{NoStop}{Pospelov:2008jd}%
\bibitem{ArkaniHamed:2008qn}%
  \BibitemOpen
  \bibfield{author}{%
  \bibinfo {author} {\bibfnamefont{N.}~\bibnamefont{Arkani-Hamed}}, \bibinfo
  {author} {\bibfnamefont{D.~P.}\ \bibnamefont{Finkbeiner}}, \bibinfo {author}
  {\bibfnamefont{T.~R.}\ \bibnamefont{Slatyer}},\ and\ \bibinfo {author}
  {\bibfnamefont{N.}~\bibnamefont{Weiner}},\ }%
  \bibfield{journal}{%
  \Doi{10.1103/PhysRevD.79.015014}{\bibinfo {journal} {Phys. Rev.}}\ }%
  \textbf{\bibinfo {volume} {D79}},\ \bibinfo {pages} {015014} (\bibinfo {year}
  {2009}),\ \Eprint{http://arxiv.org/abs/0810.0713}{arXiv:0810.0713 [hep-ph]}%
  \bibAnnoteFile{NoStop}{ArkaniHamed:2008qn}%
\bibitem{Cholis:2008qq}%
  \BibitemOpen
  \bibfield{author}{%
  \bibinfo {author} {\bibfnamefont{I.}~\bibnamefont{Cholis}}, \bibinfo {author}
  {\bibfnamefont{D.~P.}\ \bibnamefont{Finkbeiner}}, \bibinfo {author}
  {\bibfnamefont{L.}~\bibnamefont{Goodenough}},\ and\ \bibinfo {author}
  {\bibfnamefont{N.}~\bibnamefont{Weiner}},\ }%
  \bibfield{journal}{%
  \Doi{10.1088/1475-7516/2009/12/007}{\bibinfo {journal} {JCAP}}\ }%
  \textbf{\bibinfo {volume} {0912}},\ \bibinfo {pages} {007} (\bibinfo {year}
  {2009}),\ \Eprint{http://arxiv.org/abs/0810.5344}{arXiv:0810.5344
  [astro-ph]}%
  \bibAnnoteFile{NoStop}{Cholis:2008qq}%
\bibitem{Morrissey:2009ur}%
  \BibitemOpen
  \bibfield{author}{%
  \bibinfo {author} {\bibfnamefont{D.~E.}\ \bibnamefont{Morrissey}}, \bibinfo
  {author} {\bibfnamefont{D.}~\bibnamefont{Poland}},\ and\ \bibinfo {author}
  {\bibfnamefont{K.~M.}\ \bibnamefont{Zurek}},\ }%
  \bibfield{journal}{%
  \Doi{10.1088/1126-6708/2009/07/050}{\bibinfo {journal} {JHEP}}\ }%
  \textbf{\bibinfo {volume} {07}},\ \bibinfo {pages} {050} (\bibinfo {year}
  {2009}),\ \Eprint{http://arxiv.org/abs/0904.2567}{arXiv:0904.2567 [hep-ph]}%
  \bibAnnoteFile{NoStop}{Morrissey:2009ur}%
\bibitem{Meade:2009rb}%
  \BibitemOpen
  \bibfield{author}{%
  \bibinfo {author} {\bibfnamefont{P.}~\bibnamefont{Meade}}, \bibinfo {author}
  {\bibfnamefont{M.}~\bibnamefont{Papucci}},\ and\ \bibinfo {author}
  {\bibfnamefont{T.}~\bibnamefont{Volansky}}}%
   (\bibinfo {year} {2009}),\
  \Eprint{http://arxiv.org/abs/0901.2925}{arXiv:0901.2925 [hep-ph]}%
  \bibAnnoteFile{NoStop}{Meade:2009rb}%
\bibitem{Cohen:2010kn}%
  \BibitemOpen
  \bibfield{author}{%
  \bibinfo {author} {\bibfnamefont{T.}~\bibnamefont{Cohen}}, \bibinfo {author}
  {\bibfnamefont{D.~J.}\ \bibnamefont{Phalen}}, \bibinfo {author}
  {\bibfnamefont{A.}~\bibnamefont{Pierce}},\ and\ \bibinfo {author}
  {\bibfnamefont{K.~M.}\ \bibnamefont{Zurek}},\ }%
  \bibfield{journal}{%
  \Doi{10.1103/PhysRevD.82.056001}{\bibinfo {journal} {Phys. Rev.}}\ }%
  \textbf{\bibinfo {volume} {D82}},\ \bibinfo {pages} {056001} (\bibinfo {year}
  {2010}),\ \Eprint{http://arxiv.org/abs/1005.1655}{arXiv:1005.1655 [hep-ph]}%
  \bibAnnoteFile{NoStop}{Cohen:2010kn}%
\bibitem{Hooper:2012cw}%
  \BibitemOpen
  \bibfield{author}{%
  \bibinfo {author} {\bibfnamefont{D.}~\bibnamefont{Hooper}}, \bibinfo {author}
  {\bibfnamefont{N.}~\bibnamefont{Weiner}},\ and\ \bibinfo {author}
  {\bibfnamefont{W.}~\bibnamefont{Xue}},\ }%
  \bibfield{journal}{%
  \Doi{10.1103/PhysRevD.86.056009}{\bibinfo {journal} {Phys.Rev.}}\ }%
  \textbf{\bibinfo {volume} {D86}},\ \bibinfo {pages} {056009} (\bibinfo {year}
  {2012}),\ \Eprint{http://arxiv.org/abs/1206.2929}{arXiv:1206.2929 [hep-ph]}%
  \bibAnnoteFile{NoStop}{Hooper:2012cw}%
\bibitem{Berlin:2014pya}%
  \BibitemOpen
  \bibfield{author}{%
  \bibinfo {author} {\bibfnamefont{A.}~\bibnamefont{Berlin}}, \bibinfo {author}
  {\bibfnamefont{P.}~\bibnamefont{Gratia}}, \bibinfo {author}
  {\bibfnamefont{D.}~\bibnamefont{Hooper}},\ and\ \bibinfo {author}
  {\bibfnamefont{S.~D.}\ \bibnamefont{McDermott}},\ }%
  \bibfield{journal}{%
  \Doi{10.1103/PhysRevD.90.015032}{\bibinfo {journal} {Phys.Rev.}}\ }%
  \textbf{\bibinfo {volume} {D90}},\ \bibinfo {pages} {015032} (\bibinfo {year}
  {2014}),\ \Eprint{http://arxiv.org/abs/1405.5204}{arXiv:1405.5204 [hep-ph]}%
  \bibAnnoteFile{NoStop}{Berlin:2014pya}%
\bibitem{Ko:2014gha}%
  \BibitemOpen
  \bibfield{author}{%
  \bibinfo {author} {\bibfnamefont{P.}~\bibnamefont{Ko}}, \bibinfo {author}
  {\bibfnamefont{W.-I.}\ \bibnamefont{Park}},\ and\ \bibinfo {author}
  {\bibfnamefont{Y.}~\bibnamefont{Tang}},\ }%
  \bibfield{journal}{%
  \Doi{10.1088/1475-7516/2014/09/013}{\bibinfo {journal} {JCAP}}\ }%
  \textbf{\bibinfo {volume} {1409}},\ \bibinfo {pages} {013} (\bibinfo {year}
  {2014}),\ \Eprint{http://arxiv.org/abs/1404.5257}{arXiv:1404.5257 [hep-ph]}%
  \bibAnnoteFile{NoStop}{Ko:2014gha}%
\bibitem{Abdullah:2014lla}%
  \BibitemOpen
  \bibfield{author}{%
  \bibinfo {author} {\bibfnamefont{M.}~\bibnamefont{Abdullah}}, \bibinfo
  {author} {\bibfnamefont{A.}~\bibnamefont{DiFranzo}}, \bibinfo {author}
  {\bibfnamefont{A.}~\bibnamefont{Rajaraman}}, \bibinfo {author}
  {\bibfnamefont{T.~M.~P.}\ \bibnamefont{Tait}}, \bibinfo {author}
  {\bibfnamefont{P.}~\bibnamefont{Tanedo}},\ and\ \bibinfo {author}
  {\bibfnamefont{A.~M.}\ \bibnamefont{Wijangco}},\ }%
  \bibfield{journal}{%
  \Doi{10.1103/PhysRevD.90.035004}{\bibinfo {journal} {Phys. Rev.}}\ }%
  \textbf{\bibinfo {volume} {D90}},\ \bibinfo {pages} {035004} (\bibinfo {year}
  {2014}),\ \Eprint{http://arxiv.org/abs/1404.6528}{arXiv:1404.6528 [hep-ph]}%
  \bibAnnoteFile{NoStop}{Abdullah:2014lla}%
\bibitem{Boehm:2014bia}%
  \BibitemOpen
  \bibfield{author}{%
  \bibinfo {author} {\bibfnamefont{C.}~\bibnamefont{Boehm}}, \bibinfo {author}
  {\bibfnamefont{M.~J.}\ \bibnamefont{Dolan}},\ and\ \bibinfo {author}
  {\bibfnamefont{C.}~\bibnamefont{McCabe}},\ }%
  \bibfield{journal}{%
  \Doi{10.1103/PhysRevD.90.023531}{\bibinfo {journal} {Phys. Rev.}}\ }%
  \textbf{\bibinfo {volume} {D90}},\ \bibinfo {pages} {023531} (\bibinfo {year}
  {2014}),\ \Eprint{http://arxiv.org/abs/1404.4977}{arXiv:1404.4977 [hep-ph]}%
  \bibAnnoteFile{NoStop}{Boehm:2014bia}%
\bibitem{Fonseca:2015rwa}%
  \BibitemOpen
  \bibfield{author}{%
  \bibinfo {author} {\bibfnamefont{N.}~\bibnamefont{Fonseca}}, \bibinfo
  {author} {\bibfnamefont{L.}~\bibnamefont{Necib}},\ and\ \bibinfo {author}
  {\bibfnamefont{J.}~\bibnamefont{Thaler}}}%
   (\bibinfo {year} {2015}),\
  \Eprint{http://arxiv.org/abs/1507.08295}{arXiv:1507.08295 [hep-ph]}%
  \bibAnnoteFile{NoStop}{Fonseca:2015rwa}%
\bibitem{McDermott:2014rqa}%
  \BibitemOpen
  \bibfield{author}{%
  \bibinfo {author} {\bibfnamefont{S.~D.}\ \bibnamefont{McDermott}},\ }%
  \bibfield{journal}{%
  \Doi{10.1016/j.dark.2015.05.001}{\bibinfo {journal} {Phys. Dark Univ.}}\ }%
  \textbf{\bibinfo {volume} {7-8}},\ \bibinfo {pages} {12} (\bibinfo {year}
  {2014}),\ \Eprint{http://arxiv.org/abs/1406.6408}{arXiv:1406.6408 [hep-ph]}%
  \bibAnnoteFile{NoStop}{McDermott:2014rqa}%
\bibitem{Fornengo:2002db}%
  \BibitemOpen
  \bibfield{author}{%
  \bibinfo {author} {\bibfnamefont{N.}~\bibnamefont{Fornengo}}, \bibinfo
  {author} {\bibfnamefont{A.}~\bibnamefont{Riotto}},\ and\ \bibinfo {author}
  {\bibfnamefont{S.}~\bibnamefont{Scopel}},\ }%
  \bibfield{journal}{%
  \Doi{10.1103/PhysRevD.67.023514}{\bibinfo {journal} {Phys. Rev.}}\ }%
  \textbf{\bibinfo {volume} {D67}},\ \bibinfo {pages} {023514} (\bibinfo {year}
  {2003}),\ \Eprint{http://arxiv.org/abs/hep-ph/0208072}{arXiv:hep-ph/0208072
  [hep-ph]}%
  \bibAnnoteFile{NoStop}{Fornengo:2002db}%
\bibitem{Gelmini:2006pq}%
  \BibitemOpen
  \bibfield{author}{%
  \bibinfo {author} {\bibfnamefont{G.}~\bibnamefont{Gelmini}}, \bibinfo
  {author} {\bibfnamefont{P.}~\bibnamefont{Gondolo}}, \bibinfo {author}
  {\bibfnamefont{A.}~\bibnamefont{Soldatenko}},\ and\ \bibinfo {author}
  {\bibfnamefont{C.~E.}\ \bibnamefont{Yaguna}},\ }%
  \bibfield{journal}{%
  \Doi{10.1103/PhysRevD.74.083514}{\bibinfo {journal} {Phys. Rev.}}\ }%
  \textbf{\bibinfo {volume} {D74}},\ \bibinfo {pages} {083514} (\bibinfo {year}
  {2006}),\ \Eprint{http://arxiv.org/abs/hep-ph/0605016}{arXiv:hep-ph/0605016
  [hep-ph]}%
  \bibAnnoteFile{NoStop}{Gelmini:2006pq}%
\bibitem{Hooper:2013nia}%
  \BibitemOpen
  \bibfield{author}{%
  \bibinfo {author} {\bibfnamefont{D.}~\bibnamefont{Hooper}},\ }%
  \bibfield{journal}{%
  \Doi{10.1103/PhysRevD.88.083519}{\bibinfo {journal} {Phys.Rev.}}\ }%
  \textbf{\bibinfo {volume} {D88}},\ \bibinfo {pages} {083519} (\bibinfo {year}
  {2013}),\ \Eprint{http://arxiv.org/abs/1307.0826}{arXiv:1307.0826 [hep-ph]}%
  \bibAnnoteFile{NoStop}{Hooper:2013nia}%
\bibitem{Kane:2015jia}%
  \BibitemOpen
  \bibfield{author}{%
  \bibinfo {author} {\bibfnamefont{G.}~\bibnamefont{Kane}}, \bibinfo {author}
  {\bibfnamefont{K.}~\bibnamefont{Sinha}},\ and\ \bibinfo {author}
  {\bibfnamefont{S.}~\bibnamefont{Watson}},\ }%
  \bibfield{journal}{%
  \Doi{10.1142/S0218271815300220}{\bibinfo {journal} {Int. J. Mod. Phys.}}\ }%
  \textbf{\bibinfo {volume} {D24}},\ \bibinfo {pages} {1530022} (\bibinfo
  {year} {2015}),\ \Eprint{http://arxiv.org/abs/1502.07746}{arXiv:1502.07746
  [hep-th]}%
  \bibAnnoteFile{NoStop}{Kane:2015jia}%
\bibitem{Patwardhan:2015kga}%
  \BibitemOpen
  \bibfield{author}{%
  \bibinfo {author} {\bibfnamefont{A.~V.}\ \bibnamefont{Patwardhan}}, \bibinfo
  {author} {\bibfnamefont{G.~M.}\ \bibnamefont{Fuller}}, \bibinfo {author}
  {\bibfnamefont{C.~T.}\ \bibnamefont{Kishimoto}},\ and\ \bibinfo {author}
  {\bibfnamefont{A.}~\bibnamefont{Kusenko}}}%
   (\bibinfo {year} {2015}),\
  \Eprint{http://arxiv.org/abs/1507.01977}{arXiv:1507.01977 [astro-ph.CO]}%
  \bibAnnoteFile{NoStop}{Patwardhan:2015kga}%
\bibitem{Lyth:1995ka}%
  \BibitemOpen
  \bibfield{author}{%
  \bibinfo {author} {\bibfnamefont{D.~H.}\ \bibnamefont{Lyth}}\ and\ \bibinfo
  {author} {\bibfnamefont{E.~D.}\ \bibnamefont{Stewart}},\ }%
  \bibfield{journal}{%
  \Doi{10.1103/PhysRevD.53.1784}{\bibinfo {journal} {Phys. Rev.}}\ }%
  \textbf{\bibinfo {volume} {D53}},\ \bibinfo {pages} {1784} (\bibinfo {year}
  {1996}),\ \Eprint{http://arxiv.org/abs/hep-ph/9510204}{arXiv:hep-ph/9510204
  [hep-ph]}%
  \bibAnnoteFile{NoStop}{Lyth:1995ka}%
\bibitem{Cohen:2008nb}%
  \BibitemOpen
  \bibfield{author}{%
  \bibinfo {author} {\bibfnamefont{T.}~\bibnamefont{Cohen}}, \bibinfo {author}
  {\bibfnamefont{D.~E.}\ \bibnamefont{Morrissey}},\ and\ \bibinfo {author}
  {\bibfnamefont{A.}~\bibnamefont{Pierce}},\ }%
  \bibfield{journal}{%
  \Doi{10.1103/PhysRevD.78.111701}{\bibinfo {journal} {Phys. Rev.}}\ }%
  \textbf{\bibinfo {volume} {D78}},\ \bibinfo {pages} {111701} (\bibinfo {year}
  {2008}),\ \Eprint{http://arxiv.org/abs/0808.3994}{arXiv:0808.3994 [hep-ph]}%
  \bibAnnoteFile{NoStop}{Cohen:2008nb}%
\bibitem{Boeckel:2011yj}%
  \BibitemOpen
  \bibfield{author}{%
  \bibinfo {author} {\bibfnamefont{T.}~\bibnamefont{Boeckel}}\ and\ \bibinfo
  {author} {\bibfnamefont{J.}~\bibnamefont{Schaffner-Bielich}},\ }%
  \bibfield{journal}{%
  \Doi{10.1103/PhysRevD.85.103506}{\bibinfo {journal} {Phys. Rev.}}\ }%
  \textbf{\bibinfo {volume} {D85}},\ \bibinfo {pages} {103506} (\bibinfo {year}
  {2012}),\ \Eprint{http://arxiv.org/abs/1105.0832}{arXiv:1105.0832
  [astro-ph.CO]}%
  \bibAnnoteFile{NoStop}{Boeckel:2011yj}%
\bibitem{Boeckel:2009ej}%
  \BibitemOpen
  \bibfield{author}{%
  \bibinfo {author} {\bibfnamefont{T.}~\bibnamefont{Boeckel}}\ and\ \bibinfo
  {author} {\bibfnamefont{J.}~\bibnamefont{Schaffner-Bielich}},\ }%
  \bibfield{journal}{%
  \Doi{10.1103/PhysRevLett.105.041301, 10.1103/PhysRevLett.106.069901}{\bibinfo
  {journal} {Phys. Rev. Lett.}}\ }%
  \textbf{\bibinfo {volume} {105}},\ \bibinfo {pages} {041301} (\bibinfo {year}
  {2010}),\ \bibinfo {note} {[Erratum: Phys. Rev. Lett.106,069901(2011)]},\
  \Eprint{http://arxiv.org/abs/0906.4520}{arXiv:0906.4520 [astro-ph.CO]}%
  \bibAnnoteFile{NoStop}{Boeckel:2009ej}%
\bibitem{Davoudiasl:2015vba}%
  \BibitemOpen
  \bibfield{author}{%
  \bibinfo {author} {\bibfnamefont{H.}~\bibnamefont{Davoudiasl}}, \bibinfo
  {author} {\bibfnamefont{D.}~\bibnamefont{Hooper}},\ and\ \bibinfo {author}
  {\bibfnamefont{S.~D.}\ \bibnamefont{McDermott}}}%
   (\bibinfo {year} {2015}),\
  \Eprint{http://arxiv.org/abs/1507.08660}{arXiv:1507.08660 [hep-ph]}%
  \bibAnnoteFile{NoStop}{Davoudiasl:2015vba}%
\bibitem{Drees:1992rr}%
  \BibitemOpen
  \bibfield{author}{%
  \bibinfo {author} {\bibfnamefont{M.}~\bibnamefont{Drees}}\ and\ \bibinfo
  {author} {\bibfnamefont{M.~M.}\ \bibnamefont{Nojiri}},\ }%
  \bibfield{journal}{%
  \Doi{10.1103/PhysRevD.47.4226}{\bibinfo {journal} {Phys.Rev.}}\ }%
  \textbf{\bibinfo {volume} {D47}},\ \bibinfo {pages} {4226} (\bibinfo {year}
  {1993}),\ \Eprint{http://arxiv.org/abs/hep-ph/9210272}{arXiv:hep-ph/9210272
  [hep-ph]}%
  \bibAnnoteFile{NoStop}{Drees:1992rr}%
\bibitem{Hisano:2010ct}%
  \BibitemOpen
  \bibfield{author}{%
  \bibinfo {author} {\bibfnamefont{J.}~\bibnamefont{Hisano}}, \bibinfo {author}
  {\bibfnamefont{K.}~\bibnamefont{Ishiwata}},\ and\ \bibinfo {author}
  {\bibfnamefont{N.}~\bibnamefont{Nagata}},\ }%
  \bibfield{journal}{%
  \Doi{10.1103/PhysRevD.82.115007}{\bibinfo {journal} {Phys.Rev.}}\ }%
  \textbf{\bibinfo {volume} {D82}},\ \bibinfo {pages} {115007} (\bibinfo {year}
  {2010}),\ \Eprint{http://arxiv.org/abs/1007.2601}{arXiv:1007.2601 [hep-ph]}%
  \bibAnnoteFile{NoStop}{Hisano:2010ct}%
\bibitem{Hisano:2011cs}%
  \BibitemOpen
  \bibfield{author}{%
  \bibinfo {author} {\bibfnamefont{J.}~\bibnamefont{Hisano}}, \bibinfo {author}
  {\bibfnamefont{K.}~\bibnamefont{Ishiwata}}, \bibinfo {author}
  {\bibfnamefont{N.}~\bibnamefont{Nagata}},\ and\ \bibinfo {author}
  {\bibfnamefont{T.}~\bibnamefont{Takesako}},\ }%
  \bibfield{journal}{%
  \Doi{10.1007/JHEP07(2011)005}{\bibinfo {journal} {JHEP}}\ }%
  \textbf{\bibinfo {volume} {1107}},\ \bibinfo {pages} {005} (\bibinfo {year}
  {2011}),\ \Eprint{http://arxiv.org/abs/1104.0228}{arXiv:1104.0228 [hep-ph]}%
  \bibAnnoteFile{NoStop}{Hisano:2011cs}%
\bibitem{Hisano:2015rsa}%
  \BibitemOpen
  \bibfield{author}{%
  \bibinfo {author} {\bibfnamefont{J.}~\bibnamefont{Hisano}}, \bibinfo {author}
  {\bibfnamefont{K.}~\bibnamefont{Ishiwata}},\ and\ \bibinfo {author}
  {\bibfnamefont{N.}~\bibnamefont{Nagata}}}%
   (\bibinfo {year} {2015}),\
  \Eprint{http://arxiv.org/abs/1504.00915}{arXiv:1504.00915 [hep-ph]}%
  \bibAnnoteFile{NoStop}{Hisano:2015rsa}%
\bibitem{Hisano:2015bma}%
  \BibitemOpen
  \bibfield{author}{%
  \bibinfo {author} {\bibfnamefont{J.}~\bibnamefont{Hisano}}, \bibinfo {author}
  {\bibfnamefont{R.}~\bibnamefont{Nagai}},\ and\ \bibinfo {author}
  {\bibfnamefont{N.}~\bibnamefont{Nagata}},\ }%
  \bibfield{journal}{%
  \Doi{10.1007/JHEP05(2015)037}{\bibinfo {journal} {JHEP}}\ }%
  \textbf{\bibinfo {volume} {1505}},\ \bibinfo {pages} {037} (\bibinfo {year}
  {2015}),\ \Eprint{http://arxiv.org/abs/1502.02244}{arXiv:1502.02244
  [hep-ph]}%
  \bibAnnoteFile{NoStop}{Hisano:2015bma}%
\bibitem{Ibarra:2015fqa}%
  \BibitemOpen
  \bibfield{author}{%
  \bibinfo {author} {\bibfnamefont{A.}~\bibnamefont{Ibarra}}\ and\ \bibinfo
  {author} {\bibfnamefont{S.}~\bibnamefont{Wild}},\ }%
  \bibfield{journal}{%
  \Doi{10.1088/1475-7516/2015/05/047}{\bibinfo {journal} {JCAP}}\ }%
  \textbf{\bibinfo {volume} {1505}},\ \bibinfo {pages} {047} (\bibinfo {year}
  {2015}),\ \Eprint{http://arxiv.org/abs/1503.03382}{arXiv:1503.03382
  [hep-ph]}%
  \bibAnnoteFile{NoStop}{Ibarra:2015fqa}%
\bibitem{Hill:2014yka}%
  \BibitemOpen
  \bibfield{author}{%
  \bibinfo {author} {\bibfnamefont{R.~J.}\ \bibnamefont{Hill}}\ and\ \bibinfo
  {author} {\bibfnamefont{M.~P.}\ \bibnamefont{Solon}},\ }%
  \bibfield{journal}{%
  \Doi{10.1103/PhysRevD.91.043504}{\bibinfo {journal} {Phys.Rev.}}\ }%
  \textbf{\bibinfo {volume} {D91}},\ \bibinfo {pages} {043504} (\bibinfo {year}
  {2015}),\ \Eprint{http://arxiv.org/abs/1401.3339}{arXiv:1401.3339 [hep-ph]}%
  \bibAnnoteFile{NoStop}{Hill:2014yka}%
\bibitem{Hill:2014yxa}%
  \BibitemOpen
  \bibfield{author}{%
  \bibinfo {author} {\bibfnamefont{R.~J.}\ \bibnamefont{Hill}}\ and\ \bibinfo
  {author} {\bibfnamefont{M.~P.}\ \bibnamefont{Solon}},\ }%
  \bibfield{journal}{%
  \Doi{10.1103/PhysRevD.91.043505}{\bibinfo {journal} {Phys.Rev.}}\ }%
  \textbf{\bibinfo {volume} {D91}},\ \bibinfo {pages} {043505} (\bibinfo {year}
  {2015}),\ \Eprint{http://arxiv.org/abs/1409.8290}{arXiv:1409.8290 [hep-ph]}%
  \bibAnnoteFile{NoStop}{Hill:2014yxa}%
\bibitem{Agrawal:2011ze}%
  \BibitemOpen
  \bibfield{author}{%
  \bibinfo {author} {\bibfnamefont{P.}~\bibnamefont{Agrawal}}, \bibinfo
  {author} {\bibfnamefont{S.}~\bibnamefont{Blanchet}}, \bibinfo {author}
  {\bibfnamefont{Z.}~\bibnamefont{Chacko}},\ and\ \bibinfo {author}
  {\bibfnamefont{C.}~\bibnamefont{Kilic}},\ }%
  \bibfield{journal}{%
  \Doi{10.1103/PhysRevD.86.055002}{\bibinfo {journal} {Phys.Rev.}}\ }%
  \textbf{\bibinfo {volume} {D86}},\ \bibinfo {pages} {055002} (\bibinfo {year}
  {2012}),\ \Eprint{http://arxiv.org/abs/1109.3516}{arXiv:1109.3516 [hep-ph]}%
  \bibAnnoteFile{NoStop}{Agrawal:2011ze}%
\bibitem{Goodenough:2009gk}%
  \BibitemOpen
  \bibfield{author}{%
  \bibinfo {author} {\bibfnamefont{L.}~\bibnamefont{Goodenough}}\ and\ \bibinfo
  {author} {\bibfnamefont{D.}~\bibnamefont{Hooper}}}%
   (\bibinfo {year} {2009}),\
  \Eprint{http://arxiv.org/abs/0910.2998}{arXiv:0910.2998 [hep-ph]}%
  \bibAnnoteFile{NoStop}{Goodenough:2009gk}%
\bibitem{Hooper:2010mq}%
  \BibitemOpen
  \bibfield{author}{%
  \bibinfo {author} {\bibfnamefont{D.}~\bibnamefont{Hooper}}\ and\ \bibinfo
  {author} {\bibfnamefont{L.}~\bibnamefont{Goodenough}},\ }%
  \bibfield{journal}{%
  \Doi{10.1016/j.physletb.2011.02.029}{\bibinfo {journal} {Phys.Lett.}}\ }%
  \textbf{\bibinfo {volume} {B697}},\ \bibinfo {pages} {412} (\bibinfo {year}
  {2011}),\ \Eprint{http://arxiv.org/abs/1010.2752}{arXiv:1010.2752 [hep-ph]}%
  \bibAnnoteFile{NoStop}{Hooper:2010mq}%
\bibitem{Boyarsky:2010dr}%
  \BibitemOpen
  \bibfield{author}{%
  \bibinfo {author} {\bibfnamefont{A.}~\bibnamefont{Boyarsky}}, \bibinfo
  {author} {\bibfnamefont{D.}~\bibnamefont{Malyshev}},\ and\ \bibinfo {author}
  {\bibfnamefont{O.}~\bibnamefont{Ruchayskiy}},\ }%
  \bibfield{journal}{%
  \Doi{10.1016/j.physletb.2011.10.014}{\bibinfo {journal} {Phys.Lett.}}\ }%
  \textbf{\bibinfo {volume} {B705}},\ \bibinfo {pages} {165} (\bibinfo {year}
  {2011}),\ \Eprint{http://arxiv.org/abs/1012.5839}{arXiv:1012.5839 [hep-ph]}%
  \bibAnnoteFile{NoStop}{Boyarsky:2010dr}%
\bibitem{Hooper:2011ti}%
  \BibitemOpen
  \bibfield{author}{%
  \bibinfo {author} {\bibfnamefont{D.}~\bibnamefont{Hooper}}\ and\ \bibinfo
  {author} {\bibfnamefont{T.}~\bibnamefont{Linden}},\ }%
  \bibfield{journal}{%
  \Doi{10.1103/PhysRevD.84.123005}{\bibinfo {journal} {Phys.Rev.}}\ }%
  \textbf{\bibinfo {volume} {D84}},\ \bibinfo {pages} {123005} (\bibinfo {year}
  {2011}),\ \Eprint{http://arxiv.org/abs/1110.0006}{arXiv:1110.0006
  [astro-ph.HE]}%
  \bibAnnoteFile{NoStop}{Hooper:2011ti}%
\bibitem{Abazajian:2012pn}%
  \BibitemOpen
  \bibfield{author}{%
  \bibinfo {author} {\bibfnamefont{K.~N.}\ \bibnamefont{Abazajian}}\ and\
  \bibinfo {author} {\bibfnamefont{M.}~\bibnamefont{Kaplinghat}},\ }%
  \bibfield{journal}{%
  \Doi{10.1103/PhysRevD.86.083511}{\bibinfo {journal} {Phys.Rev.}}\ }%
  \textbf{\bibinfo {volume} {D86}},\ \bibinfo {pages} {083511} (\bibinfo {year}
  {2012}),\ \Eprint{http://arxiv.org/abs/1207.6047}{arXiv:1207.6047
  [astro-ph.HE]}%
  \bibAnnoteFile{NoStop}{Abazajian:2012pn}%
\bibitem{Gordon:2013vta}%
  \BibitemOpen
  \bibfield{author}{%
  \bibinfo {author} {\bibfnamefont{C.}~\bibnamefont{Gordon}}\ and\ \bibinfo
  {author} {\bibfnamefont{O.}~\bibnamefont{Macias}},\ }%
  \bibfield{journal}{%
  \Doi{10.1103/PhysRevD.88.083521}{\bibinfo {journal} {Phys.Rev.}}\ }%
  \textbf{\bibinfo {volume} {D88}},\ \bibinfo {pages} {083521} (\bibinfo {year}
  {2013}),\ \Eprint{http://arxiv.org/abs/1306.5725}{arXiv:1306.5725
  [astro-ph.HE]}%
  \bibAnnoteFile{NoStop}{Gordon:2013vta}%
\bibitem{Hooper:2013rwa}%
  \BibitemOpen
  \bibfield{author}{%
  \bibinfo {author} {\bibfnamefont{D.}~\bibnamefont{Hooper}}\ and\ \bibinfo
  {author} {\bibfnamefont{T.~R.}\ \bibnamefont{Slatyer}},\ }%
  \bibfield{journal}{%
  \Doi{10.1016/j.dark.2013.06.003}{\bibinfo {journal} {Phys.Dark Univ.}}\ }%
  \textbf{\bibinfo {volume} {2}},\ \bibinfo {pages} {118} (\bibinfo {year}
  {2013}),\ \Eprint{http://arxiv.org/abs/1302.6589}{arXiv:1302.6589
  [astro-ph.HE]}%
  \bibAnnoteFile{NoStop}{Hooper:2013rwa}%
\bibitem{Huang:2013pda}%
  \BibitemOpen
  \bibfield{author}{%
  \bibinfo {author} {\bibfnamefont{W.-C.}\ \bibnamefont{Huang}}, \bibinfo
  {author} {\bibfnamefont{A.}~\bibnamefont{Urbano}},\ and\ \bibinfo {author}
  {\bibfnamefont{W.}~\bibnamefont{Xue}}}%
   (\bibinfo {year} {2013}),\
  \Eprint{http://arxiv.org/abs/1307.6862}{arXiv:1307.6862 [hep-ph]}%
  \bibAnnoteFile{NoStop}{Huang:2013pda}%
\bibitem{Abazajian:2014fta}%
  \BibitemOpen
  \bibfield{author}{%
  \bibinfo {author} {\bibfnamefont{K.~N.}\ \bibnamefont{Abazajian}}, \bibinfo
  {author} {\bibfnamefont{N.}~\bibnamefont{Canac}}, \bibinfo {author}
  {\bibfnamefont{S.}~\bibnamefont{Horiuchi}},\ and\ \bibinfo {author}
  {\bibfnamefont{M.}~\bibnamefont{Kaplinghat}}}%
   (\bibinfo {year} {2014}),\
  \Eprint{http://arxiv.org/abs/1402.4090}{arXiv:1402.4090 [astro-ph.HE]}%
  \bibAnnoteFile{NoStop}{Abazajian:2014fta}%
\bibitem{Daylan:2014rsa}%
  \BibitemOpen
  \bibfield{author}{%
  \bibinfo {author} {\bibfnamefont{T.}~\bibnamefont{Daylan}}, \bibinfo {author}
  {\bibfnamefont{D.~P.}\ \bibnamefont{Finkbeiner}}, \bibinfo {author}
  {\bibfnamefont{D.}~\bibnamefont{Hooper}}, \bibinfo {author}
  {\bibfnamefont{T.}~\bibnamefont{Linden}}, \bibinfo {author}
  {\bibfnamefont{S.~K.~N.}\ \bibnamefont{Portillo}}, \emph{et~al.}}%
   (\bibinfo {year} {2014}),\
  \Eprint{http://arxiv.org/abs/1402.6703}{arXiv:1402.6703 [astro-ph.HE]}%
  \bibAnnoteFile{NoStop}{Daylan:2014rsa}%
\bibitem{Calore:2014xka}%
  \BibitemOpen
  \bibfield{author}{%
  \bibinfo {author} {\bibfnamefont{F.}~\bibnamefont{Calore}}, \bibinfo {author}
  {\bibfnamefont{I.}~\bibnamefont{Cholis}},\ and\ \bibinfo {author}
  {\bibfnamefont{C.}~\bibnamefont{Weniger}}}%
   (\bibinfo {year} {2014}),\
  \Eprint{http://arxiv.org/abs/1409.0042}{arXiv:1409.0042 [astro-ph.CO]}%
  \bibAnnoteFile{NoStop}{Calore:2014xka}%
\bibitem{fermigc}%
  \BibitemOpen
  \bibfield{author}{%
  \bibinfo {author} {\bibfnamefont{S.}~\bibnamefont{Murgia}} (\bibinfo
  {collaboration} {Fermi-LAT Collaboration}),\ }%
  \bibfield{journal}{%
  \bibinfo {journal} {{Talk given at the 2014 Fermi Symposium, Nagoya, Japan,
  October 20-24}}}%
   (\bibinfo {year} {2014})%
  \bibAnnoteFile{NoStop}{fermigc}%
\bibitem{Agrawal:2014oha}%
  \BibitemOpen
  \bibfield{author}{%
  \bibinfo {author} {\bibfnamefont{P.}~\bibnamefont{Agrawal}}, \bibinfo
  {author} {\bibfnamefont{B.}~\bibnamefont{Batell}}, \bibinfo {author}
  {\bibfnamefont{P.~J.}\ \bibnamefont{Fox}},\ and\ \bibinfo {author}
  {\bibfnamefont{R.}~\bibnamefont{Harnik}},\ }%
  \bibfield{journal}{%
  \Doi{10.1088/1475-7516/2015/05/011}{\bibinfo {journal} {JCAP}}\ }%
  \textbf{\bibinfo {volume} {1505}},\ \bibinfo {pages} {011} (\bibinfo {year}
  {2015}),\ \Eprint{http://arxiv.org/abs/1411.2592}{arXiv:1411.2592 [hep-ph]}%
  \bibAnnoteFile{NoStop}{Agrawal:2014oha}%
\bibitem{Cheung:2012qy}%
  \BibitemOpen
  \bibfield{author}{%
  \bibinfo {author} {\bibfnamefont{C.}~\bibnamefont{Cheung}}, \bibinfo {author}
  {\bibfnamefont{L.~J.}\ \bibnamefont{Hall}}, \bibinfo {author}
  {\bibfnamefont{D.}~\bibnamefont{Pinner}},\ and\ \bibinfo {author}
  {\bibfnamefont{J.~T.}\ \bibnamefont{Ruderman}},\ }%
  \bibfield{journal}{%
  \Doi{10.1007/JHEP05(2013)100}{\bibinfo {journal} {JHEP}}\ }%
  \textbf{\bibinfo {volume} {1305}},\ \bibinfo {pages} {100} (\bibinfo {year}
  {2013}),\ \Eprint{http://arxiv.org/abs/1211.4873}{arXiv:1211.4873 [hep-ph]}%
  \bibAnnoteFile{NoStop}{Cheung:2012qy}%
\bibitem{Huang:2014xua}%
  \BibitemOpen
  \bibfield{author}{%
  \bibinfo {author} {\bibfnamefont{P.}~\bibnamefont{Huang}}\ and\ \bibinfo
  {author} {\bibfnamefont{C.~E.~M.}\ \bibnamefont{Wagner}},\ }%
  \bibfield{journal}{%
  \Doi{10.1103/PhysRevD.90.015018}{\bibinfo {journal} {Phys. Rev.}}\ }%
  \textbf{\bibinfo {volume} {D90}},\ \bibinfo {pages} {015018} (\bibinfo {year}
  {2014}),\ \Eprint{http://arxiv.org/abs/1404.0392}{arXiv:1404.0392 [hep-ph]}%
  \bibAnnoteFile{NoStop}{Huang:2014xua}%
\bibitem{Cheung:2014lqa}%
  \BibitemOpen
  \bibfield{author}{%
  \bibinfo {author} {\bibfnamefont{C.}~\bibnamefont{Cheung}}, \bibinfo {author}
  {\bibfnamefont{M.}~\bibnamefont{Papucci}}, \bibinfo {author}
  {\bibfnamefont{D.}~\bibnamefont{Sanford}}, \bibinfo {author}
  {\bibfnamefont{N.~R.}\ \bibnamefont{Shah}},\ and\ \bibinfo {author}
  {\bibfnamefont{K.~M.}\ \bibnamefont{Zurek}},\ }%
  \bibfield{journal}{%
  \Doi{10.1103/PhysRevD.90.075011}{\bibinfo {journal} {Phys.Rev.}}\ }%
  \textbf{\bibinfo {volume} {D90}},\ \bibinfo {pages} {075011} (\bibinfo {year}
  {2014}),\ \Eprint{http://arxiv.org/abs/1406.6372}{arXiv:1406.6372 [hep-ph]}%
  \bibAnnoteFile{NoStop}{Cheung:2014lqa}%
\bibitem{Caron:2015wda}%
  \BibitemOpen
  \bibfield{author}{%
  \bibinfo {author} {\bibfnamefont{A.}~\bibnamefont{Achterberg}}, \bibinfo
  {author} {\bibfnamefont{S.}~\bibnamefont{Caron}}, \bibinfo {author}
  {\bibfnamefont{L.}~\bibnamefont{Hendriks}}, \bibinfo {author}
  {\bibfnamefont{R.}~\bibnamefont{Ruiz~de Austri}},\ and\ \bibinfo {author}
  {\bibfnamefont{C.}~\bibnamefont{Weniger}}}%
   (\bibinfo {year} {2015}),\
  \Eprint{http://arxiv.org/abs/1502.05703}{arXiv:1502.05703 [hep-ph]}%
  \bibAnnoteFile{NoStop}{Caron:2015wda}%
\bibitem{Young:2009zb}%
  \BibitemOpen
  \bibfield{author}{%
  \bibinfo {author} {\bibfnamefont{R.~D.}\ \bibnamefont{Young}}\ and\ \bibinfo
  {author} {\bibfnamefont{A.~W.}\ \bibnamefont{Thomas}},\ }%
  \bibfield{journal}{%
  \Doi{10.1103/PhysRevD.81.014503}{\bibinfo {journal} {Phys. Rev.}}\ }%
  \textbf{\bibinfo {volume} {D81}},\ \bibinfo {pages} {014503},\
  \Eprint{http://arxiv.org/abs/0901.3310}{arXiv:0901.3310 [hep-lat]}%
  \bibAnnoteFile{NoStop}{Young:2009zb}%
\bibitem{Oksuzian:2012rzb}%
  \BibitemOpen
  \bibfield{author}{%
  \bibinfo {author} {\bibfnamefont{H.}~\bibnamefont{Ohki}}, \bibinfo {author}
  {\bibfnamefont{K.}~\bibnamefont{Takeda}}, \bibinfo {author}
  {\bibfnamefont{S.}~\bibnamefont{Aoki}}, \bibinfo {author}
  {\bibfnamefont{S.}~\bibnamefont{Hashimoto}}, \bibinfo {author}
  {\bibfnamefont{T.}~\bibnamefont{Kaneko}}, \bibinfo {author}
  {\bibfnamefont{H.}~\bibnamefont{Matsufuru}}, \bibinfo {author}
  {\bibfnamefont{J.}~\bibnamefont{Noaki}},\ and\ \bibinfo {author}
  {\bibfnamefont{T.}~\bibnamefont{Onogi}} (\bibinfo {collaboration} {JLQCD}),\
  }%
  \bibfield{journal}{%
  \Doi{10.1103/PhysRevD.87.034509}{\bibinfo {journal} {Phys. Rev.}}\ }%
  \textbf{\bibinfo {volume} {D87}},\ \bibinfo {pages} {034509} (\bibinfo {year}
  {2013}),\ \Eprint{http://arxiv.org/abs/1208.4185}{arXiv:1208.4185 [hep-lat]}%
  \bibAnnoteFile{NoStop}{Oksuzian:2012rzb}%
\bibitem{Durr:2011mp}%
  \BibitemOpen
  \bibfield{author}{%
  \bibinfo {author} {\bibfnamefont{S.}~\bibnamefont{Durr}} \emph{et~al.},\ }%
  \bibfield{journal}{%
  \Doi{10.1103/PhysRevD.85.014509}{\bibinfo {journal} {Phys. Rev.}}\ }%
  \textbf{\bibinfo {volume} {D85}},\ \bibinfo {pages} {014509} (\bibinfo {year}
  {2012}),\ \Eprint{http://arxiv.org/abs/1109.4265}{arXiv:1109.4265 [hep-lat]}%
  \bibAnnoteFile{NoStop}{Durr:2011mp}%
\bibitem{Junnarkar:2013ac}%
  \BibitemOpen
  \bibfield{author}{%
  \bibinfo {author} {\bibfnamefont{P.}~\bibnamefont{Junnarkar}}\ and\ \bibinfo
  {author} {\bibfnamefont{A.}~\bibnamefont{Walker-Loud}},\ }%
  \bibfield{journal}{%
  \Doi{10.1103/PhysRevD.87.114510}{\bibinfo {journal} {Phys.Rev.}}\ }%
  \textbf{\bibinfo {volume} {D87}},\ \bibinfo {pages} {114510} (\bibinfo {year}
  {2013}),\ \Eprint{http://arxiv.org/abs/1301.1114}{arXiv:1301.1114 [hep-lat]}%
  \bibAnnoteFile{NoStop}{Junnarkar:2013ac}%
\bibitem{Martin:2009iq}%
  \BibitemOpen
  \bibfield{author}{%
  \bibinfo {author} {\bibfnamefont{A.}~\bibnamefont{Martin}}, \bibinfo {author}
  {\bibfnamefont{W.}~\bibnamefont{Stirling}}, \bibinfo {author}
  {\bibfnamefont{R.}~\bibnamefont{Thorne}},\ and\ \bibinfo {author}
  {\bibfnamefont{G.}~\bibnamefont{Watt}},\ }%
  \bibfield{journal}{%
  \Doi{10.1140/epjc/s10052-009-1072-5}{\bibinfo {journal} {Eur.Phys.J.}}\ }%
  \textbf{\bibinfo {volume} {C63}},\ \bibinfo {pages} {189} (\bibinfo {year}
  {2009}),\ \Eprint{http://arxiv.org/abs/0901.0002}{arXiv:0901.0002 [hep-ph]}%
  \bibAnnoteFile{NoStop}{Martin:2009iq}%
\bibitem{Cheng:2012qr}%
  \BibitemOpen
  \bibfield{author}{%
  \bibinfo {author} {\bibfnamefont{H.-Y.}\ \bibnamefont{Cheng}}\ and\ \bibinfo
  {author} {\bibfnamefont{C.-W.}\ \bibnamefont{Chiang}},\ }%
  \bibfield{journal}{%
  \Doi{10.1007/JHEP07(2012)009}{\bibinfo {journal} {JHEP}}\ }%
  \textbf{\bibinfo {volume} {07}},\ \bibinfo {pages} {009} (\bibinfo {year}
  {2012}),\ \Eprint{http://arxiv.org/abs/1202.1292}{arXiv:1202.1292 [hep-ph]}%
  \bibAnnoteFile{NoStop}{Cheng:2012qr}%
\bibitem{Akerib:2013tjd}%
  \BibitemOpen
  \bibfield{author}{%
  \bibinfo {author} {\bibfnamefont{D.}~\bibnamefont{Akerib}} \emph{et~al.}
  (\bibinfo {collaboration} {LUX Collaboration})}%
   (\bibinfo {year} {2013}),\
  \Eprint{http://arxiv.org/abs/1310.8214}{arXiv:1310.8214 [astro-ph.CO]}%
  \bibAnnoteFile{NoStop}{Akerib:2013tjd}%
\bibitem{Cushman:2013zza}%
  \BibitemOpen
  \bibfield{author}{%
  \bibinfo {author} {\bibfnamefont{P.}~\bibnamefont{Cushman}} \emph{et~al.},\
  }%
  in\ \emph{\bibinfo {booktitle} {{Community Summer Study 2013: Snowmass on the
  Mississippi (CSS2013) Minneapolis, MN, USA, July 29-August 6, 2013}}}\
  (\bibinfo {year} {2013})\
  \Eprint{http://arxiv.org/abs/1310.8327}{arXiv:1310.8327 [hep-ex]},\
  \url{http://inspirehep.net/record/1262767/files/arXiv:1310.8327.pdf}%
  \bibAnnoteFile{NoStop}{Cushman:2013zza}%
\bibitem{Ackermann:2015zua}%
  \BibitemOpen
  \bibfield{author}{%
  \bibinfo {author} {\bibfnamefont{M.}~\bibnamefont{Ackermann}} \emph{et~al.}
  (\bibinfo {collaboration} {Fermi-LAT})}%
   (\bibinfo {year} {2015}),\
  \Eprint{http://arxiv.org/abs/1503.02641}{arXiv:1503.02641 [astro-ph.HE]}%
  \bibAnnoteFile{NoStop}{Ackermann:2015zua}%
\bibitem{Ruppin:2014bra}%
  \BibitemOpen
  \bibfield{author}{%
  \bibinfo {author} {\bibfnamefont{F.}~\bibnamefont{Ruppin}}, \bibinfo {author}
  {\bibfnamefont{J.}~\bibnamefont{Billard}}, \bibinfo {author}
  {\bibfnamefont{E.}~\bibnamefont{Figueroa-Feliciano}},\ and\ \bibinfo {author}
  {\bibfnamefont{L.}~\bibnamefont{Strigari}},\ }%
  \bibfield{journal}{%
  \Doi{10.1103/PhysRevD.90.083510}{\bibinfo {journal} {Phys.Rev.}}\ }%
  \textbf{\bibinfo {volume} {D90}},\ \bibinfo {pages} {083510} (\bibinfo {year}
  {2014}),\ \Eprint{http://arxiv.org/abs/1408.3581}{arXiv:1408.3581 [hep-ph]}%
  \bibAnnoteFile{NoStop}{Ruppin:2014bra}%
\bibitem{pdg}%
  \BibitemOpen
  \bibfield{author}{%
  \bibinfo {author} {\bibfnamefont{K.}~\bibnamefont{Olive}} \emph{et~al.}
  (\bibinfo {collaboration} {Particle Data Group}),\ }%
  \bibfield{journal}{%
  \Doi{10.1088/1674-1137/38/9/090001}{\bibinfo {journal} {Chin.Phys.}}\ }%
  \textbf{\bibinfo {volume} {C38}},\ \bibinfo {pages} {090001} (\bibinfo {year}
  {2014})%
  \bibAnnoteFile{NoStop}{pdg}%
\end{thebibliography}%

\end{document}